\documentstyle[preprint,aps]{revtex}
\begin{document}
\draft

\title
{Spinning Particles on Spacelike Hypersurfaces and their Rest-Frame 
Description.}

\author{Francesco Bigazzi}

\address
{Dipartimento di Fisica\\
Universita' di Milano\\
Via G.Celoria 16\\
20133 Milano, Italy}

\author{and}

\author{Luca Lusanna}

\address
{Sezione INFN di Firenze\\
L.go E.Fermi 2 (Arcetri)\\
50125 Firenze, Italy\\
E-mail LUSANNA@FI.INFN.IT}

\maketitle

\begin{abstract}

A new spinning particle with a definite sign of the energy is defined
on spacelike hypersurfaces after a critical discussion of the standard spinning 
particles. They are the pseudoclassical basis of the positive energy
$({1\over 2},0)$ [or negative energy $(0,{1\over 2})$] parts of the 
$({1\over 2},{1\over 2})$ solutions of the Dirac equation. The
study of the isolated system of N such spinning charged particles plus the 
electromagnetic field leads to their description in
the rest-frame Wigner-covariant instant form of
dynamics on the Wigner hyperplanes orthogonal to the total 4-momentum of the
isolated system (when it is timelike). We find that on such hyperplanes these 
spinning particles have a nonminimal coupling only of the type ``spin-magnetic 
field" like the nonrelativistic Pauli particles to which they tend in the
nonrelativistic limit. The Lienard-Wiechert potentials associated with
these charged spinning particles are found.
Then, a comment on how to quantize the spinning particles 
respecting their fibered structure describing the spin structure is done.

\vskip 1truecm
\noindent July 1998
\vskip 1truecm

\end{abstract}
\vfill\eject

\section
{Introduction}

A series of papers\cite{dira,lusa,lv1,lv2,lv3} 
utilizing the Shanmugadhasan canonical transformation \cite{sha}
was needed to arrive at the classical noncovariant generalized Coulomb
gauge for the standard SU(3)xSU(2)xU(1) model with Grassmann-valued fermion 
fields [see Refs.\cite{re} for reviews]. In this gauge there is no gauge freedom
left and the physical fields are the Dirac observables of the model. In order
to covariantize the result, the isolated system of N scalar massive particles
with Grassmann-valued electric charges plus the electromagnetic field
\cite{lus1,lus2} was reformulated on spacelike hypersurfaces
\cite{dirac1,kuchar},
the leaves of a foliation of Minkowski spacetime defining an arbitrary 3+1
splitting of it. The degrees of freedom of the spacelike hypersurfaces 
$z^{\mu}(\tau ,\vec \sigma )$ [$\tau$ labels the leaves $\Sigma_{\tau}$ of the
foliation, while $\vec \sigma$ are curvilinear coordinates on $\Sigma_{\tau}$]
are extra configurational variables, but there are 4 first class constraints 
in each point ensuring that the description is independent from the foliation. 
Therefore, one can restrict oneself to spacelike hyperplanes, so that these
extra degrees of freedom are restricted to only 10: i) an origin $x^{\mu}
_s(\tau )$; ii) an orthonormal tetrad on this point, adapted to $\Sigma_{\tau}$
containing the normal to $\Sigma_{\tau}$. Only 10 global first class 
constraints are left implying the independence of the description from the
choice of the hyperplanes. For the dense subset of system configurations having 
a total timelike 4-momentum (the boundary conditions on the fields must be such
that the 10 conserved Poincar\'e generators are finite) one can make the 
further restriction to those special hyperplanes (denoted Wigner hyperplanes),
which are orthogonal to the total 4-momentum of the isolated system. This is 
done by requiring that the orthonormal tetrad coincides with the standard
Wigner boost associated with the timelike total 4-momentum, after having 
boosted at rest all the variables with it. The extra degrees of freedom are
reduced to only 4 and only 4 global first class constraints are left. The 4
degrees of freedom are a noncovariant canonical variable ${\tilde x}^{\mu}
_s(\tau )$ [with conjugate momentum $p^{\mu}_s$], which behaves as a classical
analogue of the Newton-Wigner position operator with the covariance of the
little group of timelike Poincar\'e orbits. All the variables describing
the isolated system are forced to become Wigner covariant\cite{wigner,longhi}.
This construction defines a new kind of instant form of the dynamics
\cite{dirac2}, the ``rest-frame 1-time Wigner-covariant instant form", which
is the special relativistic generalization of the Newtonian separation of the 
center of mass [$H={{{\vec P}^2}\over {2M}}+H_{rel}$]. One of the 4 left
constraints [$\epsilon_s=\pm \sqrt{p^2_s}\approx \pm M_{system}$] says that the
analogue of the relative nonrelativistic Hamiltonian $H_{rel}$ is the
invariant mass $M_{system}$ of the isolated system, so that the natural 
evolution parameter is obtained by identifying the label $\tau$ of the leaves
$\Sigma_{\tau}$ of the foliation with the rest-frame center-of-mass time
$T_s=p_s\cdot {\tilde x}_s/\epsilon_s=p_s\cdot x_s/\epsilon_s$. The other 
three constraints say that the total Wigner spin 1${}$ 3-momentum 
${\vec \kappa}_{+\, system}$ of the 
isolated system inside the Wigner hyperplane vanishes, so that the Wigner
hyperplane is the intrinsic rest-frame of the isolated system. Since the
position of the origin of the hyperplane is arbitrary [like the unit vector
$p^{\mu}_s/\epsilon_s$, describing the orientation of the Wigner hyperplane 
with respect to an arbitrary Lorentz frame in Minkowski spacetime], one 
imposes as gauge-fixings to these 3 constraints the condition that the center 
of mass of the isolated system coincides with the origin: $X^{\mu}_{system}
(\tau )=z^{\mu}(\tau ,{\vec \eta}_{+\, system}(\tau ))
\approx x^{\mu}_s(\tau )=z^{\mu}(\tau ,\vec \sigma =0)$ or ${\vec \eta}
_{+\, system}(\tau )\approx 0$. However, till 
now $X^{\mu}_{system}$ is known only for systems of particles; there are
preliminary results on its identification for Klein-Gordon fields
\cite{lm}.

If one makes the canonical reduction of the gauge degrees of freedom of the
isolated system in the rest-frame instant form on the Wigner hyperplane, one 
gets the rest-frame Wigner-covariant generalized Coulomb gauge in which the
universal breaking of covariance is restricted to the decoupled center-of-mass
variable. However, as shown in Refs.\cite{lusa,lus1,lus3}, the region of 
spacetime, over which this noncovariance is spread, is finite in spacelike 
directions and identifies a classical intrinsic unit of length, the M\o ller 
radius $\rho =\sqrt{-W^2}/P^2=|{\hat {\vec S}}| /\sqrt{P^2}$, where $P^2 > 0$ 
and $W^2=-P^2 {\hat {\vec S}}^2$ are the Poincar\'e Casimirs and ${\hat {\vec 
S}}$ the Thomas rest-frame spin of the isolated system respectively. This
unit of length gives rise to a physical intrinsic ultraviolet cutoff at the
quantum level in the spirit of Dirac and Yukawa.

After the canonical reduction of the isolated system of N charged particles
plus the electromagnetic field (for this system also the Lienard-Wiechert
potential in the Coulomb gauge has been evaluated \cite{lus2}) one gets the
emergence of the Coulomb potential from field theory and a regularization
of the Coulomb self-energy due to the Grassmann character of the electric
charge of the particles. Then, the isolated system of N scalar massive
particles with Grassmann-valued color charges plus the SU(3) color Yang-Mills 
field has been studied till to find the rest-frame Wigner-covariant Coulomb
gauge\cite{lus3}, obtaining a pseudoclassical description of the relativistic 
quark model. The physical invariant mass of the system is
given in terms of the Dirac observables. From the reduced Hamilton equations,  
the second order equations of motion both for the reduced transverse color 
field and the particles are extracted. Then, one studies  the N=2 
(meson) case. A special form of the requirement of having only color singlets, 
suited for a field-independent quark model, produces a ``pseudoclassical 
asymptotic freedom" and a regularization of the quark self-energy. With these
results one can covariantize the bosonic part of the standard model given in
Ref.\cite{lv3}.

The limitation of the description of relativistic particles on
spacelike hypersurfaces is that they must have a well defined sign of the 
energy. This is due to the fact that the intersection of a timelike worldline 
with a spacelike hypersurface is defined by 3 numbers $\vec \sigma ={\vec \eta}
_i(\tau )$ and not 4: $x^{\mu}_i(\tau )=z^{\mu}(\tau ,{\vec \eta}_i(\tau ))$,
i=1,..,N. This means that the mass shell constraints $p_i^2-m^2_i\approx 0$
(in the free case) have been solved, $p^o_i\approx \pm 
\sqrt{{\vec p}_i^2+m_i^2}$,
and that a choice of the sign $\eta_i=\pm$ of the energy [namely on which of
the two topologically disjoined branches of the mass-hyperboloid is the
particle] has been done. These models describe only one of the $2^N$
branches of the mass spectrum of the N particle system (there is a different 
model for each branch), before that the interactions become so strong to cause
branch-crossing (zero energy gap; pair production at the quantum level). 
Therefore, these models allow a genuine consistent formulation of
``relativistic particle mechanics" with a definite sign of the energy of the 
particles, with relativistic kinetic energies $\sqrt{m^2_i+{\vec \kappa}_i^2}$
and already oriented to the canonical formulation of general relativity.
Since the mass spectrum has only one branch, there is no problem with pair
production. In a sense this is a consistent classical realization of the 
quenched approximation in lattice gauge theory: due to the Grassmann character
of the charges one does not only ignore the fermion loops but also all the 
effects of the same order in the charges. 
The only problem, present with scalar interactions modifying the mass
($\sqrt{m^2_i+{\vec \kappa}^2_i}\, \mapsto \sqrt{m^2_i+V+{\vec \kappa}^2_i}$),
is that the modified kinetic energy must remain real. All the consistent
couplings to the electromagnetic field on spacelike hypersurfaces are in accord 
with quenched approximation (no pair production). See
Ref.\cite{lus2} for the connection of this description of N scalar particle
systems with the Feshbach-Villars diagonalization of the Klein-Gordon equation
\cite{fv}, which confirms what has been said about the couplings to the 
electromagnetic field.

What is still lacking is the quantization of this approach to get a consistent
``quantum relativistic mechanics", due to the complications of the square
root in the kinetic term. However, the starting point, i.e. the quantization
of the free particle, has been realized in Ref.\cite{lamm} by using 
pseudodifferential operators in a scheme consistent with the reformulation
on spacelike hypersurfaces [see Refs.\cite{sqroot} for what is known with
a Coulomb potential added to the kinetic term]: the resulting nonlocal
quantum theory may also be second quantized with microcausality replaced by
macrocausality [the commutator of two field operators at spacelike separation
decreases exponentially to zero when the spacelike separation exceeds the
Compton wavelength in a qualitative accord with the problematic of the
M\"oller radius].

To get the description of the standard model on spacelike hypersurfaces one
still needs the formulation of fermions on them. This is also needed for
treating the fermions in general relativity: given their coupling to tetrad
gravity [see for instance Refs.\cite{wei,naka}; the canonical reduction of 
tetrad gravity in under investigation\cite{dplr}] one needs this formulation
to arrive at the ADM canonical formalism based on 3+1 splittings of the globally
hyperbolic asymptotically flat spacetime [see Refs.\cite{hen} for what is
known on the subject].

Before studying Dirac fermion fields, in this paper we shall give the
description of the positive energy spinning particle on spacelike
hypersurfaces, with the spin described by Grassmann variables. It corresponds
to study the pseudoclassical basis of either the positive or negative
solutions of the Dirac equation in first quantization. The nonrelativistic
limit corresponds to a Pauli spinning particle\cite{levy,gomis}.

In Section II we review the standard spinning particles with the spin
described by Grassmann variables, giving some properties of their 
pseudoclassical description and a discussion of how to separate the two 
branches of their mass spectrum.

In Section III we study N spinning particles with Grassmann-valued electric 
charges plus the electromagnetic field on spacelike hypersurfaces and we find
the form of the lacking odd second class constraints needed to get 
that their spin structure is described by Pauli 
matrices after quantization.

In Section IV the previous system is restricted firstly to spacelike 
hyperplanes and secondly to the Wigner ones, obtaining the rest-frame 
description in which the spinning particles have nonminimal couplings only of 
the type ``spin - magnetic field" like the nonrelativistic Pauli particles.
The nonrelativistic limit is shown explicitly.

In Section V the Dirac observables with respect to electromagnetic gauge 
transformations and their equations of motion are obtained.

In Section VI the Lienard-Wiechert potentials produced by these charged 
spinning particles in the rest-frame description are evaluated.

In the Conclusions we introduce the problem of how to quantize spinning 
particles preserving their fibered spin structure, namely a spin structure
(described by Grassmann variables) over a scalar particle tracing a worldline
in Minkowski spacetime [in analogy to the pseudoclassical photon \cite{photon}].
It seems that there is an incompleteness in our quantum description of fermions
(either first quantized Dirac wave functions or second quantized Dirac fields)
regarding their ability to generate not only a spin structure but also
trajectories in Minkowski spacetime like the ones of charged leptons in 
particle detectors (or more in general the spacetime trajectories of electric 
currents).  This problem will be treated elsewhere.

In the Appendix there is a review of the Foldy-Wouthuysen, Cini-Touschek and 
chiral representations of Dirac matrices from the pseudoclassical point of
view.

\vfill\eject

\section{Old Massive Spinning Particles.}

The use of Grassmann variables for the spin of the spinning particles started 
with Berezin and Marinov\cite{bm}. In the case of spin 1/2 one speaks of 
pseudoclassical description of the electron. Then, the spinning particle was 
described in various forms \cite{casal,spinn,spinn1} all utilizing Grassmann 
variables. We shall not consider either other descriptions of spin 1/2 with 
tools different from Grassmann variables or models for spinning particles using 
bosonic variables like rotators (only integer spins, usually towers of spins,
are obtained at the quantum level).

In Ref.\cite{casal}, a Lagrangian was given depending on a bosonic
position $x^{\mu}(\tau )$ and on 5 real Grassmann variables $\xi^{\mu}(\tau )$ 
and $\xi_5(\tau )$: 

\begin{equation}
L = - \frac i2 \xi_5 \dot{\xi}_5 - \frac i2 \xi_\mu 
\dot{\xi}^\mu -m \sqrt{\Big( \dot{x}_\mu -\frac im \xi_\mu \dot{\xi}_5
\Big)^2},\quad S=\int d\tau L,
\label{II1}
\end{equation}

\noindent where ${\dot x}^{\mu}={{\partial}\over {\partial \tau}}x^{\mu}$; 
the canonical momenta are $p_\mu= - \frac{\partial L}{\partial \dot{x}^\mu} =
m \frac{\dot{x}_\mu -  \frac im \xi_\mu \dot{\xi}_5}
{\sqrt{\Big( \dot{x}_\mu -\frac im \xi_\mu \dot{\xi}_5
\Big)^2}}$,
$\pi_\mu =  \frac{\partial L}{\partial \dot{\xi}^\mu} =
\frac i2 \xi_\mu$,
$\pi_5 =  \frac{\partial L}{\partial \dot{\xi}_5} =
\frac i2 \xi_5 -\frac im p_\mu \xi^\mu$.
There is a local supersymmetry invariance of the action (the local
transformations are 
$\xi_\mu \longrightarrow \xi_\mu +\epsilon_5(\tau )p_{\mu}/m$,
$\xi_5 \longrightarrow  \xi_5 +\epsilon_5(\tau )$,
$x_\mu \longrightarrow x_\mu+{i\over {m^2}}\epsilon_5(\tau )p_{\mu}\xi_5  
- \frac{i}{m} \epsilon_5(\tau )[\xi_{\mu}-{1\over m}p_{\mu}\xi_5]$ ) but
no manifest world-line supersymmetry. The geometrical interpretation is 
that one has a fibered structure: over each point of the timelike worldline of 
a scalar point particle in Minkowski spacetime there is a Grassmann algebra 
$G_5$ with generators $\xi^{\mu}$, $\xi_5$, as a standard fiber describing the
spin structure.

In phase space, after the elimination of the second class constraints
(the Lagrangian is of first order in Grassmann velocities) $\chi_{\mu}=\pi
_{\mu}-{i\over 2}\xi_{\mu}\approx 0$ and $\chi_5=\pi_5+{i\over 2}\xi_5\approx
0$ (this last constraint does not appear explicitly but it is a constant of
the motion which has to be put equal zero by hand 
as required by the number of zero eigenvalues of the Hessian matrix and to get
the same set of solutions from the Euler-Lagrange and Hamilton equations), one
arrives at the Dirac brackets $\{ x^{\mu},p^{\nu} \} {}^{*}=-\eta^{\mu\nu}$,
$\{ \xi^{\mu},\xi^{\nu} \} {}^{*}=i\eta^{\mu\nu}$, $\{ \xi_5,\xi_5 \} {}^{*}=
-i$, from the original Poisson brackets $\{ x^{\mu},p^{\nu} \} =\{ \xi^{\mu},
\pi^{\nu} \} =-\eta^{\mu\nu}$, $\{ \pi_5,\xi_5 \} =-1$ [the odd-odd Poisson 
brackets are symmetric]. After quantization\cite{ca}, these Dirac brackets 
become $\big[ \xi_\mu , \xi_\nu \big]_+ = - \hbar \eta_{\mu \nu}$, 
$\big[ \xi_\mu , \xi_5 \big]_+ = 0$, $\big[ \xi_5 , \xi_5 \big]_+ = \hbar$, so
that the Grassmann variables give rise to the  Dirac matrices 
$\sqrt{{{\hbar}\over 2}}\gamma_5\gamma^{\mu}$ and $\sqrt{{{\hbar}\over 2}}
\gamma_5$ respectively.

There are two first class constraints: i) $\chi =p^2-
m^2\approx 0$ (from repametrization invariance); ii) $\chi_D=p_{\mu}\xi^{\mu}-
m\xi_5\approx 0$ (from the local supersymmetry invariance); with the only
nonvanishing Poisson bracket $\{ \chi_D,\chi_D \} {}^{*}=i\chi$ (in this sense
$\chi_D$ is the square root of $\chi$). The Dirac Hamiltonian is $H_D=\lambda 
(\tau ) \chi +i\lambda_D(\tau ) \chi_D$, with $\lambda_D(\tau )$ an odd real 
Dirac multiplier and one has ${\dot \xi}^{\mu}\, {\buildrel \circ \over =}\, 
\{ \xi^{\mu},H_D\}{}^{*} =\lambda_D(\tau ) p^{\mu}$, ${\dot \xi}_5\, {\buildrel 
\circ \over =}\, \lambda_D(\tau ) m$ showing that both $\xi^{\mu}$ and $\xi_5$ 
are gauge dependent (${\buildrel \circ 
\over =}$ means evaluated on the solutions).
The conserved Poincar\'e generators are
$p_{\mu}$, $J_{\mu \nu} = L_{\mu \nu} + S_{\mu \nu}$,
$L_{\mu \nu} = - p_\mu x_\nu + p_\nu x_\mu$,
$S_{\mu \nu} = -\pi_\mu \xi_\nu + \pi_\nu \xi_\mu
\stackrel{\chi_\mu \equiv \chi_5 \equiv 0} = -i \xi_\mu \xi_\nu$
[the components of the Lorentz generators satisfy 
$\{ L^{\mu \nu} , L^{\alpha \beta} \} {}^{*}= C^{\mu \nu \alpha \beta}_{\gamma
\delta} L^{\gamma \delta}$,
$\{ S^{\mu \nu} , S^{\alpha \beta} \} {}^{*} = C^{\mu \nu \alpha \beta}_{\gamma
\delta} S^{\gamma \delta}$, where $C^{\mu \nu \alpha \beta}_{\gamma
\delta}$ are the structure constants of the Lorentz algebra] with the spin
3-vector $S_i = \frac 12 \epsilon_{ijk} S_{jk} \Rightarrow
\vec{S} = - \frac i2 \vec{\xi} \wedge \vec{\xi}$. Since ${\dot S}^{\mu\nu}\,
{\buildrel \circ \over =}\, i\lambda_D(\tau ) (\xi^{\mu}p^{\nu}-\xi^{\nu}p
^{\mu})$, the spin tensor is gauge dependent, as it should because its boost 
part cannot be independent from the spin part for a point particle with a 
definite spin. 

Concerning the position $x^{\mu}$, its equations of motion are ${\dot x}^{\mu}\,
{\buildrel \circ \over =}\, \{ x^{\mu},H_D \} {}^{*}= -2\lambda (\tau ) p^{\mu}
-i \lambda_D(\tau ) \xi^{\mu}$. This shows that superimposed to the free motion
with velocity proportional to the 4-momentum there is a zitterbewegung 
proportional to $\xi^{\mu}$. The pseudoclassical Foldy-Wouthuysen mean position
\cite{fw} (see also Ref.\cite{ellis})
is $x^o_M=x^o$, $x^i_M=x^i-{{p^ip_k+\delta^i_k \sqrt{{\vec p}^2+m^2}
(\sqrt{{\vec p}^2+m^2}+m)}\over {({\vec p}^2+m^2)(\sqrt{{\vec p}^2+m^2}+m)}}
i\xi^k\xi_5- {{i\xi^i\vec p\cdot \vec \xi}\over {\sqrt{{\vec p}^2+m^2}
(\sqrt{{\vec p}^2+m^2}+m)}}$, because ${\dot x}^i_M\, {\buildrel \circ \over 
=}\, -[2\lambda (\tau ) +i\lambda_D(\tau ){{m\xi_5+\vec p\cdot \vec \xi}\over 
{{\vec p}^2+m^2}}] p^i$, ${\dot x}^o_M\, {\buildrel \circ \over =}\, 
-2\lambda (\tau ) p^o-i\lambda_D(\tau )\xi^o$; if we choose $\lambda_D(\tau )=
\alpha (\tau )\xi^o$, so that $\xi^o(p_{\mu}\xi^{\mu}-m\xi_5) \mapsto \hat H=
\vec p\cdot \vec \alpha +m\beta$ [see Refs.\cite{casal,sp3}], 
we have ${\dot x}^o
_M\, {\buildrel \circ \over =}\, -2\lambda (\tau )p^o$. The spin tensor has 
the following two decompositions $J^{\mu\nu}=L^{\mu\nu}+S^{\mu\nu}=L_M^{\mu\nu}
+S^{\mu\nu}_M$ with $L_M^{\mu\nu}=x^{\mu}_Mp^{\nu}-x^{\nu}_Mp^{\mu}$ and
$S^{oi}_M=S^{oi}-p^o[{{p^ip_k+\delta^i_k \sqrt{{\vec p}^2+m^2}
(\sqrt{{\vec p}^2+m^2}+m)}\over {({\vec p}^2+m^2)(\sqrt{{\vec p}^2+m^2}+m)}}
i\xi^k\xi_5+ {{i\xi^i\vec p\cdot \vec \xi}\over {\sqrt{{\vec p}^2+m^2}
(\sqrt{{\vec p}^2+m^2}+m)}}]$, $S^{ij}_M=S^{ij}+{{\delta^i_kp^j-\delta^j_kp^i}
\over {\sqrt{{\vec p}^2+m^2}}} i\xi^k\xi_5+{{i(\xi^ip^j-\xi^jp^i) \vec p\cdot
\vec \xi}\over {\sqrt{{\vec p}^2+m^2}(\sqrt{{\vec p}^2+m^2}+m)}}$
(it is the Foldy-Wouthuysen mean spin tensor). While $L_M
^{\mu\nu}$ and $S_M^{\mu\nu}$ are both constants of the motion 
[for $L^{oi}_M$ and $S^{oi}_M$  the choice $\lambda_D(\tau )=\alpha (\tau )
\xi^o$ is needed], neither
$L^{\mu\nu}$ nor $S^{\mu\nu}$ are constants due to zitterbewegung. See chapter
2 section C of Ref.\cite{grandy}, chapter 1 sections 1.6, 1.7, 1.8 of Ref.
\cite{thaller} and Ref.\cite{cost}.

At the quantum level the 
constraint $\chi_D$ becomes the Dirac equation $\gamma_5(\gamma^{\mu}{\hat p}
_{\mu}-m) \psi (x)=0$  [${\hat p}_{\mu}=i\partial_{\mu}$; from now on $\hat {}$
will denote operators], whose square 
$({\hat p}^2-m^2)\psi (x)=0$ is consistent with the quantization of the
constraint $\chi$. However, it is disturbing that no trace of the 
pseudoclassical fibration describing the spin structure is present in the
quantum description: the Dirac spinor naturally describes only the spin 
structure and does not seem able to trace a worldline in Minkowski
spacetime.

Since on spacelike hypersurfaces (and, in particular, on the Wigner hyperplanes
orthogonal to the 4-momentum and defining the rest frame)
one must choose the sign of the energy of the
spinning particle and since to describe a $({1\over 2},0)$ particle one needs
to reduce the Clifford algebra $C_5$ associated with the Dirac matrices to a 
Clifford algebra $C_3$ associated with Pauli matrices, let us look for a 
canonical transformation to the rest frame which could help in this job.
Since we can restrict ourselves to timelike $p_{\mu}$ due to $\chi \approx 0$,
we can use the standard Wigner boost for timelike Poincar\'e orbits
\cite{longhi} $L^{\mu}{}_{\nu}(p,{\buildrel o \over p})=\eta^{\mu}_{\nu}+
2{{p^{\mu}{\buildrel o \over p}_{\nu}}\over {\epsilon^2}}-{{(p^{\mu}+
{\buildrel o \over p}^{\mu})(p_{\nu}+{\buildrel o \over p}_{\nu})}\over 
{\epsilon (\epsilon +p_o)}}$, where ${\buildrel o \over p}^{\mu}=(\epsilon ,
\vec 0)$, $\epsilon =\eta \sqrt{p^2}$, $\eta =sign\, p_o=\pm 1$. If we denote
$\epsilon^{\mu}_A(u(p))=L^{\mu}{}_A(p,{\buildrel o \over p})$ with 
$\epsilon^{\mu}_o(u(p))=u^{\mu}(p)=
p^{\mu}/\epsilon$, this new canonical transformation maps the canonical basis
$x^{\mu}$, $p^{\mu}$, $\xi^{\mu}$, $\xi_5$, in the new canonical basis
(see Refs.\cite{longhi,lus1} for the definition of ${\tilde x}^{\mu}$)

\begin{eqnarray}
&&{\tilde x}^{\mu}=x^{\mu}+{1\over 2}\epsilon^{\nu}_A(u(p)) \eta^{AB} {{\partial
\epsilon^{\rho}_B(u(p))}\over {\partial p_{\mu}}} S_{\rho\nu}=\nonumber \\
&&=x^{\mu}-{1\over {\epsilon (p_o+\epsilon )}}[p_{\nu}S^{\nu\mu}+\epsilon 
(S^{o\mu}-S^{o\nu}{{p_{\nu}p^{\mu}}\over {\epsilon^2}})],\nonumber \\
&&p_{\mu}=p_{\mu},\nonumber \\
&&{\tilde \xi}_{\tau}={{p_{\mu}\xi^{\mu}}\over {\epsilon}},\nonumber \\
&&{\tilde \xi}_r=\epsilon^{\mu}_r(u(p))\xi_{\mu},\nonumber \\
&&\xi_5=\xi_5,\nonumber \\
&&{}\nonumber \\
&&\{ {\tilde x}^{\mu},p^{\nu} \} {}^{*}=-\eta^{\mu\nu},\quad\quad
\{ \xi_5,\xi_5 \} =-i,\nonumber \\
&&\{ {\tilde \xi}_{\tau},{\tilde \xi}_{\tau} \} {}^{*}=i,\quad\quad
\{ {\tilde \xi}_r,{\tilde \xi}_s \} {}^{*}= -i\delta_{rs}.
\label{II2}
\end{eqnarray}

\noindent As shown in Ref.\cite{longhi}, ${\tilde \xi}_r$ is a Wigner spin 1
${}$ 3-vector. The rest-frame spin tensor \cite{lus1}
is ${\bar S}_{AB}=\epsilon^{\mu}
_A(u(p))\epsilon^{\nu}_B(u(p)) S_{\mu\nu}=-i{\tilde \xi}_A{\tilde \xi}_B{}{}$
[${\tilde \xi}_A=({\tilde \xi}_{\tau};{\tilde \xi}_r)$] with ${\bar S}_r=
-{i\over 2}\epsilon_{ruv}{\tilde \xi}_u{\tilde \xi}_v$ and ${\bar S}_{\tau r}
=-i{\tilde \xi}_{\tau}{\tilde \xi}_r$. The noncovariant variable
\cite{longhi,lus1} ${\tilde x}^{\mu}$ and its momentum $p^{\mu}$ may be
replaced by the canonical basis\cite{longhi} $\epsilon =\eta \sqrt{p^2}$,
$T=p_{\mu}{\tilde x}^{\mu}/\epsilon =p_{\mu}x^{\mu}/\epsilon$, $\vec k=\vec u
(p)$, $\vec z=\epsilon ({\vec {\tilde x}}-{{\vec p}\over {p_o}}{\tilde x}^o)$;
T is the rest-frame Lorentz scalar time and the noncovariant variable $\vec z$
is the mass multiplied by the classical analogue of the Newton-Wigner position
operator (it gives the independent Cauchy data for the 3-position).

In the new variables the constraints become $\chi =p^2-m^2=(\epsilon -m)
(\epsilon +m)\approx 0$ and $\chi_D=\epsilon ({\tilde \xi}_{\tau}-{m\over 
{\epsilon}} \xi_5)\approx 0$, so that, if $\eta =sign\, p_o$, they may be 
replaced by the two pairs

\begin{eqnarray}
&&{\tilde \chi}_{\eta}=\epsilon -\eta m \approx 0,\nonumber \\
&&{\tilde \chi}_{\eta D}={\tilde \xi}_{\tau}-\eta \xi_5 \approx 0,\nonumber \\
&&{}\nonumber \\
&&\{ {\tilde \chi}_{\eta},{\tilde \chi}_{\eta} \} {}^{*}=\{ {\tilde \chi}
_{\eta},{\tilde \chi}_{\eta D} \} {}^{*}= \{ {\tilde \chi}_{\eta D},{\tilde 
\chi}_{\eta D} \} {}^{*} =0,
\label{II3}
\end{eqnarray}

\noindent describing the two disjoined branches of the mass hyperboloid. For 
each branch the Dirac Hamiltonian is $H_{\eta D}=\lambda_{\eta}(\tau ){\tilde 
\chi}_{\eta}(\tau )+i\lambda_{\eta D}(\tau ){\tilde \chi}_{\eta D}$ with the 
following Hamilton equations for the Grassmann variables : 
${\dot {\tilde \xi}}_{\tau}\,
{\buildrel \circ \over =}\, \{ {\tilde \xi}_{\tau},H_{\eta D} \} {}^{*}=
\lambda_{\eta D}$, ${\dot \xi}_5\, {\buildrel \circ \over =}\, \eta \lambda
_{\eta D}$, ${\dot {\tilde \xi}}_r\, {\buildrel \circ \over =}\, 0$. These 
equations show that both ${\tilde \xi}_{\tau}$ and ${\xi}_5$ are gauge 
variables on a chosen branch of the mass hyperboloid (where the spin has to be
described only by 3 Grassmann variables giving rise after quantization to the
Clifford algebra $C_3$ of Pauli matrices). Therefore, one can add the 
gauge-fixing constraint $\rho_{\eta}={\tilde \xi}_{\tau}+\eta {\tilde \xi}_5
\approx 0{}{}$ [$\{ \rho_{\eta},{\tilde \chi}_{\eta D} \} {}^{*}=2i$,
$\{ \rho_{\eta},\rho_{\eta} \} {}^{*}=0$], implying 

\begin{equation}
{\tilde \xi}_{\tau}\approx 0,\quad\quad \xi_5\approx 0.
\label{II4}
\end{equation} 

\noindent  On each branch the new Dirac brackets [such that 
${\tilde \xi}_{\tau}=\xi_5\equiv 0$] are $\{ A,B\}{}^{**}_{\eta}=\{ A,B \} {}
^{*}+{i\over 2} [\{ A,{\tilde \chi}_{\eta D} \} {}^{*} \{ \rho_{\eta},B \}
{}^{*}+\{ A,\rho_{\eta} \} {}^{*} \{ {\tilde \chi}_{\eta D},B \} {}^{*}]$, so
that only the variables ${\tilde x}^{\mu}$, $p^{\mu}$, ${\tilde \xi}_r$ are left
with $\{ {\tilde x}^{\mu},p^{\nu} \} {}^{**}_{\eta}=-\eta^{\mu\nu}$, 
$\{ {\tilde \xi}_r,{\tilde \xi}_s \} {}^{**}_{\eta}=-i \delta_{rs}$ and Dirac 
Hamiltonian $H_{\eta D}=\lambda_{\eta}(\tau ){\tilde \chi}_{\eta}$.

Now one has ${\bar S}_{\tau r}\equiv 0$ [the boost part of the spin tensor 
disappears, because ${\tilde \xi}_{\tau}\approx 0$ implies $p_{\mu}S^{\mu\nu}
\approx 0$ in this gauge], ${\bar S}_{rs}=-i{\tilde \xi}_r
{\tilde \xi}_s$, ${\bar S}_r=-{i\over 2}\epsilon_{ruv}{\tilde \xi}_u{\tilde 
\xi}_v$, and then\cite{lus1} one gets: i) ${\tilde x}^{\mu}=x^{\mu}-{1\over 
{\epsilon}} [\eta^{\mu}_A({\bar S}^{\tau A}-{{{\bar S}^{Ar}p^r}\over {p_o+
\epsilon}})+{{p^{\mu}+2\epsilon \eta^{\mu o}}\over {\epsilon (p_o+\epsilon )}}
{\bar S}^{\tau r}p^r] \equiv x^{\mu}+{{\eta^{\mu}_s {\bar S}^{sr}p^r}\over
{\epsilon (p_o+\epsilon )}}=x^{\mu}+\eta^{\mu}_s {{({\vec {\bar S}}\times \vec 
p)^s}\over {\epsilon (p_o+\epsilon )}}$ and $\vec z=\epsilon [\vec x+{{ {\vec 
{\bar S}}\times \vec p}\over {\epsilon (p_o+\epsilon )}}- {{\vec p}\over 
{p_o}} x^o]$${}{}$ [${\tilde x}^o=x^o$] for the noncovariant positions; ii)
$p^{\mu}$, $J^{ij}={\tilde x}^ip^j-{\tilde x}^jp^i+\delta^{ir}\delta^{js}
{\bar S}^{rs}$, $J^{oi}={\tilde x}^op^i-{\tilde x}^ip^o -{{\delta^{ir}({\vec
{\bar S}}\times \vec p)^r}\over {p_o+\epsilon}}$ for the Poincar\'e generators.

Let us remark that in the original variables the same result can be obtained by
adding the gauge-fixing $\rho_D=p_{\mu}\xi^{\mu}+m\xi_5 \approx 0$ to the
first class constraint $\chi_D=p_{\mu}\xi^{\mu}-m\xi_5\approx 0$. One has
$\{ \chi_D,\chi_D \} {}^{*}=\{ \rho_D,\rho_D \} {}^{*}=i \chi$, $\{ \chi_D,
\rho_D \} {}^{*} =i (p^2+m^2)\approx 2im^2$, so that the elimination of these
two second class constraints implying $p_{\mu}\xi^{\mu}\approx 0$ and $\xi_5
\approx 0$, gives the Dirac brackets $\{ A,B \} {}^{**}=\{ A,B \} {}^{*}+
{{i(p^2+m^2)}\over {(p^2+m^2)^2-\chi^2}} [ \{ A,\chi_D \} {}^{*} \{ \rho_D,B 
\} {}^{*} + \{ A,\rho_D \} {}^{*} \{ \chi_D,B \} {}^{*}]- {{i\chi}\over
{(p^2+m^2)^2-\chi^2}} [ \{ A,\chi_D \} {}^{*} \{ \chi_D,B \} {}^{*} + \{ A,
\rho_D \} {}^{*} \{ \rho_D,B \} {}^{*} ]$. However, it is not possible to 
eliminate one component of $\xi^{\mu} \equiv \xi^{\mu}_{\perp}=(\eta^{\mu\nu}-
{{p^{\mu}p^{\nu}}\over {p^2}})\xi_{\nu}$ without breaking Lorentz covariance;
moreover, one gets $\{ x^{\mu},x^{\nu} \} {}^{**}\not= 0$, $\{ x^{\mu},p^{\nu} 
\} {}^{**} \not= -\eta^{\mu\nu}$.

Quantizing one gets $[{\hat {\tilde \xi}}_r,{\hat {\tilde \xi}}_s]_{+}=\hbar
\delta_{rs}$, so that ${\tilde \xi}_r \mapsto \sqrt{ {{\hbar}\over 2}} \sigma
_r$, with $\sigma_r$ being Pauli matrices (transforming as Wigner spin 1${}{}$ 
3-vectors under Lorentz transformations). Therefore, this is the 
pseudoclassical description of massive spinning particles belonging to the
$({1\over 2},0)$ (for $\eta =1$) and $(0,{1\over 2})$ (for $\eta =-1$) 
representation of SL(2,C). Since only the constraint ${\tilde \chi}_{\eta}=
\epsilon -\eta m \approx 0$ is left, at the quantum level one has only the
wave equation $(i{{\partial}\over {\partial T}}-\eta m) \psi_{\eta}(T,\vec z)=0$
with $\psi_{\eta}$ a SU(2) 2-spinor. This
equation, defining the mass shell, is not a spinor equation, because such an
equation does not exist in the massive case with massive 2-spinors [in the
massless case it would be the Weyl equation].
See Ref.\cite{longhi} for the study of the
wave functions $\phi (T,\vec z)$, $\phi (\epsilon ,\vec k)$, associated with the
replacement of ${\tilde x}^{\mu}$, $p^{\mu}$, with $T, \epsilon$, $\vec z, 
\vec k$. In the momentum basis the wave equation becomes (after multiplication 
by $\eta$) $(\eta \epsilon -m) \psi_{\eta}(\epsilon ,\vec k)=0$ or, by
introducing the $4\times 4$ matrix $\gamma^{\tau}_{(cha)}=\left( 
\begin{array}{cc} 1&0\\ 0&-1 \end{array} \right){}{}$, $[\epsilon \gamma
^{\tau}_{(cha)}-m] \psi_{(cha)}(\epsilon ,\vec k)=0$ with $\psi_{(cha)}=\left(
\begin{array}{c} \psi_{(cha)+} \\ \psi_{(cha)-} \end{array} \right)$. 
This is the
Chakrabarti's representation \cite{cha}, which, as noted in this paper,
coincides with the Shirokov's representation \cite{shi} for spin 1/2.
It is obtained from the Dirac equation in momentum space $(p_{\mu}\gamma^{\mu}
-m) \psi(p)=0$ by diagonalizing $p_{\mu}\gamma^{\mu}$ with a similarity 
transformation generated by ${\hat U}_{(cha)}=e^{i{\hat S}_{(cha)}}=
{1\over { \sqrt{2m(p_o+m)}}} [\gamma^o\gamma^{\mu}p_{\mu}+m]$
[$p^o=\sqrt{{\vec p}^2+m^2}$]; the quantum Poincar\'e generators are just the
quantization of the classical ones just given with ${\vec {\bar S}} \mapsto
{1\over 2} \vec \sigma$. In Ref.\cite{cha} it is shown that $\psi_{(cha)}(p)=
{\hat U}_{(cha)} \psi (p)= [Q(L({\buildrel \circ \over p},p)) \psi ](p)= 
Q(p,L({\buildrel \circ \over p},p))) \psi (p)$, since a homogeneous Lorentz
transformation may be written as $U(\Lambda )=Q(\Lambda )T(\Lambda )$ with
$[T(\Lambda )\psi ](p)=\psi (\Lambda^{-1}p)$ and $[Q(\Lambda )\psi ](p)=
Q(p,\Lambda ) \psi (p)$; namely $\psi_{(cha)}(p)$ is obtained by boosting at 
rest $\psi (p)$ with the Wigner boost, consistently with our previous classical
canonical transformation. The variable ${\vec {\tilde x}}$ [or $\vec z$] goes
into a quantum position operator ${\hat {\vec x}}_{(cha)}$, which can be 
obtained from the standard operator ${\hat {\vec x}}$ by means of ${\hat U}
_{(cha)}$. Since the parity operator $\gamma^{\tau}_{(cha)}$ is diagonal in
the Chakrabarti representation, $\pm \psi_{(cha)\pm}$ are parity eigenstates
and also eigenstates of spin, ${\vec \Sigma}_{(cha)}=\gamma_{5(cha)}\gamma
^{\tau}_{(cha)}{\vec \gamma}_{(cha)}=\left( \begin{array}{cc} \vec \sigma &0
\\ 0& \vec \sigma \end{array} \right)$ and helicity $h(\vec p)=\vec p\cdot
\vec \Sigma_{(cha)}/|\vec p|$ [they are mixed by charge conjugation $C_{(cha)}=
i\gamma^2_{(cha)}\gamma^{\tau}_{(cha)}$ and by chirality $\gamma_{5(cha)}=
\left( \begin{array}{cc} 0&1\\1&0 \end{array} \right)$].

It is evident that this representation is different from the Foldy-Wouthuysen
one \cite{fw}, which is reviewed in the Appendix together with other relevant
representations.

Instead in Ref.\cite{spinn}, one starts with the Lagrangian 

\begin{equation}
L^{'}={1\over 2}\{ {{{\dot x}^2}\over e}+em^2-i(\xi_{\mu}{\dot \xi}^{\mu}-
\xi_5{\dot \xi}_5)-i\varphi ({{\xi_{\mu}{\dot x}^{\mu}}\over e}-m\xi_5) \}, 
\label{II6}
\end{equation}

\noindent where 
$e$ and $\varphi$ are Lagrange multipliers ensuring the
existence of two first class constraints $p^2-m^2\approx 0$ and $p_{\mu}\xi
^{\mu}-m\xi_5\approx 0$. The canonical momenta are $\pi^{\mu}={i\over 2}\xi
^{\mu}$, $\pi_5=-{i\over 2}\xi_5$, $p^{\mu}=-{1\over e}[{\dot x}^{\mu}
-{i\over 2}\varphi \xi^{\mu}]$; elimination of 
the Grassmann second class constraints gives the same Dirac brackets and the 
same quantization as before. Now there is manifest local world-line 
supersymmetry (invariance under N=1 superdiffeomorphisms), namely the action 
is invariant under the local infinitesimal transformations $\delta x^{\mu}=
i\alpha (\tau ) \xi^{\mu}$, $\delta \xi^{\mu}=\alpha (\tau )[{{{\dot x}^{\mu}}
\over e}-{i\over {2e}}\varphi \xi^{\mu}]$, $\delta e=i\alpha (\tau ) \varphi$, 
$\delta \varphi =2\dot \alpha (\tau )$, $\delta \xi_5 =\alpha (\tau ) 
[m+{i\over {me}}\xi_5 ({\dot \xi}_5-{1\over 2}m\varphi )]$.

In Ref.\cite{sp1} it is shown that the action $S^{'}=\int d\tau L^{'}$ can be 
put in a superfield form. If one introduces the 3 superfields $X^{\mu}=x^{\mu}
+\theta \xi^{\mu}$, $E=e+\theta \varphi$ ($e$ and $\varphi$ are called the 
einbein and the gravitino respectively), $B=\xi_5+\theta b_5$ [${}{}b_5$ is an
even auxiliary field needed because the supersymmetry transformations close
on $\xi_5$ only on-shell], and if $D=\partial_{\theta}+\theta \partial_{\tau}$
[${}{}D^2X^{\mu}=\partial_{\tau}X^{\mu}={\dot X}^{\mu}$] is the spinor
derivative, one gets 

\begin{equation}
S^{'}={1\over 2}\int d\tau d\theta [E^{-1} DX_{\mu}\, 
{\dot X}^{\mu} -B\, DB+2m B E^{1/2}].
\label{II7}
\end{equation} 

The main point to be observed is that the superfield $X^{\mu}$ may be
interpreted as giving a
representation of the Grassmann fibration over $x^{\mu}$ associated with the 
spin structure [the absence of $\xi_5$ is irrelevant because one can get the
Clifford algebra $C_4$ of Dirac matrices also starting from a Grassmann algebra 
$G_4$ with generators $\xi^{\mu} \mapsto \sqrt{ {{\hbar}\over 2}}\gamma^{\mu}$].
Namely one utilizes the Noether supersymmetry transformations but with
an interpretation different from the standard one [formation of supermultiplets
of different bosonic and fermionic particles]. Let us remark that the Dirac
Hamiltonian $H_D=\lambda (\tau )\chi +i\lambda_D(\tau )\chi_D$ associated
with the Lagrangian (\ref{II1}) may be interpreted as a Hamiltonian superfield
describing the spin structure with the two first class constraints.

In Ref.\cite{sp2} one tries to eliminate the superfields E and B by putting the
solution of their Euler-Lagrange equations in $S^{'}$. Then, there is a long
discussion about the various forms of the reduced action according to possible
gauge redefinitions. However a definite final form of the action without $\xi_5$
is not obtained [it seems that $\xi_5$ can be explicitly eliminated only in 
the massless case\cite{sp4,spinn}]. Instead, there is the proposal of using a
Lagrangian which may be written in the form

\begin{equation}
L = - m \sqrt{\dot{x}^2} -{i\over 2} \xi_\mu \dot{\xi}^\mu - i\lambda 
\dot{x}^\mu \xi_\mu
\label{II8}
\end{equation}

\noindent where $\lambda (\tau )$ is an odd real Lagrange multiplier implying
the Lagrangian constraint $\dot{x}^{\mu} \xi_{\mu} \, {\buildrel \circ \over 
=}\, 0$. This constraint eliminates one of the components of $\xi^{\mu}$, so 
that the quantization of the final theory will produce Pauli matrices and SU(2) 
2-spinors. The canonical momenta are $p^{\mu}={{m{\dot x}^{\mu}}\over
{\sqrt{{\dot x}^2}}}+i\lambda \xi
^{\mu}$, $\pi_{\mu}={i\over 2} \xi_{\mu}$, $p^{\lambda}=0$ [$\{\lambda,p^
\lambda\} = -1$] and there are the primary constraints
$\chi = p^2 - m^2 - 2i \lambda p_\mu \xi^\mu \approx 0$,
$\chi_\mu = \pi_\mu - {i\over 2}\xi_\mu \approx 0$,
$p^\lambda \approx  0$ . The Dirac Hamiltonian $H_D=a(\tau ) \chi +ib^{\mu}
(\tau )\chi_{\mu}+c(\tau ) p^{\lambda}$ implies the secondary constraint
$\phi =p_{\mu}\xi^{\mu}\approx 0$ and $b^{\mu}(\tau )=-a(\tau )p^{\mu} \lambda$;
the time constancy of $\phi \approx 0$ generates the tertiary constraint
$\lambda \approx 0$ and then $c(\tau )\approx 0$. The second class constraints 
$\lambda \approx 0$, $p^{\lambda}\approx 0$, can be trivially eliminated. The
Dirac Hamiltonian becomes $H_D=a(\tau ) (p^2-m^2)$. The remaining constraints 
$\chi =p^2-m^2\approx 0$, $\chi_{\mu}=\pi_{\mu}-{i\over 2}\xi_{\mu}\approx 0$,
$\phi =p_{\mu}\xi^{\mu}\approx 0$, have only the following nonzero Poisson
brackets: $\{\chi_\mu,\chi_\nu\}=i\eta_{\mu \nu}$, $\{\chi_\mu,\phi\} = 
-p_\mu$ . While $\chi$ is first class, there are 5 second class constraints. By
introducing the projector $\Pi_{\mu \nu} = \eta_{\mu \nu}- (p_\mu p_\nu)/p^2$ ,
the 4 $\chi_{\mu}$ can be replaced by $p_{\mu}\pi^{\mu}\approx 0$ and by
$\chi_{\perp \, \mu}=\Pi_{\mu\nu} \chi^{\nu}\approx 0$ [${}{}p^{\mu}\chi
_{\perp \, \mu}\equiv 0$], with the algebra $\{\chi_{\perp \,\mu}, \chi_{\perp
\, \nu} \} = i \Pi_{\mu \nu}$, $\{ p_{\mu}\pi^{\mu} , \phi \} = -p^2$. The 
constraints  $\phi =p_{\mu}\xi^{\mu}\approx 0$ and $p_{\mu}\pi^{\mu}\approx 0$ 
form a second class pair, while each of the independent $\chi_{\perp \, \mu}
\approx 0$ is second class by itself. With the Dirac brackets
$\{A,B\}^\ast = \{A,B\} +i \{A,\chi_{\perp \, \mu} \} \Pi^{\mu \nu}
\{\chi_{\perp \, \nu} , B\} +\frac{1}{p^2} \{A, \hat\chi\} \{\phi, B\} + 
\frac{1}{p^2}\{A, \phi\} \{\hat\chi, B\}$, one gets $p_{\mu}\xi^{\mu}=
p_{\mu}\pi^{\mu}\equiv 0$, $\Pi_{\mu\nu} \pi^{\nu}\equiv {i\over 2} \xi
_{\perp \, \mu}$ with $\xi^{\mu}_{\perp}=\Pi^{\mu\nu}\xi_{\nu}\equiv \xi
^{\mu}$, and $\{\xi^\mu_{\perp}, \xi^\nu_{\perp} \}^\ast = i \Pi^{\mu \nu}$ .
However, again this reduction, implying that only the 3 Grassmann variables
$\xi^{\mu}_{\perp}$ survive, has the drawback that $\{ x^{\mu},x^{\nu} \} {}^{*}
\not= 0$, $\{ x^{\mu},p^{\nu} \} {}^{*}\not= -\eta^{\mu\nu}$.

Therefore, starting either from Eq.(\ref{II1}) or from Eq.(\ref{II8}) one
arrives at a description without $\xi_5$ and with $p_{\mu}\xi^{\mu}\approx 0$
for a spinning particle with a definite sign of the energy.

In Ref.\cite{casal} there is the coupling of the spinning particle to external
electromagnetic fields. It is based on the Lagrangian

\begin{eqnarray}
&L& = - \frac {i}{2} \xi_5 \dot{\xi}_5 - \frac {i}{2} \xi_\mu \dot{\xi}^\mu + 
\nonumber \\
&-& \sqrt{m^2 - i e F_{\mu \nu}(x) \xi^\mu \xi^\nu}
\sqrt{\Big( \dot{x}_\mu - \frac im \xi_\mu \dot{\xi}_5 \Big)^2} -
e \dot{x}_\mu A^\mu(x)=\nonumber \\
&=&  - \frac {i}{2} \xi_5 \dot{\xi}_5 - \frac {i}{2} \xi_\mu \dot{\xi}^\mu 
-e \dot{x}_\mu A^\mu(x)-\nonumber \\
&-& \Big[ m -\frac{i e}{2m} F_{\mu \nu}(x) \xi^\mu \xi^\nu
- \frac {e^2}{m^3} F_{\mu \nu}(x) F_{\rho \lambda}(x) \xi^\mu \xi^\nu
\xi^\rho \xi^\lambda \Big] \sqrt{\Big( \dot{x}_\mu - \frac im \xi_\mu 
\dot{\xi}_5 \Big)^2},
\label{II9}
\end{eqnarray}

\noindent which, besides the standard minimal coupling, has a nonminimal
``mass renormalization" $-ieF_{\mu\nu}\xi^{\mu}\xi^{\nu}=eF_{\mu\nu}S^{\mu\nu}$.
The $ie$ coefficient in front of $F_{\mu\nu}\xi^{\mu}\xi^{\nu}$, corresponding
to the absence of an anomalous magnetic moment of the electron, is the only one
ensuring that the two constraints remain first class even in presence of an 
external electromagnetic field.

Indeed, besides the second class constraints $\pi_{\mu}-{i\over 2}\xi_{\mu}
\approx 0$ and the added one (like in the free case) $\pi_5+{i\over 2}\xi_5
\approx 0$ [eliminated by going to the Dirac brackets $\{ \xi^{\mu},\xi^{\nu} \}
{}^{*}=i\eta^{\mu\nu}$, $\{ \xi_5,\xi_5 \} {}^{*}=-i$], one gets the first class
constraints

\begin{eqnarray}
\chi_D &=& (p_\mu - e A_\mu(x)) \xi^\mu - m \xi_5 \approx 0 \nonumber \\
\chi &=& (p-e A(x))^2 - m^2 + i e F_{\mu \nu}(x) \xi^\mu \xi^\nu \approx 0 ,
\nonumber \\
&&{}\nonumber \\
&&\{ \chi_D,\chi_D \} {}^{*} =i \chi ,\quad\quad \{ \chi ,\chi \} {}^{*}=
\{ \chi ,\chi_D \} {}^{*} =0.
\label{II10}
\end{eqnarray}

Following the pattern of the free case one could try to eliminate 2 Grassmann 
variables by adding the gauge-fixing $\rho_D=(p-eA(x))_{\mu}\xi^{\mu}+m\xi_5
\approx 0$ with $\{ \rho_D,\rho_D\} {}^{*} =i\chi$, $\{ \chi_D,\rho_D \} {}^{*}=
(p-eA(x))^2+m^2+ieF_{\mu\nu}(x)\xi^{\mu}\xi^{\nu}$. This would imply $\xi_5
\approx 0$ and $(p-eA(x))_{\mu}\xi^{\mu}\approx 0$. By going to Dirac brackets 
one would have $\xi^{\mu}\equiv \xi^{\mu}_{\perp (A)}=[\eta^{\mu\nu}-{{(p-eA(x))
^{\mu}(p-eA(x))^{\nu}}\over {(p-eA(x))^2}}] \xi_{\nu}$. Besides getting
$\{ x^{\mu},x^{\nu}\} {}^{**}\not= 0$, $\{ x^{\mu},p^{\nu} \} {}^{**}\not= -
\eta^{\mu\nu}$, one has a external-potential dependent transversality condition
and no control on the possible zeroes of $(p-eA(x))^2$.

One could also try to go to the Chakrabarti representation but now with a 
$p_{\mu}$ which is not conserved and gauge dependent [$A_A(x,p)=A_{\mu}(x)L
^{\mu}{}_A(p,{\buildrel \circ \over p})$; ${\tilde F}_{AB}(x,p)=F_{\mu\nu}(x)
L^{\mu}{}_A(p,{\buildrel \circ \over p})L^{\nu}{}_B(p,{\buildrel \circ \over 
p})$] one gets

\begin{eqnarray}
&&\chi =\epsilon^2-2eA_{\tau}(x,p)\epsilon+e^2[A^2_{\tau}(x,p)-{\vec A}^2(x,p)]
-m^2+ie{\tilde F}_{AB}(x,p){\tilde \xi}^A{\tilde \xi}^B=\nonumber \\
&&=(\epsilon -\epsilon_{+})(\epsilon -\epsilon_{-}) \approx 0,\nonumber \\
&&\epsilon_{\eta}=\epsilon_{\pm}=eA_{\tau}(x,p)+\eta \sqrt{m^2+e^2{\vec A}
^2(x,p)-ie{\tilde F}_{AB}(x,p){\tilde \xi}^A{\tilde \xi}^B}=\nonumber \\
&&=eA_{\tau}(x,p)+\eta \sqrt{m^2+e^2{\vec A}^2(x,p)} [1-{{ie{\tilde F}
_{AB}(x,p){\tilde \xi}^A{\tilde \xi}^B}\over {2[m^2+e^2{\vec A}^2(x,p)]}}-
\nonumber \\
&&-{{e^2{\tilde F}_{\tau r}(x,p){\tilde F}_{uv}(x,p){\tilde \xi}_{\tau}{\tilde
\xi}_r{\tilde \xi}_u{\tilde \xi}_v}\over {2[m^2+e^2{\vec A}^2(x,p)]^2}}],
\nonumber \\
&&{}\nonumber \\
&&\chi_D=[\epsilon -eA_{\tau}(x,p)]{\tilde \xi}_{\tau}+eA_r(x,p){\tilde \xi}_r
-m\xi_5 \approx 0,\nonumber \\
&&\Downarrow \nonumber \\
&&\chi_{\eta}=\epsilon -\epsilon_{\eta} \approx 0,\nonumber \\
&&\chi_{\eta D}={\tilde \xi}_{\tau}-\eta {{m\xi_5-eA_r(x,p){\tilde \xi}_r}\over
{\sqrt{m^2+e^2{\vec A}^2(x,p)}}} [1+\nonumber \\
&&+{{ie{\tilde F}_{AB}(x,p){\tilde \xi}^A{\tilde \xi}^B}\over {2[m^2+e^2{\vec 
A}^2(x,p)]}}+{{3e^2{\tilde F}_{\tau r}(x,p){\tilde F}_{uv}(x,p){\tilde \xi}
_{\tau}{\tilde \xi}_r{\tilde \xi}_u{\tilde \xi}_v}\over {2[m^2+e^2{\vec A}
^2(x,p)]^2}}],
\label{II11}
\end{eqnarray}

\noindent and the calculations become quite involved. Moreover, the separation
of $\epsilon =\eta \sqrt{p^2}$ (now not gauge invariant) is natural from the 
point of view of the particle but not natural from that of the gauge-fixing
$\rho_D\approx 0$ and one has no control on the reality of the square roots.

All these problems are connected to the fact that, in presence of external 
electromagnetic fields, one cannot diagonalize the Hamiltonian with the 
Foldy-Wouthuysen transformation, because arbitrary electric fields may create
crossings of the two branches of the mass hyperboloid (so that the square roots 
may become complex; all this is interpreted as the classical background of pair 
production). In Ref.\cite{sp3} there is a study of the pseudoclassical 
Foldy-Wouthuysen transformation in presence of external electromagnetic fields.
It is shown that at the pseudoclassical level everything works with arbitrary
external stationary non-homogeneous magnetic fields: the quantization gives
rise to a Foldy-Wouthuysen transformation which diagonalizes the Hamiltonian
if these external magnetic fields are also radiation fields (i.e. they satisfy
Maxwell's equations without sources). See chapter 5 sections 5.4, 5.5, 5.6 of
Ref.\cite{thaller} for the cases (supersymmetric Dirac operator) in which there 
exist simultaneous exact Foldy-Wouthuysen and Cini-Touschek transformations;
for an electron this requires zero electric field.

Finally also the electric charge of the spinning particle may be described at 
the pseudoclassical level (quantization of charge) with a pair $\theta$, 
$\theta^{*}$, of complex Grassmann variables [see for instance Refs.
\cite{casal,lus1}]

\begin{eqnarray}
&L& = {i\over 2} (\theta^{*}{\dot \theta}-{\dot \theta}^{*}\theta )
- \frac {i}{2} \xi_5 \dot{\xi}_5 - \frac {i}{2} \xi_\mu \dot{\xi}^\mu -
\nonumber \\
&-& \sqrt{m^2 - i Q F_{\mu \nu}(x) \xi^\mu \xi^\nu}
\sqrt{\Big( \dot{x}_\mu - \frac im \xi_\mu \dot{\xi}_5 \Big)^2} -
Q \dot{x}_\mu A^\mu(x),\nonumber \\
&&{}\nonumber \\
&&Q=e \theta^{*}\theta.
\label{II12}
\end{eqnarray}

Now there are the extra pairs of second class constraints $\pi_{\theta}+
{i\over 2}\theta^{*}\approx 0$, $\pi_{\theta^{*}}+{i\over 2}\theta \approx 0$;
the Poisson brackets $\{ \theta ,\pi_{\theta} \}= \{ \theta^{*}, \pi
_{\theta^{*}} \}=-1$ become the Dirac brackets $\{ \theta ,\theta^{*} \} {}^{*}=
-i$. The quantization sends $\theta$, $\theta^{*}$, in the annihilation and 
creation operators $\hat b$, ${\hat b}^{\dagger}$, of a Fermi oscillator, so
that the quantum electric charge $\hat Q=e{\hat b}^{\dagger}\hat b$ has two 
levels: Q=0 and Q=e.

\vfill\eject

\section{New Spinning Particle on Spacelike Hypersurfaces.}

To define a charged spinning particle with a definite sign of the energy on 
spacelike hypersurfaces, we shall start with the Lagrangian description of a 
charged scalar particle [see Refs.\cite{lus1,lus2}] with a real Grassmann
4-vector $\xi^{\mu}(\tau )$ [but without $\xi_5$] for the description of spin, 
we shall make the Legendre transformation to the Hamiltonian formalism and then 
we shall add a Hamiltonian odd second class constraint (like $p_{\mu}\xi^{\mu}
\approx 0$ of the previous Section) eliminating one of the components of
$\xi^{\mu}$. At the end of the Section we shall find the Lagrangian for this 
system, by making the 
inverse Legendre transformation, only in the free case, because the full 
Lagrangian with also the electromagnetic field is too complicated and not 
illuminating. As usual in relativistic particle mechanics, only the Hamiltonian 
description is tractable, because the Lagrangian one is too involved and very 
often it is impossible to get it in closed form.

Let us first review some preliminary results from Refs.\cite{lus1,lus2}
needed in the description of physical systems on spacelike hypersurfaces, 
integrating it with the definitions needed to describe the isolated system of N
scalar particles with pseudoclassical Grassmann-valued spin and electric 
charges plus the electromagnetic field\cite{lus1}.

Let $\lbrace \Sigma_{\tau}\rbrace$ be a one-parameter family of spacelike
hypersurfaces foliating Minkowski spacetime $M^4$ and giving a 3+1 decomposition
of it. At fixed $\tau$, let 
$z^{\mu}(\tau ,\vec \sigma )$ be the coordinates of the points on $\Sigma
_{\tau }$ in $M^4$, $\lbrace \vec \sigma \rbrace$ a system of coordinates on
$\Sigma_{\tau}$. If $\sigma^{\check A}=(\sigma^{\tau}=\tau ;\vec \sigma 
=\lbrace \sigma^{\check r}\rbrace)$ [the notation ${\check A}=(\tau ,
{\check r})$ with ${\check r}=1,2,3$ will be used; note that ${\check A}=
\tau$ and ${\check A}={\check r}=1,2,3$ are Lorentz-scalar indices] and 
$\partial_{\check A}=\partial /\partial \sigma^{\check A}$, 
one can define the vierbeins

\begin{equation}
z^{\mu}_{\check A}(\tau ,\vec \sigma )=\partial_{\check A}z^{\mu}(\tau ,\vec 
\sigma ),\quad\quad
\partial_{\check B}z^{\mu}_{\check A}-\partial_{\check A}z^{\mu}_{\check B}=0,
\label {III1}
\end{equation}

\noindent so that the metric on $\Sigma_{\tau}$ is

\begin{eqnarray}
&&g_{{\check A}{\check B}}(\tau ,\vec \sigma )=z^{\mu}_{\check A}(\tau ,\vec 
\sigma )\eta_{\mu\nu}z^{\nu}_{\check B}(\tau ,\vec \sigma ),\quad\quad 
g_{\tau\tau}(\tau ,\vec \sigma ) > 0,\nonumber \\
&&g(\tau ,\vec \sigma )=-det\, ||\, g_{{\check A}{\check B}}(\tau ,\vec 
\sigma )\, || ={(det\, ||\, z^{\mu}_{\check A}(\tau ,\vec \sigma )\, ||)}^2,
\nonumber \\
&&\gamma (\tau ,\vec \sigma )=-det\, ||\, g_{{\check r}{\check s}}(\tau ,\vec 
\sigma )\, ||.
\label{III2}
\end{eqnarray}

If $\gamma^{{\check r}{\check s}}(\tau ,\vec \sigma )$ is the inverse of the 
3-metric $g_{{\check r}{\check s}}(\tau ,\vec \sigma )$ [$\gamma^{{\check r}
{\check u}}(\tau ,\vec \sigma )g_{{\check u}{\check s}}(\tau ,\vec 
\sigma )=\delta^{\check r}_{\check s}$], the inverse $g^{{\check A}{\check B}}
(\tau ,\vec \sigma )$ of $g_{{\check A}{\check B}}(\tau ,\vec \sigma )$ 
[$g^{{\check A}{\check C}}(\tau ,\vec \sigma )g_{{\check c}{\check b}}(\tau ,
\vec \sigma )=\delta^{\check A}_{\check B}$] is given by

\begin{eqnarray}
&&g^{\tau\tau}(\tau ,\vec \sigma )={{\gamma (\tau ,\vec \sigma )}\over
{g(\tau ,\vec \sigma )}},\nonumber \\
&&g^{\tau {\check r}}(\tau ,\vec \sigma )=-[{{\gamma}\over g} g_{\tau {\check 
u}}\gamma^{{\check u}{\check r}}](\tau ,\vec \sigma ),\nonumber \\
&&g^{{\check r}{\check s}}(\tau ,\vec \sigma )=\gamma^{{\check r}{\check s}}
(\tau ,\vec \sigma )+[{{\gamma}\over g}g_{\tau {\check u}}g_{\tau {\check v}}
\gamma^{{\check u}{\check r}}\gamma^{{\check v}{\check s}}](\tau ,\vec \sigma ),
\label{III3}
\end{eqnarray}

\noindent so that $1=g^{\tau {\check C}}(\tau ,\vec \sigma )g_{{\check C}\tau}
(\tau ,\vec \sigma )$ is equivalent to

\begin{equation}
{{g(\tau ,\vec \sigma )}\over {\gamma (\tau ,\vec \sigma )}}=g_{\tau\tau}
(\tau ,\vec \sigma )-\gamma^{{\check r}{\check s}}(\tau ,\vec \sigma )
g_{\tau {\check r}}(\tau ,\vec \sigma )g_{\tau {\check s}}(\tau ,\vec \sigma ).
\label{III4}
\end{equation}

We have

\begin{equation}
z^{\mu}_{\tau}(\tau ,\vec \sigma )=(\sqrt{ {g\over {\gamma}} }l^{\mu}+
g_{\tau {\check r}}\gamma^{{\check r}{\check s}}z^{\mu}_{\check s})(\tau ,
\vec \sigma ),
\label{III5}
\end{equation}

\noindent and

\begin{eqnarray}
\eta^{\mu\nu}&=&z^{\mu}_{\check A}(\tau ,\vec \sigma )g^{{\check A}{\check B}}
(\tau ,\vec \sigma )z^{\nu}_{\check B}(\tau ,\vec \sigma )=\nonumber \\
&=&(l^{\mu}l^{\nu}+z^{\mu}_{\check r}\gamma^{{\check r}{\check s}}
z^{\nu}_{\check s})(\tau ,\vec \sigma ),
\label{III6}
\end{eqnarray}

\noindent where

\begin{eqnarray}
l^{\mu}(\tau ,\vec \sigma )&=&({1\over {\sqrt{\gamma}} }\epsilon^{\mu}{}_{\alpha
\beta\gamma}z^{\alpha}_{\check 1}z^{\beta}_{\check 2}z^{\gamma}_{\check 3})
(\tau ,\vec \sigma ),\nonumber \\
&&l^2(\tau ,\vec \sigma )=1,\quad\quad l_{\mu}(\tau ,\vec \sigma )z^{\mu}
_{\check r}(\tau ,\vec \sigma )=0,
\label{III7}
\end{eqnarray}

\noindent is the unit (future pointing) normal to $\Sigma_{\tau}$ at
$z^{\mu}(\tau ,\vec \sigma )$.

For the volume element in Minkowski spacetime we have

\begin{eqnarray}
d^4z&=&z^{\mu}_{\tau}(\tau ,\vec \sigma )d\tau d^3\Sigma_{\mu}=d\tau [z^{\mu}
_{\tau}(\tau ,\vec \sigma )l_{\mu}(\tau ,\vec \sigma )]\sqrt{\gamma
(\tau ,\vec \sigma )}d^3\sigma=\nonumber \\
&=&\sqrt{g(\tau ,\vec \sigma )} d\tau d^3\sigma.
\label{III8}
\end{eqnarray}

Let us remark that according to the geometrical approach of 
Ref.\cite{kuchar},one 
can use Eq.(\ref{III5}) in the form $z^{\mu}_{\tau}(\tau ,\vec \sigma )=N(\tau ,
\vec \sigma )l^{\mu}(\tau ,\vec \sigma )+N^{\check r}(\tau ,\vec \sigma )
z^{\mu}_{\check r}(\tau ,\vec \sigma )$, where $N=\sqrt{g/\gamma}=\sqrt{g
_{\tau\tau}-\gamma^{{\check r}{\check s}}g_{\tau{\check r}}g_{\tau{\check s}}}$ 
and $N^{\check r}=g_{\tau \check s}\gamma^{\check s\check r}$ are the 
standard lapse and shift functions, so that $g_{\tau \tau}=N^2+
g_{\check r\check s}N^{\check r}N^{\check s}, g_{\tau \check r}=
g_{\check r\check s}N^{\check s},
g^{\tau \tau}=N^{-2}, g^{\tau \check r}=-N^{\check r}/N^2, g^{\check r\check
s}=\gamma^{\check r\check s}+{{N^{\check r}N^{\check s}}\over {N^2}}$,
${{\partial}\over {\partial z^{\mu}_{\tau}}}=l_{\mu}\, {{\partial}\over
{\partial N}}+z_{{\check s}\mu}\gamma^{{\check s}{\check r}} {{\partial}\over
{\partial N^{\check r}}}$, $d^4z=N\sqrt{\gamma}d\tau d^3\sigma$.

The rest frame form of a timelike fourvector $p^{\mu}$ is $\stackrel
{\circ}{p}{}^{\mu}=\eta \sqrt{p^2} (1;\vec 0)= \eta^{\mu o}\eta \sqrt{p^2}$,
$\stackrel{\circ}{p}{}^2=p^2$, where $\eta =sign\, p^o$.
The standard Wigner boost transforming $\stackrel{\circ}{p}{}^{\mu}$ into
$p^{\mu}$ is

\begin{eqnarray}
L^{\mu}{}_{\nu}(p,\stackrel{\circ}{p})&=&\epsilon^{\mu}_{\nu}(u(p))=
\nonumber \\
&=&\eta^{\mu}_{\nu}+2{ {p^{\mu}{\stackrel{\circ}{p}}_{\nu}}\over {p^2}}-
{ {(p^{\mu}+{\stackrel{\circ}{p}}^{\mu})(p_{\nu}+{\stackrel{\circ}{p}}_{\nu})}
\over {p\cdot \stackrel{\circ}{p} +p^2} }=\nonumber \\
&=&\eta^{\mu}_{\nu}+2u^{\mu}(p)u_{\nu}(\stackrel{\circ}{p})-{ {(u^{\mu}(p)+
u^{\mu}(\stackrel{\circ}{p}))(u_{\nu}(p)+u_{\nu}(\stackrel{\circ}{p}))}
\over {1+u^o(p)} },\nonumber \\
&&{} \nonumber \\
\nu =0 &&\epsilon^{\mu}_o(u(p))=u^{\mu}(p)=p^{\mu}/\eta \sqrt{p^2}, \nonumber \\
\nu =r &&\epsilon^{\mu}_r(u(p))=(-u_r(p); \delta^i_r-{ {u^i(p)u_r(p)}\over
{1+u^o(p)} }).
\label{III9}
\end{eqnarray}

The inverse of $L^{\mu}{}_{\nu}(p,\stackrel{\circ}{p})$ is $L^{\mu}{}_{\nu}
(\stackrel{\circ}{p},p)$, the standard boost to the rest frame, defined by

\begin{equation}
L^{\mu}{}_{\nu}(\stackrel{\circ}{p},p)=L_{\nu}{}^{\mu}(p,\stackrel{\circ}{p})=
L^{\mu}{}_{\nu}(p,\stackrel{\circ}{p}){|}_{\vec p\rightarrow -\vec p}.
\label{III10}
\end{equation}

Therefore, we can define the following vierbeins [the $\epsilon^{\mu}_r(u(p))$'s
are also called polarization vectors; the indices r, s will be used for A=1,2,3
and $\bar o$ for A=0]

\begin{eqnarray}
&&\epsilon^{\mu}_A(u(p))=L^{\mu}{}_A(p,\stackrel{\circ}{p}),\nonumber \\
&&\epsilon^A_{\mu}(u(p))=L^A{}_{\mu}(\stackrel{\circ}{p},p)=\eta^{AB}\eta
_{\mu\nu}\epsilon^{\nu}_B(u(p)),\nonumber \\
&&{} \nonumber \\
&&\epsilon^{\bar o}_{\mu}(u(p))=\eta_{\mu\nu}\epsilon^{\nu}_o(u(p))=u_{\mu}(p),
\nonumber \\
&&\epsilon^r_{\mu}(u(p))=-\delta^{rs}\eta_{\mu\nu}\epsilon^{\nu}_r(u(p))=
(\delta^{rs}u_s(p);\delta^r_j-\delta^{rs}\delta_{jh}{{u^h(p)u_s(p)}\over
{1+u^o(p)} }),\nonumber \\
&&\epsilon^A_o(u(p))=u_A(p),
\label{III11}
\end{eqnarray}

\noindent which satisfy

\begin{eqnarray}
&&\epsilon^A_{\mu}(u(p))\epsilon^{\nu}_A(u(p))=\eta^{\mu}_{\nu},\nonumber \\
&&\epsilon^A_{\mu}(u(p))\epsilon^{\mu}_B(u(p))=\eta^A_B,\nonumber \\
&&\eta^{\mu\nu}=\epsilon^{\mu}_A(u(p))\eta^{AB}\epsilon^{\nu}_B(u(p))=u^{\mu}
(p)u^{\nu}(p)-\sum_{r=1}^3\epsilon^{\mu}_r(u(p))\epsilon^{\nu}_r(u(p)),
\nonumber \\
&&\eta_{AB}=\epsilon^{\mu}_A(u(p))\eta_{\mu\nu}\epsilon^{\nu}_B(u(p)),\nonumber 
\\
&&p_{\alpha}{{\partial}\over {\partial p_{\alpha}} }\epsilon^{\mu}_A(u(p))=
p_{\alpha}{{\partial}\over {\partial p_{\alpha}} }\epsilon^A_{\mu}(u(p))
=0.
\label{III12}
\end{eqnarray}

The Wigner rotation corresponding to the Lorentz transformation $\Lambda$ is

\begin{eqnarray}
R^{\mu}{}_{\nu}(\Lambda ,p)&=&{[L(\stackrel{\circ}{p},p)\Lambda^{-1}L(\Lambda
p,\stackrel{\circ}{p})]}^{\mu}{}_{\nu}=\left(
\begin{array}{cc}
1 & 0 \\
0 & R^i{}_j(\Lambda ,p)
\end{array}  
\right) ,\nonumber \\
{} && {}\nonumber \\
R^i{}_j(\Lambda ,p)&=&{(\Lambda^{-1})}^i{}_j-{ {(\Lambda^{-1})^i{}_op_{\beta}
(\Lambda^{-1})^{\beta}{}_j}\over {p_{\rho}(\Lambda^{-1})^{\rho}{}_o+\eta 
\sqrt{p^2}} }-\nonumber \\
&-&{{p^i}\over {p^o+\eta \sqrt{p^2}} }[(\Lambda^{-1})^o{}_j- { {((\Lambda^{-1})^o
{}_o-1)p_{\beta}(\Lambda^{-1})^{\beta}{}_j}\over {p_{\rho}(\Lambda^{-1})^{\rho}
{}_o+\eta \sqrt{p^2}} }].
\label{III13}
\end{eqnarray}

The polarization vectors transform under the 
Poincar\'e transformations $(a,\Lambda )$ in the following way

\begin{equation}
\epsilon^{\mu}_r(u(\Lambda p))=(R^{-1})_r{}^s\, \Lambda^{\mu}{}_{\nu}\, 
\epsilon^{\nu}_s(u(p)).
\label{III14}
\end{equation}

In Ref.\cite{lus1}, the system of N charged scalar 
particles was considered. As said in the Introduction, on the hypersurface
$\Sigma_{\tau}$ the particles are described by variables ${\vec \eta}_i(\tau )$
such that $x^{\mu}_i(\tau )=z^{\mu}(\tau ,{\vec \eta}_i(\tau ))$.
The electric charge of each particle is 
described in a pseudoclassical way\cite{casal} by means of a pair of
complex conjugate Grassmann variables $\theta_i(\tau ), \theta
^{*}_i(\tau )$ satisfying [$I_i=I^{*}_i=\theta^{*}_i\theta_i$ is the generator 
of the $U_{em}(1)$ group of particle i]

\begin{eqnarray}
&&\theta^2_i=\theta_i^{{*}2}=0,\quad\quad \theta_i\theta^{*}_i+\theta^{*}_i
\theta_i=0,\nonumber \\
&&\theta_i\theta_j=\theta_j\theta_i,\quad\quad \theta_i\theta^{*}_j=
\theta_j^{*}\theta_i,\quad\quad \theta^{*}_i\theta^{*}_j=\theta^{*}_j\theta
^{*}_i,\quad\quad i\not= j.
\label{III15}
\end{eqnarray}

\noindent This amounts to assume that the electric charges $Q_i=e_i\theta^{*}_i
\theta_i$, $Q^2_i=0$, 
are quantized with levels 0 and $e_i$\cite{casal} as already said. 

Let $\xi^\mu_i(\tau)$ $i=(1,...,N)$ be the Grassmann variables describing
the spin of the spinning particles (they are assumed to commute with the
Grassmann variables describing the electric charges)

\begin{equation}
\xi^\mu_i \xi^\nu_i+\xi^\nu_i\xi^\mu_i = 0, ~~~~~
\xi^\mu_i \xi^\nu_j=\xi^\nu_i\xi^\mu_j, ~~~i\neq j .
\label{III16}
\end{equation}

On the hypersurface $\Sigma_{\tau}$, we describe the electromagnetic potential
and field strength with Lorentz-scalar variables $A_{\check A}(\tau ,\vec \sigma
)$ and $F_{{\check A}{\check B}}(\tau ,\vec \sigma )$ respectively, defined by

\begin{eqnarray}
&&A_{\check A}(\tau ,\vec \sigma )=z^{\mu}_{\check A}(\tau ,\vec \sigma )
A_{\mu}(z(\tau ,\vec \sigma )),\nonumber \\
&&F_{{\check A}{\check B}}(\tau ,\vec \sigma )={\partial}_{\check A}A_{\check
B}(\tau ,\vec \sigma )-{\partial}_{\check B}A_{\check A}(\tau ,\vec \sigma )=
z^{\mu}_{\check A}(\tau ,\vec \sigma )z^{\nu}_{\check B}(\tau ,\vec \sigma )
F_{\mu\nu}(z(\tau ,\vec \sigma )).
\label{III17}
\end{eqnarray}

Momentarily disregarding the problem to reduce the $\xi^{\mu}_i$ variables to 
only three for each particle, we shall assume the following Lagrangian density 
for the isolated system of N spinning charged particles plus the 
electromagnetic field on spacelike hypersurfaces, since it generalizes the 
Lagrangian for N charged scalar particles\cite{lus1} incorporating 
simultaneously the main properties of the Lagrangian (\ref{II9})]

\begin{eqnarray}
&{\cal L}& (\tau, \vec{\sigma}) = \sum_{i=1}^N
\delta^3 (\vec{\sigma} -\vec{\eta}_i(\tau)) \Big\{ \frac i2
\big( \theta^\ast_i(\tau) \dot{\theta}_i(\tau) -
\dot{\theta}^\ast_i(\tau) \theta_i(\tau) \big) - \frac i2 \xi_{\mu i}(\tau)
\dot{\xi}^\mu_i(\tau) - \nonumber \\
&-& \eta_i \sqrt{m_i^2 - i Q_i(\tau) \xi_i^\mu (\tau )\xi_i^\nu (\tau )
z^{\breve{A}}_\mu (\tau, \vec{\sigma}) z^{\breve{B}}_\nu (\tau, \vec{\sigma})
F_{\breve{A} \breve{B}} (\tau, \vec{\sigma}) }\cdot \nonumber \\
&&\sqrt{g_{\tau \tau}(\tau, \vec{\sigma}) +2 g_{\tau \breve{r}}(\tau, 
\vec{\sigma}) \dot{\eta}_i^{\breve{r}}(\tau ) +g_{\breve{r} \breve{s}}(\tau, 
\vec{\sigma}) \dot{\eta}_i^{\breve{r}}(\tau ) \dot{\eta}_i^{\breve{s}}(\tau )}
 - \nonumber \\
&-& Q_i(\tau) \Big( A_\tau (\tau, \vec{\sigma}) + \dot{\eta}_i^{\breve{r}}
A_{\breve{r}} (\tau, \vec{\sigma}) \Big) \Big\} -
\frac{\sqrt{g(\tau, \vec{\sigma})}}{4}
F_{\breve{A} \breve{B}} (\tau, \vec{\sigma})
F_{\breve{C}\breve{D}} (\tau, \vec{\sigma})
g^{\breve{A} \breve{C}} (\tau, \vec{\sigma})
g^{\breve{B}\breve{D}} (\tau, \vec{\sigma}) .
\label{III18}
\end{eqnarray}

The main point is to see whether this Lagrangian gives rise to a consistent set 
of constraints, reducing to the constraints for the scalar case of Ref.
\cite{lus1} by eliminating the spin.

The canonical momenta are [$E_{\check r}=F_{{\check r}\tau}$ and $B_{\check r}
={1\over 2}\epsilon_{{\check r}{\check s}{\check t}}F_{{\check s}{\check t}}$ 
($\epsilon_{{\check r}{\check s}{\check t}}=\epsilon^{{\check r}{\check s}
{\check t}}$) are the electric and magnetic fields respectively; for
$g_{\check A\check B}\rightarrow \eta_{\check A\check B}$ one gets 
$\pi^{\check r}=-E_{\check r}=E^{\check r}$]

\begin{eqnarray}
\pi_{\theta_i}(\tau)  &=& \frac {\partial^L L(\tau)}{\partial \dot{\theta}_i(\tau)}
= - \frac i2 \theta^\ast_i(\tau) \nonumber \\
\pi_{\theta^\ast_i}(\tau) &=&
\frac {\partial^L L(\tau)}{\partial \dot{\theta^\ast}_i(\tau)}
= - \frac i2 \theta_i(\tau) \nonumber \\
\pi^\mu_i(\tau)&=&
 \frac {\partial^L L(\tau)}{\partial \dot{\xi}_{i \mu}(\tau)}
=  \frac i2 \xi^\mu_i(\tau) \nonumber \\
&&{}\nonumber \\
\rho_{\mu}(\tau ,\vec \sigma )&=&-{ {\partial {\cal L}(\tau ,\vec \sigma )}
\over {\partial z^{\mu}_{\tau}(\tau ,\vec \sigma )} }=\sum_{i=1}^N\delta^3
(\vec \sigma -{\vec \eta}_i(\tau ))\eta_im_i\nonumber \\
&&{ {z_{\tau\mu}(\tau ,\vec \sigma )+z_{{\check r}\mu}(\tau ,\vec \sigma )
{\dot \eta}_i^{\check r}(\tau )}\over {\sqrt{g_{\tau\tau}(\tau ,\vec \sigma )+
2g_{\tau {\check r}}(\tau ,\vec \sigma ){\dot \eta}_i^{\check r}(\tau )+
g_{{\check r}{\check s}}(\tau ,\vec \sigma ){\dot \eta}_i^{\check r}(\tau ){\dot
\eta}_i^{\check s}(\tau ) }} }-\nonumber \\
&&-\sum_{i=1}^N\delta^3(\vec \sigma -{\vec \eta}_i(\tau )) \nonumber \\
&&\Big( {{i\eta_iQ_i}\over {2m_i}} \xi_i^{\rho}(\tau )\xi^{\nu}_i(\tau )
z^{\check A}_{\rho}(\tau ,\vec \sigma )z^{\check B}_{\nu}(\tau ,\vec \sigma )
F_{\check A\check B}(\tau ,\vec \sigma )\cdot \nonumber \\
&&{ {z_{\tau\mu}(\tau ,\vec \sigma )+z_{{\check r}\mu}(\tau ,\vec \sigma )
{\dot \eta}_i^{\check r}(\tau )}\over {\sqrt{g_{\tau\tau}(\tau ,\vec \sigma )+
2g_{\tau {\check r}}(\tau ,\vec \sigma ){\dot \eta}_i^{\check r}(\tau )+
g_{{\check r}{\check s}}(\tau ,\vec \sigma ){\dot \eta}_i^{\check r}(\tau ){\dot
\eta}_i^{\check s}(\tau ) }} }+\nonumber \\
&&+{{i\eta_iQ_i}\over {m_i}} \{ \{ \xi^{\rho}_i(\tau )\xi^{\nu}_i(\tau )
[ z_{\tau \mu}(\tau ,\vec \sigma ) (g^{\check A\tau}g^{\tau \check C}g
^{\check B\check D}+g^{\check A\check C}g^{\check B\tau}g^{\tau \check D})
+\nonumber \\
&&+z_{\check r \mu} (g^{\check A\check r}g^{\tau \check C}+g
^{\check A\tau}g^{\check r \check C})g^{\check B\check D}]
z_{\check C\rho}z_{\check D\nu}+\xi_{i\mu}(\tau )\xi_{i\rho}(\tau )
g^{\check A\tau}g^{\check B\check D}z^{\rho}_{\check D} \} F_{\check A\check B}
\} (\tau ,\vec \sigma )\nonumber \\
&&\sqrt{g_{\tau\tau}(\tau ,\vec \sigma )+
2g_{\tau {\check r}}(\tau ,\vec \sigma ){\dot \eta}_i^{\check r}(\tau )+
g_{{\check r}{\check s}}(\tau ,\vec \sigma ){\dot \eta}_i^{\check r}(\tau ){\dot
\eta}_i^{\check s}(\tau ) }  \Big) +\nonumber \\
&&+{ {\sqrt {g(\tau ,\vec \sigma )}}\over 4}[(g^{\tau \tau}z_{\tau \mu}+
g^{\tau {\check r}}z_{\check r\mu})(\tau ,\vec \sigma )g^{{\check A}{\check C}}
(\tau ,\vec \sigma )g^{{\check B}{\check D}}(\tau ,\vec \sigma )F_{{\check A}
{\check B}}(\tau ,\vec \sigma )F_{{\check C}{\check D}}(\tau ,\vec \sigma )
-\nonumber \\
&-&2[z_{\tau \mu}(\tau ,\vec \sigma )(g^{\check A\tau}g^{\tau \check C}
g^{{\check B}{\check D}}+g^{{\check A}{\check C}}g^{\check B\tau}g^{\tau 
\check D})(\tau ,\vec \sigma )+\nonumber \\
&+&z_{\check r\mu}(\tau ,\vec \sigma )
(g^{{\check A}{\check r}}g^{\tau {\check C}}+g^{{\check A}\tau}g^{{\check r}
{\check C}})(\tau ,\vec \sigma )g^{{\check B}{\check D}}
(\tau ,\vec \sigma )]F_{{\check A}{\check B}}(\tau ,\vec \sigma )
F_{{\check C}{\check D}}(\tau ,\vec \sigma )]=
\nonumber \\
&&=[(\rho_{\nu}l^{\nu})l_{\mu}+(\rho_{\nu}z^{\nu}_{\check r})\gamma^{{\check r}
{\check s}}z_{{\check s}\mu}](\tau ,\vec \sigma ),\nonumber \\
&&{}\nonumber \\
\kappa_{i{\check r}}(\tau )&=&-{ {\partial L(\tau )}\over {\partial {\dot
\eta}_i^{\check r}(\tau )} }=\nonumber \\
&=&\eta_i\sqrt{m_i^2-iQ_i(\tau )\xi^{\mu}_i(\tau )\xi^{\nu}_i(\tau )z^{\check 
A}_{\mu}(\tau ,{\vec \eta}_i(\tau ))z^{\check B}_{\nu}(\tau ,{\vec \eta}
_i(\tau ))F_{\check A\check B}(\tau ,{\vec \eta}_i(\tau ))}\nonumber \\
&&{ {g_{\tau {\check r}}(\tau ,{\vec \eta}_i(\tau ))+g_{{\check r}
{\check s}}(\tau ,{\vec \eta}_i(\tau )){\dot \eta}_i^{\check s}(\tau )}\over
{ \sqrt{g_{\tau\tau}(\tau ,{\vec \eta}_i(\tau ))+
2g_{\tau {\check r}}(\tau ,{\vec \eta}_i(\tau )){\dot \eta}_i^{\check r}(\tau )+
g_{{\check r}{\check s}}(\tau ,{\vec \eta}_i(\tau )){\dot \eta}_i^{\check r}
(\tau ){\dot \eta}_i^{\check s}(\tau ) }} }+\nonumber \\
&+&e_i\theta^{*}_i(\tau )\theta_i(\tau )A_{\check r}(\tau ,{\vec \eta}_i
(\tau )),\nonumber \\
&&{}\nonumber \\
\pi^\tau (\tau, \vec{\sigma}) &=& \frac{\partial {\cal L} (\tau, \vec{\sigma})}
{\partial (\partial_\tau A_\tau (\tau, \vec{\sigma}))} = 0 \nonumber \\
\pi^{\breve{r}}(\tau, \vec{\sigma}) &=&
 \frac {\partial {\cal L}(\tau, \vec{\sigma})}
 {\partial (\partial_\tau A_{\breve{r}} (\tau, \vec{\sigma}))} =
- \frac {\gamma (\tau, \vec{\sigma})}{\sqrt{ g(\tau, \vec{\sigma})}}
\gamma^{\breve{r} \breve{s}}(\tau, \vec{\sigma})
\Big( F_{\tau \breve{s}} - g_{\tau \breve{v}}\gamma^{\breve{v} \breve{u}}
F_{\breve{u} \breve{s}} \Big)(\tau, \vec{\sigma}) + \nonumber \\
&+& \sum_{i=1}^N \delta^3 (\vec{\sigma} -\vec{\eta}_i(\tau)) i \eta_i
\frac{Q_i(\tau) \xi^\mu_i(\tau) \xi^\nu_i(\tau) l_\mu (\tau, \vec{\sigma})
z_{\breve{s} \nu} (\tau, \vec{\sigma})}{\sqrt{m_i^2-
 \gamma^{\breve{r} \breve{s}}(\tau, \vec{\sigma})k_{i\breve{r}}(\tau)
 k_{i\breve{s}}(\tau)}} \gamma^{\breve{r} \breve{s}}(\tau, \vec{\sigma})=
\nonumber \\
&=&{ {\gamma (\tau ,\vec \sigma )}\over {\sqrt {g(\tau ,\vec \sigma )}} }
\gamma^{{\check r}{\check s}}(\tau ,\vec \sigma )(E_{\check s}(\tau ,\vec 
\sigma )+g_{\tau {\check v}}(\tau ,\vec \sigma )\gamma^{{\check v}{\check u}}
(\tau ,\vec \sigma )\epsilon_{{\check u}{\check s}{\check t}} B_{\check t}
(\tau ,\vec \sigma ))+\nonumber \\
&+&\sum_{i=1}^N \delta^3 (\vec{\sigma} -\vec{\eta}_i(\tau)) i \eta_i
\frac{Q_i(\tau) \xi^\mu_i(\tau) \xi^\nu_i(\tau) l_\mu (\tau, \vec{\sigma})
z_{\breve{s} \nu} (\tau, \vec{\sigma})}{\sqrt{m_i^2-
 \gamma^{\breve{r} \breve{s}}(\tau, \vec{\sigma})k_{i\breve{r}}(\tau)
 k_{i\breve{s}}(\tau)}} \gamma^{\breve{r} \breve{s}}(\tau, \vec{\sigma}).
\label{III19}
\end{eqnarray}

\noindent Let us note that, due to the interaction of the electromagnetic field
with the spin, the electromagnetic momentum $\pi^{\check r}(\tau ,\vec \sigma )$
has an extra term concentrated on the particles.

The following Poisson brackets are assumed

\begin{eqnarray}
&&\lbrace z^{\mu}(\tau ,\vec \sigma ),\rho_{\nu}(\tau ,{\vec \sigma}^{'}\rbrace
=-\eta^{\mu}_{\nu}\delta^3(\vec \sigma -{\vec \sigma}^{'}),\nonumber \\
&&\lbrace A_{\check A}(\tau ,\vec \sigma ),\pi^{\check B}(\tau ,\vec 
\sigma^{'} )\rbrace =\eta^{\check B}_{\check A}
\delta^3(\vec \sigma -\vec \sigma^{'}),\nonumber \\
&&\lbrace \eta^{\check r}_i(\tau ),\kappa_{j{\check s}}(\tau )\rbrace =-
\delta_{ij}\delta^{\check r}_{\check s},\nonumber \\
&&\lbrace \theta_i(\tau ),\pi_{\theta \, j}(\tau )\rbrace =-\delta_{ij},
\nonumber \\
&&\lbrace \theta^{*}_i(\tau ),\pi_{\theta^{*} \, j}(\tau )\rbrace =-\delta_{ij},
\nonumber \\
&&\lbrace \xi^{\mu}_i,\pi_j^{\nu}\rbrace =-\delta_{ij}\eta^{\mu\nu}.
\label{III20}
\end{eqnarray}

The Grassmann momenta associated with the Grassmann variables describing
electric charges give rise to the second class constraints $\pi_{\theta
\, i}+{i\over 2}\theta^{*}_i\approx 0$, $\pi_{\theta^{*}\, i}+{i\over 2}
\theta_i\approx 0$ [$\lbrace \pi_{\theta \, i}+{i\over 2}\theta^{*}_i,
\pi_{\theta^{*}\, j}+{i\over 2}\theta_j\rbrace =-i\delta_{ij}$]; $\pi
_{\theta \, i}$ and $\pi_{\theta^{*}\, i}$ are then eliminated with the
help of Dirac brackets

\begin{equation}
\lbrace A,B\rbrace {}^{*}=\lbrace A,B\rbrace -i[\lbrace A,\pi_{\theta \, i}+
{i\over 2}\theta^{*}_i\rbrace \lbrace \pi_{\theta^{*}\, i}+{i\over 2}
\theta_i,B\rbrace  +\lbrace A,\pi_{\theta^{*}\, i}+{i\over 2}
\theta_i \rbrace \lbrace \pi_{\theta \, i}+{i\over 2}\theta^{*}_i,B\rbrace ]
\label{III21}
\end{equation}

\noindent so that the remaining Grassmann variables have the fundamental
Dirac brackets [which we will still denote $\lbrace .,.\rbrace$ for the sake of
simplicity]

\begin{eqnarray}
&&\lbrace \theta_i(\tau ),\theta_j(\tau )\rbrace = \lbrace \theta_i^{*}(\tau ),
\theta_j^{*}(\tau )\rbrace =0,\nonumber \\
&&\lbrace \theta_i(\tau ),\theta_j^{*}(\tau )\rbrace =-i\delta_{ij}.
\label{III22}
\end{eqnarray}

Moreover, we also have the following primary constraints

\begin{eqnarray}
\chi_i^\mu (\tau) &=& \pi_i^\mu (\tau) - \frac i2 \xi_i^\mu (\tau) \approx 0
~~~~~~~~~~~~~~~i=1,...,N \nonumber \\
\pi^\tau (\tau, \vec{\sigma}) &\approx & 0 \nonumber \\
{\cal H}_\mu (\tau, \vec{\sigma}) &=& \rho_\mu (\tau, \vec{\sigma}) -
l_\mu (\tau, \vec{\sigma}) \Big[ -\frac{1}{2\sqrt{\gamma(\tau, \vec{\sigma})}}
\pi^{\breve{r}}(\tau, \vec{\sigma}) g_{\breve{r} \breve{s}}(\tau, \vec{\sigma})
\pi^s (\tau, \vec{\sigma})+ \nonumber \\
&+& \frac{\sqrt{\gamma(\tau, \vec{\sigma})}}{4}
 \gamma^{\breve{r} \breve{s}}(\tau, \vec{\sigma})
 \gamma^{\breve{u} \breve{v}}(\tau, \vec{\sigma})
F_{\breve{r} \breve{u}}(\tau, \vec{\sigma})
F_{\breve{s} \breve{v}}(\tau, \vec{\sigma}) + \nonumber \\
&+& \sum_{i=1}^N \delta^3 (\vec{\sigma} -\vec{\eta}_i(\tau)) \eta_i \cdot 
\nonumber \\
&\cdot &\sqrt{m_i^2 - \gamma^{\breve{r} \breve{s}}(\tau, \vec{\sigma})
\big( k_{i \breve{r}}(\tau) - Q_i(\tau) A_{\breve{r}}(\tau, \vec{\sigma}) \big)
\big( k_{i \breve{s}}(\tau) -
 Q_i(\tau) A_{\breve{s}}(\tau, \vec{\sigma}) \big)} + \nonumber \\
&+& \frac{1}{2\sqrt{\gamma(\tau, \vec{\sigma})}}
\sum_{i,j=1}^N \delta^3 (\vec{\sigma} -\vec{\eta}_i(\tau))
\delta^3 (\vec{\sigma} -\vec{\eta}_j(\tau)) \cdot \nonumber \\
&\cdot &\frac{\eta_i \eta_j Q_i(\tau) Q_j(\tau) \xi^\mu_i(\tau)\xi^\nu_i(\tau)
l_\mu (\tau, \vec{\sigma}) \eta_{\nu \beta} \xi^\alpha_j(\tau)\xi^\beta_j(\tau)
l_\alpha (\tau, \vec{\sigma})}
{\sqrt{m_i^2 - \gamma^{\breve{r} \breve{s}}(\tau, \vec{\sigma})
k_{i \breve{r}}(\tau) k_{i \breve{s}}(\tau)}
\sqrt{m_j^2 - \gamma^{\breve{r} \breve{s}}(\tau, \vec{\sigma})
k_{j \breve{r}}(\tau) k_{j \breve{s}}(\tau)}} + \nonumber \\
&+& \frac{i}{\sqrt{\gamma(\tau, \vec{\sigma})}}
 \sum_{i=1}^N \delta^3 (\vec{\sigma} -\vec{\eta}_i(\tau)) \cdot \nonumber \\
&\cdot & \frac{\eta_i Q_i(\tau) \xi^\mu_i(\tau)\xi^\nu_i(\tau)}
{\sqrt{m_i^2 - \gamma^{\breve{r} \breve{s}}(\tau, \vec{\sigma})
k_{i \breve{r}}(\tau) k_{i \breve{s}}(\tau)}}
l_\mu (\tau, \vec{\sigma}) z_{\breve{s} \nu}(\tau, \vec{\sigma})
\pi^{\breve{s}} (\tau, \vec{\sigma}) + \nonumber \\
&-& \frac i2 \sum_{i=1}^N \delta^3 (\vec{\sigma} -\vec{\eta}_i(\tau))
 \frac{\eta_i Q_i(\tau) \xi^\mu_i(\tau)\xi^\nu_i(\tau)}
{\sqrt{m_i^2 - \gamma^{\breve{r} \breve{s}}(\tau, \vec{\sigma})
k_{i \breve{r}}(\tau) k_{i \breve{s}}(\tau)}}\cdot \nonumber \\
&\cdot & z_{\breve{u} \mu}(\tau, \vec{\sigma})
 z_{\breve{v} \nu}(\tau, \vec{\sigma})
\gamma^{\breve{r} \breve{u}}(\tau, \vec{\sigma})
\gamma^{\breve{s} \breve{v}}(\tau, \vec{\sigma})
F_{\breve{r} \breve{s}}(\tau, \vec{\sigma}) \Big] + \nonumber \\
&-& \gamma^{\breve{r} \breve{s}}(\tau, \vec{\sigma})
z_{\breve{s} \mu}(\tau, \vec{\sigma})
\Big[F_{\breve{r} \breve{u}}(\tau, \vec{\sigma})
\pi^{\breve{u}}(\tau, \vec{\sigma}) +\nonumber \\
&+&\sum_{i=1}^N \delta^3 (\vec{\sigma} -\vec{\eta}_i(\tau))
\big(k_{\breve{r} i} - Q_i(\tau) A_{\breve{r}}(\tau, \vec{\sigma})\big)
\Big] \approx 0.
\label{III23}
\end{eqnarray}

\noindent As we see, the component of ${\cal H}_\mu(\tau, \vec{\sigma})$ along
$l^{\mu}(\tau ,\vec \sigma )$ [i.e. orthogonal to $\Sigma_{\tau}$] contains the
electromagnetic energy density and also ``spin-spin", ``spin-electric field"
and ``spin-magnetic field" interactions. All of them are necessary to get 
the first class property for these constraints. Instead the components of
${\cal H}_\mu(\tau, \vec{\sigma})$ along $z^{\mu}_{\check r}(\tau ,\vec \sigma 
)$ [i.e. tangent to $\Sigma_{\tau}$] contain only the electromagnetic Poynting 
vector as with scalar particles \cite{lus1}.

The canonical Hamiltonian is

\begin{equation}
H_c = - \int{d^3\sigma A_\tau (\tau, \vec{\sigma}) \Gamma (\tau, \vec{\sigma})}
\label{III24}
\end{equation}

\noindent with $\Gamma (\tau, \vec{\sigma}) = \partial_{\breve{r}}
 \pi^{\breve{r}}  (\tau, \vec{\sigma}) - \sum_{i=1}^N
\delta^3(\vec{\sigma}- \vec{\eta}_i(\tau)) Q_i(\tau)$. The Dirac Hamiltonian is

\begin{equation}
H_D=H_c+\sum_{i=1}^N \mu_{i\mu}(\tau )\chi^{\mu}_i(\tau )+\int d^3\sigma
[\lambda^{\mu}{\cal H}_{\mu}+\mu_{\tau} \pi^{\tau}](\tau ,\vec \sigma );
\label{III25}
\end{equation}

\noindent since $[\chi_i^{\mu}(\tau )]^{*}=-\chi_i^{\mu}(\tau )$ are immaginary
odd quantities, $H_D$ is real if the odd multipliers are real: $\mu^{*}
_{i\mu}(\tau )=\mu_{i\mu}(\tau )$.

Since we have

\begin{eqnarray}
&&\{\chi^\mu_i(\tau) , \chi^\nu_j(\tau) \} = i \eta^{\mu \nu} \delta_{ij},
\nonumber \\
&&\{ {\cal H}_{\mu}(\tau ,\vec \sigma ), \chi^{\nu}_i(\tau ) \}=-l_{\mu}(\tau
,\vec \sigma ) [\, {{\delta^3(\vec \sigma -{\vec \eta}_i(\tau )) Q_i(\tau )}
\over {\sqrt{\gamma} \eta_i \sqrt{m^2_i-\gamma^{\check r\check s}\kappa
_{ir}(\tau )\kappa_{is}(\tau )} }}\cdot \nonumber \\
&&\sum_{j=1}^N{{\delta^3(\vec \sigma -{\vec \eta}_j(\tau )) Q_j(\tau )l_{\alpha}
\xi^{\alpha}_j(\tau )}\over {\sqrt{m^2_j-\gamma^{\check u\check v}\kappa
_{ju}(\tau )\kappa_{jv}(\tau )} }} (\xi_{j\beta}(\tau )\xi^{\beta}_i(\tau )
l^{\nu}-\xi_j^{\nu}(\tau )\xi^{\beta}_i(\tau )l_{\beta})+\nonumber \\
&+&{i\over {\sqrt{\gamma}}} {{\delta^3(\vec \sigma -{\vec \eta}_i(\tau )) 
Q_i(\tau )}\over {\eta_i\sqrt{m^2_i-\gamma^{\check r\check s}\kappa
_{ir}(\tau )\kappa_{is}(\tau )} }} \pi^{\check s}(l^{\nu}z_{\check s\alpha}\xi
^{\alpha}_i(\tau )-l_{\alpha} z^{\nu}_{\check s} \xi^{\alpha}_i(\tau ) )-
\nonumber \\
&-&i {{\delta^3(\vec \sigma -{\vec \eta}_i(\tau )) Q_i(\tau )}\over {\eta_i
\sqrt{m^2_i-\gamma^{\check r\check s}\kappa_{ir}(\tau )\kappa_{is}(\tau )} }}
z^{\nu}_{\check u} z_{\check v\beta} \gamma^{\check r\check u}\gamma
^{\check s\check v} F_{\check r\check s}\xi^{\beta}_i(\tau )\, ](\tau ,\vec 
\sigma )\not= 0,
\label{III26}
\end{eqnarray}

\noindent we see that the constraints $\chi^{\mu}_i\approx 0$ are second class.
By replacing the constraints ${\cal H}_{\mu}(\tau ,\vec \sigma )\approx 0$
with the new ones

\begin{equation}
{\cal H}^{'}_\mu(\tau, \vec{\sigma}) = {\cal H}_\mu(\tau, \vec{\sigma}) +
i \sum_{i=1}^N \{{\cal H}_\mu(\tau, \vec{\sigma}), \chi^\nu_i(\tau)\}
\chi_{i \nu} (\tau)\approx 0,
\label{III27}
\end{equation}

\noindent we have that the new constraints are first class [the first line
vanishes weakly by definition of ${\cal H}_{\mu}^{'}(\tau ,\vec \sigma )$]

\begin{eqnarray}
&&\{{\cal H}^{'}_\mu(\tau, \vec{\sigma}), \chi^\nu_i(\tau)\}\approx 0,
\nonumber \\
&&\{ {\cal H}^{'}_\mu (\tau, \vec{\sigma}) ,
 {\cal H}^{'}_\nu(\tau, \vec{\sigma}^\prime) \} \approx
\Big[ \Big( l_\mu (\tau, \vec{\sigma}) z_{\breve{r} \nu} (\tau, \vec{\sigma}) -
l_\nu (\tau, \vec{\sigma})
z_{\breve{r} \mu} (\tau, \vec{\sigma})\Big) \Big]
{\displaystyle\frac{\big(\pi^{\breve{r}} (\tau, \vec{\sigma})
- \pi^{\breve{r}}_\xi (\tau, \vec{\sigma}) \big)}
{\sqrt{\gamma (\tau, \vec{\sigma}) }}} + \nonumber \\
&&- z_{\breve{u} \mu} (\tau, \vec{\sigma})
\gamma^{\breve{u} \breve{r}} (\tau, \vec{\sigma})
F_{\breve{r} \breve{s}} (\tau, \vec{\sigma})
\gamma^{\breve{s} \breve{v}} (\tau, \vec{\sigma})
z_{\breve{v} \nu} (\tau, \vec{\sigma})\Big]\cdot
  \Gamma  (\tau, \vec{\sigma})
\delta^3 (\vec{\sigma} - \vec{\sigma}^\prime) \approx 0 ,
\label{III28}
\end{eqnarray}

\noindent with

\begin{equation}
\pi^{\breve{r}}_\xi (\tau, \vec{\sigma}) \equiv
i \sum_{i=1}^N \delta^3 (\vec{\sigma} -\vec{\eta}_i(\tau)) \eta_i
 \frac{ Q_i(\tau) \xi^\mu_i(\tau)\xi^\nu_i(\tau)
l_\mu (\tau, \vec{\sigma}) z_{\breve{s} \nu}(\tau, \vec{\sigma})
\gamma^{\breve{s} \breve{r}} (\tau, \vec{\sigma})}
{\sqrt{m_i^2 - \gamma^{\breve{s} \breve{r}}(\tau, \vec{\sigma})
k_{i \breve{r}}(\tau) k_{i \breve{s}}(\tau)}} .
\label{III29}
\end{equation}

In the previous equation we have anticipated the fact that
the time constancy of the primary constraints implies $\mu
_{i\mu}(\tau )\approx \int d^3\sigma \lambda^{\nu}(\tau ,\vec \sigma ) \{ {\cal
H}_{\nu}(\tau ,\vec \sigma ),\chi_{i \mu}(\tau ) \}$ and the Gauss law 
secondary constraint

\begin{equation}
\Gamma(\tau, \vec{\sigma}) = \partial_{\breve{r}}
\pi^{\breve{r}}(\tau, \vec{\sigma}) -
\sum_{i=1}^N \delta^3 (\vec{\sigma} -\vec{\eta}_i(\tau)) Q_i(\tau) \approx 0,
\label{III30}
\end{equation}

Therefore, we get a consistent set of second and first class constraints. Then
we can see what happens if we add by hand the following constraints (assumed to 
belong to the set of primary constraints of a modified Lagrangian), which 
generalize $p_{\mu}\xi^{\mu}\approx 0$ of the previous Section to spacelike 
hypersurfaces, to the set of Eqs.(\ref{III23})

\begin{equation}
\phi_i(\tau) = \big(\pi_i^\mu (\tau) + \frac i2 \xi_i^\mu(\tau)\big)
\int{d^3\sigma \rho_\mu (\tau, \vec{\sigma})
\equiv \big(\pi_i^\mu(\tau) + \frac i2 \xi_i^\mu(\tau)\big){p_s}_\mu} \approx 0,
\label{III31}
\end{equation}

\noindent
with $p^{\mu}_s=\int d^3\sigma \rho^{\mu}(\tau ,\vec \sigma )$. Now one has
the algebra

\begin{eqnarray}
\{\chi^\mu_i(\tau) , \chi^\nu_j(\tau) \} &=& i \eta^{\mu \nu} \delta_{ij},
~~~~~\{ \chi^\mu_i(\tau) , \phi_j(\tau) \} = 0 \nonumber \\
\{\phi_i(\tau) , \phi_j(\tau) \} &=& -i \delta_{ij} p_s^2,\nonumber \\
\{ {\cal H}_{\mu}(\tau ,\vec \sigma ),\chi^{\nu}_i(\tau ) \} &=&-l_{\mu}(\tau
,\vec \sigma ) [\, {{\delta^3(\vec \sigma -{\vec \eta}_i(\tau )) Q_i(\tau )}
\over {\sqrt{\gamma} \eta_i \sqrt{m^2_i-\gamma^{\check r\check s}\kappa
_{ir}(\tau )\kappa_{is}(\tau )} }}\cdot \nonumber \\
&&\sum_{j=1}^N{{\delta^3(\vec \sigma -{\vec \eta}_j(\tau )) Q_j(\tau )l_{\alpha}
\xi^{\alpha}_j(\tau )}\over {\sqrt{m^2_j-\gamma^{\check u\check v}\kappa
_{ju}(\tau )\kappa_{jv}(\tau )} }} (\xi_{j\beta}(\tau )\xi^{\beta}_i(\tau )
l^{\nu}-\xi_j^{\nu}(\tau )\xi^{\beta}_i(\tau )l_{\beta})+\nonumber \\
&+&{i\over {\sqrt{\gamma}}} {{\delta^3(\vec \sigma -{\vec \eta}_i(\tau )) 
Q_i(\tau )}\over {\eta_i\sqrt{m^2_i-\gamma^{\check r\check s}\kappa
_{ir}(\tau )\kappa_{is}(\tau )} }} \pi^{\check s}(l^{\nu}z_{\check s\alpha}\xi
^{\alpha}_i(\tau )-l_{\alpha} z^{\nu}_{\check s} \xi^{\alpha}_i(\tau ) )-
\nonumber \\
&-&i {{\delta^3(\vec \sigma -{\vec \eta}_i(\tau )) Q_i(\tau )}\over {\eta_i
\sqrt{m^2_i-\gamma^{\check r\check s}\kappa_{ir}(\tau )\kappa_{is}(\tau )} }}
z^{\nu}_{\check u} z_{\check v\beta} \gamma^{\check r\check u}\gamma
^{\check s\check v} F_{\check r\check s}\xi^{\beta}_i(\tau )\, ](\tau ,\vec 
\sigma ),\nonumber \\
\{ {\cal H}_{\mu}(\tau ,\vec \sigma ),\phi_i(\tau ) \} &=& p_{s\nu} \{ {\cal
H}_{\mu}(\tau ,\vec \sigma ),\chi^{\nu}_i(\tau ) \} .   
\label{III32}
\end{eqnarray}

If we define

\begin{eqnarray}
&{\cal H}^\ast_\mu(\tau, \vec{\sigma})& = {\cal H}_\mu(\tau, \vec{\sigma}) +
i \sum_{i=1}^N \{{\cal H}_\mu(\tau, \vec{\sigma}), \chi^\nu_i(\tau)\}
\chi_{i \nu} (\tau) + \nonumber \\
&-& \frac {i}{p_s^2} \sum_{i=1}^N \{ {\cal H}_\mu(\tau, \vec{\sigma}),
 \phi_i(\tau)\}\phi_i(\tau) \approx 0
\label{III33}
\end{eqnarray}

\noindent we get

\begin{eqnarray}
&&\{{\cal H}^\ast_\mu(\tau, \vec{\sigma}), \chi^\nu_i(\tau)\}\approx 0,~~~~~
\{{\cal H}^\ast_\mu(\tau, \vec{\sigma}), \phi_i(\tau)\} \approx 0,\nonumber \\
&&{}\nonumber \\
&&\{ {\cal H}^\ast_\mu (\tau, \vec{\sigma}) ,
 {\cal H}^\ast_\nu(\tau, \vec{\sigma}^\prime) \} \approx
\Big[ \Big( l_\mu (\tau, \vec{\sigma}) z_{\breve{r} \nu} (\tau, \vec{\sigma}) -
l_\nu (\tau, \vec{\sigma})
z_{\breve{r} \mu} (\tau, \vec{\sigma})\Big) \Big]
{\displaystyle\frac{\big(\pi^{\breve{r}} (\tau, \vec{\sigma})
- \pi^{\breve{r}}_\xi (\tau, \vec{\sigma}) \big)}
{\sqrt{\gamma (\tau, \vec{\sigma}) }}} + \nonumber \\
&&- z_{\breve{u} \mu} (\tau, \vec{\sigma})
\gamma^{\breve{u} \breve{r}} (\tau, \vec{\sigma})
F_{\breve{r} \breve{s}} (\tau, \vec{\sigma})
\gamma^{\breve{s} \breve{v}} (\tau, \vec{\sigma})
z_{\breve{v} \nu} (\tau, \vec{\sigma})\Big]\cdot
  \Gamma  (\tau, \vec{\sigma})
\delta^3 (\vec{\sigma} - \vec{\sigma}^\prime) \approx 0 ,
\label{III34}
\end{eqnarray}

\noindent with the same $\pi^{\breve{r}}_\xi (\tau, \vec{\sigma})$ of Eq.
(\ref{III29}).

By introducing the Dirac Hamiltonian [$\rho^{*}_i(\tau )=\rho_i(\tau )$ are
real odd multipliers, since $\phi^{*}_i(\tau )=-\phi_i(\tau )$]

\begin{eqnarray}
H_D &=& \int{d^3\sigma \Big[\lambda^\mu (\tau, \vec{\sigma})
{\cal H}_\mu(\tau, \vec{\sigma}) - A_{\tau}(\tau, \vec{\sigma})
\Gamma (\tau, \vec{\sigma})  + \mu_\tau (\tau, \vec{\sigma})
  \pi^\tau (\tau, \vec{\sigma})   \Big] } +\nonumber \\
&+& \sum_{i=1}^N [\rho_i(\tau )\phi_i(\tau )+\mu_{i \mu}(\tau ) \chi^{\mu}
_i(\tau )],
\label{III35}
\end{eqnarray}

\noindent we have that the time constancy of the primary constraints implies: 
i) $\mu_{i \mu}(\tau )\approx \int d^3\sigma \lambda^{\nu}(\tau ,\vec \sigma )
 \{ {\cal H}_{\nu}(\tau ,\vec \sigma ), \chi_{i \mu}(\tau ) \}$; ii)
$\rho_i(\tau ) \approx \int d^3\sigma \lambda^{\nu}(\tau ,\vec \sigma ) \{ 
{\cal H}_{\nu}(\tau ,\vec \sigma ), \phi_i(\tau ) \}$; iii) $\Gamma (\tau ,\vec
\sigma ) \approx 0$; iv) no further condition is implied by the time constancy 
of the Gauss law constraint. Therefore, we have that the constraints 
$\phi_i(\tau) \approx 0$, $\chi^\mu_i(\tau) \approx 0$  are second class,
while ${\cal H}^{*}_{\mu}(\tau ,\vec \sigma )\approx 0$, $\pi^{\tau}
(\tau ,\vec \sigma )\approx 0$ and $\Gamma (\tau ,\vec \sigma )\approx 0$
are first class as expected. The final Dirac Hamiltonian is

\begin{equation}
H^F_D=\int{d^3\sigma \Big[\lambda^\mu (\tau, \vec{\sigma})
{\cal H}^\ast_\mu(\tau, \vec{\sigma}) - A_{\tau}(\tau, \vec{\sigma})
\Gamma (\tau, \vec{\sigma})  + \mu_\tau (\tau, \vec{\sigma})
  \pi^\tau (\tau, \vec{\sigma})   \Big] } .
\label{III36}
\end{equation}

The conserved Poincar\'e generators are

\begin{eqnarray}
p_s^\mu &=& \int{d^3\sigma \rho^\mu (\tau, \vec{\sigma}) } \nonumber \\
J^{\mu \nu} &=& \int{d^3\sigma \Big[ z^\mu (\tau, \vec{\sigma})
\rho^\nu (\tau, \vec{\sigma}) - z^\nu (\tau, \vec{\sigma})
\rho^\mu (\tau, \vec{\sigma}) \Big] - \sum_{i=1}^N
\Big[ \xi_i^\mu (\tau) \pi_i^\nu (\tau) - \xi_i^\nu (\tau) \pi_i^\mu (\tau)
\Big]} . 
\label{III37}
\end{eqnarray}

Since $p^{\mu}_s$ is a constant of the motion independently from the isolated
system under investigation, we have the possibility of reducing the $\xi^{\mu}
_i$'s from 4 to 3 for each particle independently from the interactions but at
the price that the whole hypersurface $\Sigma_{\tau}$ is involved in the 
reduction [it weakly depends on the total 4-momentum of the isolated system
by using ${\cal H}_{\mu}(\tau, \vec{\sigma})\approx 0$].

Since this Hamiltonian definition of spinning particles plus the 
electromagnetic field  on spacelike hypersurfaces gives a perfectly 
consistent set of constraints, we can ask about the Lagrangian generating it.

In the case N=1 and in absence of the electromagnetic field, we have the
primary constraints (besides the ones associated with the Grassmann variables 
for the  electric charge)

\begin{eqnarray}
&&{\cal H}_{\mu}(\tau ,\vec \sigma )=\rho_{\mu}(\tau ,\vec \sigma )-\delta
^3(\vec \sigma -{\vec \eta}(\tau ))[l_{\mu}(\tau ,\vec \sigma ) \eta
\sqrt{m^2-\gamma^{\check r\check s}(\tau ,\vec \sigma )\kappa_{\check r}(\tau )
\kappa_{\check s}(\tau )}+\nonumber \\
&&+\gamma^{\check r\check s}(\tau ,\vec \sigma )z_{\check s \mu}(\tau ,\vec 
\sigma ) \kappa_{\check r}(\tau ) ] \approx 0,\nonumber \\
&&\chi_{\mu}(\tau )=\pi_{\mu}(\tau ) -{i\over 2} \xi_{\mu}(\tau ) \approx 0,
\nonumber \\
&&\phi (\tau )=(\pi_{\mu}(\tau )+{i\over 2}\xi_{\mu}(\tau ))\int d^3\sigma
\rho^{\mu}(\tau ,\vec \sigma ) \approx 0,
\label{III38}
\end{eqnarray}

\noindent which give the Hamiltonian definition of the spinning particle with 
definite sign of the energy. The derived vector $p^{\mu}(\tau )=l^{\mu}(\tau ,
\vec \eta (\tau )) \eta \sqrt{m^2-\gamma^{\check r\check s}(\tau ,\vec \eta 
(\tau ))\kappa_{\check r}(\tau )\kappa_{\check s}(\tau )}+\gamma^{\check 
r\check s}(\tau ,\vec \eta (\tau ))z^{\mu}_{\check s}(\tau ,\vec \eta (\tau ))
\kappa_{\check r}(\tau )$ is a solution of the mass-shell constraint $p^2-
m^2\approx 0$ with $sign\, p^o=\eta$. The Dirac Hamiltonian is $H_D=\int 
d^3\sigma \lambda^{\mu}(\tau ,\vec \sigma ){\cal H}_{\mu}(\tau ,\vec \sigma )
+\mu_{\mu}(\tau ) \chi^{\mu}(\tau )+\lambda (\tau )\phi (\tau )$.
One has $z^{\mu}_{\tau}(\tau ,\vec \sigma )\, {\buildrel \circ \over =}\,
\{ z^{\mu}(\tau ,\vec \sigma ),H_D \} =-\lambda^{\mu}(\tau ,\vec \sigma )-
\lambda (\tau ) i \xi^{\mu}(\tau )$. This equation determines $\lambda^{\mu}
(\tau ,\vec \sigma )$.

The inverse Legendre transformation, after having used the constraints
$\chi_{\mu}\approx 0$, ${\cal H}_{\mu}(\tau ,\vec \sigma )\approx 0$, to
eliminate the momenta $\pi_{\mu}(\tau )$, $\rho_{\mu}(\tau ,\vec \sigma )$, and 
the Hamilton equations ${\dot {\vec \eta}}(\tau )\, {\buildrel \circ \over =}\,
\{ \vec \eta (\tau ),H_D \} =\lambda_{\mu}(\tau ,\vec \eta (\tau )) \{ l^{\mu}
(\tau ,\vec \eta (\tau )) \eta {{-\gamma^{\check r\check s}(\tau ,\vec \eta 
(\tau ))\kappa_{\check s}(\tau )}\over {\sqrt{m^2-\gamma^{\check r\check s}
(\tau ,\vec \eta (\tau ))\kappa_{\check r}(\tau )\kappa_{\check s}(\tau )} }}+
\gamma^{\check r\check s}(\tau ,\vec \eta (\tau ))z^{\mu}_{\check s}(\tau 
,\vec \eta (\tau )) \}$ to eliminate $\vec \kappa (\tau )$, gives
the Lagrangian

\begin{eqnarray}
L&=&{\dot \xi}_{\mu}\pi^{\mu}-{\dot \eta}^{\check r}\kappa_{\check r}-\int
d^3\sigma z^{\mu}_{\tau}(\tau ,\vec \sigma )\rho_{\mu}(\tau ,\vec \sigma )-
H_D=\nonumber \\
&=&\int d^3\sigma
\delta^3(\vec \sigma -\vec \eta (\tau )) \{ -{i\over 2} \xi_{\mu}(\tau )
{\dot \xi}^{\mu}(\tau )-\nonumber \\
&-&\eta m\sqrt{g_{\tau\tau}(\tau ,\vec \sigma )+2g_{\tau \check r}(\tau 
,\vec \sigma ){\dot \eta}^{\check r}(\tau )+g_{\check r\check s}(\tau ,\vec 
\sigma ){\dot \eta}^{\check r}(\tau ){\dot \eta}^{\check s}(\tau )}-
\nonumber \\
&-& \lambda (\tau ) i\xi_{\mu}(\tau ) \eta m {{l^{\mu} \sqrt{g_{\tau\tau}-
\gamma^{\check r\check s}g_{\tau \check r}g_{\tau \check s}}+\gamma
^{\check r\check s}z^{\mu}_{\check s} (g_{\tau \check r}+g_{\check 
r\check u}{\dot \eta}^{\check u})(\tau )}\over 
{\sqrt{g_{\tau\tau}+2g_{\tau \check r}
{\dot \eta}^{\check r}(\tau )+g_{\check r\check s}{\dot \eta}^{\check r}(\tau )
{\dot \eta}^{\check s}(\tau )} }} (\tau ,\vec \sigma )  \} =\nonumber \\
&=&\int d^3\sigma
\delta^3(\vec \sigma -\vec \eta (\tau )) \{ -{i\over 2} \xi_{\mu}(\tau )
{\dot \xi}^{\mu}(\tau )-\nonumber \\
&-&\eta m\sqrt{g_{\tau\tau}(\tau ,\vec \sigma )+2g_{\tau \check r}(\tau 
,\vec \sigma ){\dot \eta}^{\check r}(\tau )+g_{\check r\check s}(\tau ,\vec 
\sigma ){\dot \eta}^{\check r}(\tau ){\dot \eta}^{\check s}(\tau )}-
\nonumber \\
&-& \lambda (\tau ) i\xi_{\mu}(\tau ) \eta m {{z_{\tau}^{\mu} 
+z^{\mu}_{\check r} {\dot \eta}^{\check r}(\tau )}\over 
{\sqrt{g_{\tau\tau}+2g_{\tau \check r}
{\dot \eta}^{\check r}(\tau )+g_{\check r\check s}{\dot \eta}^{\check r}(\tau )
{\dot \eta}^{\check s}(\tau )} }} (\tau ,\vec \sigma )  \} ,
\label{III39}
\end{eqnarray}

By considering $\lambda (\tau )$ an independent variable, this Lagrangian 
generates  the primary constraints $\chi^{\mu}(\tau )\approx 0$, ${\cal H}
_{\mu}(\tau ,\vec \sigma )\approx 0$ and $\pi_{\lambda}(\tau )\approx 0$ and
the Dirac Hamiltonian $H_D=\lambda (\tau )\phi (\tau )+\int d^3\sigma 
\lambda^{\mu}(\tau ,\vec \sigma ){\cal H}_{\mu}(\tau ,\vec \sigma )
+\mu_{\mu}(\tau ) \chi^{\mu}(\tau )+\zeta (\tau )\pi_{\lambda}(\tau )$. The
time constancy of this last constraint generates the secondary constraint
$\phi (\tau )\approx 0$ and its $\tau$-constancy implies the vanishing of the
multiplier $\zeta (\tau )$, so that we get the desired Dirac Hamiltonian with
$\lambda (\tau )$ an arbitrary gauge variable.

It is interesting to note that this Lagrangian is the reformulation on 
spacelike hypersurfaces (therefore with only one sign of the energy) of the
manifestly covariant Lagrangian $L(\tau )= -{i\over 2}\xi_{\mu}(\tau )
{\dot \xi}^{\mu}(\tau ) -\eta m \sqrt{{\dot x}^2(\tau )}-i\lambda (\tau )
\xi_{\mu}(\tau ) \eta m {{{\dot x}^{\mu}(\tau )}\over {\sqrt{{\dot x}^2(\tau 
)}}}$ to be compared with Eq.(7).

With the same technique one could try to get the Lagrangian generating Eqs.
(\ref{III23}) and (\ref{III31}) as primary constraints. However, we shall not 
do it, because the resulting Lagrangian would be very complicated (due to
non-minimal couplings) and not very illuminating.

\vfill\eject

\section{The restriction to Wigner's hyperplanes}

As shown in Ref.\cite{lus1}, the restriction from arbitrary hypersurfaces to 
hyperplanes is done by introducing the gauge-fixings

\begin{eqnarray}
&&\zeta^\mu (\tau, \vec{\sigma}) = z^\mu (\tau, \vec{\sigma})
-x_s^\mu(\tau) -b_{\breve{r}}^\mu (\tau) \sigma^{\breve{r}} \approx 0,
\nonumber \\
&&\{ \zeta^\mu (\tau, \vec{\sigma}) ,  {\cal H}^\ast_\nu(\tau,
\vec{\sigma}^\prime) \} =
- \eta^\mu_\nu \delta^3 (\vec{\sigma} - \vec{\sigma}^\prime),
\label{IV1}
\end{eqnarray}

\noindent and the Dirac brackets

\begin{equation}
\{ A, B\}^\ast = \{ A , B \} - \int{d^3\sigma \{ A ,
\zeta^\mu (\tau, \vec{\sigma}) \} \{ {\cal H}^\ast_\mu(\tau, \vec{\sigma}) ,
B \} } + \int{d^3\sigma \{ A ,
{\cal H}^\ast_\mu (\tau, \vec{\sigma}) \}
\{ \zeta^\mu(\tau, \vec{\sigma}) , B \} }   .
\label{IV2}
\end{equation}

The hyperplane $\Sigma_{\tau H}$ is described by 10 configuration variables: 
an origin $x^{\mu}_s(\tau )$ and the 6 independent degrees of freedom in an 
orthonormal tetrad $b^{\mu}_{\check A}(\tau )$ [${}b^{\mu}_{\check A}\, \eta
_{\mu\nu} b^{\nu}_{\check B}=\eta_{\check A\check B}$] with $b^{\mu}_{\tau}=l
^{\mu}$, where $l^{\mu}$ is the $\tau$-independent normal to the hyperplane.
Now, we have $z^{\mu}_{\check r}(\tau ,\vec \sigma )\equiv b^{\mu}_{\check r}
(\tau )$, $z^{\mu}_{\tau}(\tau ,\vec \sigma )\equiv {\dot x}^{\mu}_s(\tau )+
b^{\mu}_{\check r}(\tau ) \sigma^{\check r}$, $g_{\check r\check s}(\tau ,\vec 
\sigma )\equiv -\delta_{\check r\check s}$, $\gamma^{\check r\check s}
(\tau ,\vec \sigma )\equiv -\delta^{\check r\check s}$, $\gamma (\tau ,\vec 
\sigma )=det\, g_{\check r\check s}(\tau ,\vec \sigma )\equiv 1$. The
nonvanishing Dirac brackets of the variables $x^\mu_s$,
$p_s^\mu$, $b^\mu_{\breve{A}}$, $S^{\mu \nu}_s$, $A_{\breve{A}}$,
$\pi^{\breve{A}}$, $\xi_i^\mu$, $\pi^\nu_j$ are

\begin{eqnarray}
\{ x_s^\mu (\tau), p_s^\nu \}^\ast &=& -\eta^{\mu \nu}, \nonumber \\
\{\eta_i^{\breve{r}}(\tau), k^{\breve{s}}_j (\tau) \}^\ast &=&
\delta_{ij} \delta^{\breve{r} \breve{s}}, \nonumber \\
\{ S^{\mu \nu}_s(\tau) , b^\rho_{\breve{A}}\}^\ast &=&
\eta^{\rho \nu} b^\mu_{\breve{A}} (\tau) -
\eta^{\rho \mu} b^\nu_{\breve{A}} (\tau), \nonumber \\
\{ S^{\mu \nu}_s(\tau) , S^{\alpha \beta}_s(\tau) \}^\ast &=&
C^{\mu \nu \alpha \beta}_{\gamma \delta} S^{\gamma \delta}_s (\tau), 
\nonumber \\
\{ \xi^\mu_i (\tau) , \pi^\nu_j (\tau) \}^\ast &=&
- \eta^{\mu \nu} \delta_{ij}.
\label{IV3}
\end{eqnarray}

While $p^{\mu}_s$ is the momentum conjugate to $x^{\mu}_s$, the 6 independent
momenta conjugate to the 6 degrees of freedom in the $b^{\mu}_{\check A}$'s
are hidden in $S_s^{\mu\nu}$, which is a component of the angular momentum 
tensor

\begin{eqnarray}
J^{\mu \nu} &=& L^{\mu \nu}_s + S^{\mu \nu}_s + S^{\mu \nu}_\xi , 
\nonumber \\
L^{\mu \nu}_s &=& x^\mu_s(\tau) p^\nu_s - x^\nu_s(\tau) p^\mu_s , 
\nonumber \\
S^{\mu \nu}_s &=& b^\mu_{\breve{r}} (\tau) \int{d^3\sigma \sigma^{\breve{r}}
\rho^\nu (\tau, \vec{\sigma})} -
b^\nu_{\breve{r}} (\tau) \int{d^3\sigma \sigma^{\breve{r}}
\rho^\mu (\tau, \vec{\sigma})} , \nonumber \\
S^{\mu \nu}_\xi  &=& - \sum_{i=1}^N
\Big( \xi_i^\mu (\tau) \pi_i^\nu (\tau) - \xi_i^\nu (\tau) \pi_i^\mu (\tau)
\Big) ,\nonumber \\
&&{}\nonumber \\
&&\{ J^{\mu\nu},J^{\alpha\beta} \}^{*} = C^{\mu\nu\alpha\beta}_{\gamma\delta} 
J^{\gamma\delta},\quad\quad \{ L_s^{\mu\nu},L_s^{\alpha\beta} \}^{*} = C
^{\mu\nu\alpha\beta}_{\gamma\delta} L_s^{\gamma\delta},\nonumber \\
&&\{ S_s^{\mu\nu},S_s^{\alpha\beta} \}^{*} 
= C^{\mu\nu\alpha\beta}_{\gamma\delta} 
S_s^{\gamma\delta},\quad\quad \{ S_{\xi}^{\mu\nu},S_{\xi}^{\alpha\beta} \}^{*} 
= C^{\mu\nu\alpha\beta}_{\gamma\delta} S_{\xi}^{\gamma\delta},\nonumber \\
&&C^{\mu\nu\alpha\beta}_{\gamma\delta}=\eta^{\nu}_{\gamma}\eta^{\alpha}
_{\delta}\eta^{\mu\beta}+\eta^{\mu}_{\gamma}\eta^{\beta}_{\delta}\eta
^{\nu\alpha}-\eta^{\nu}_{\gamma}\eta^{\beta}_{\delta}\eta^{\mu\alpha}-
\eta^{\mu}_{\gamma}\eta^{\alpha}_{\delta}\eta^{\nu\beta}.
\label{IV4}
\end{eqnarray}

Then, we eliminate the second class constraints $\chi^\mu_i \approx 0$, 
$\phi_i \approx 0$ with the new Dirac brackets

\begin{equation}
\{A,B\}^\ast_D = \{A,B\}^\ast + i \sum_{i=1}^N \{ A, \chi^\mu_i(\tau) \}^\ast
\eta_{\mu \nu} \{\chi_i^\nu (\tau), B \}^\ast
- \frac{i}{p_s^2} \sum^N_{i=1} \{ A, \phi_i(\tau) \}^\ast
\{\phi_i(\tau) , B\}^\ast.
\label{IV5}
\end{equation}

Now we have

\begin{eqnarray}
&&\pi_i^\mu(\tau) \equiv \frac i2 \xi^\mu_i (\tau),\nonumber \\
&&\xi_i^\mu(\tau) {p_s}_\mu \equiv 0 ,\nonumber \\
&&\Rightarrow \, \xi^\mu_i(\tau) \equiv \xi^\mu_{i \perp}(\tau) \equiv
\Pi^{\mu \nu} \xi_{i \nu}(\tau)=\Big( \eta^{\mu \nu} - 
\frac{p_s^\mu p_s^\nu}{p_s^2}\Big) \xi_{i\nu}(\tau ),\nonumber \\
&&S^{\mu\nu}_{\xi}\equiv -i \sum_{i=1}^N \xi^{\mu}_i \xi^{\nu}_i,\quad\quad 
p_{s \mu} S^{\mu\nu}_{\xi}\equiv 0.
\label{IV6}
\end{eqnarray}

However, now we get the following non canonical Dirac brackets on $\Sigma_{\tau
H}$

\begin{eqnarray}
\{ x_s^\mu(\tau), x_s^\nu(\tau) \}^\ast_D &=&
- i \sum_{i=1}^N \frac{\xi^\mu_i(\tau) \xi^\nu_i(\tau)}{p_s^2}
\equiv \frac{S^{\mu \nu}_\xi (\tau)}{p_s^2}, \nonumber \\
\{ x_s^\mu(\tau), \xi_i^\nu(\tau) \}^\ast_D &=&
\frac{\xi^\mu_i(\tau) p_s^\nu}{p_s^2}, \nonumber \\
\{ \xi_i^\mu(\tau), \xi_j^\nu(\tau) \}^\ast_D &=&
i \Big( \eta^{\mu \nu} - \frac{p_s^\mu p_s^\nu}{p_s^2}\Big) \delta_{ij}
\equiv i \Pi^{\mu \nu} \delta_{ij}.
\label{IV7}
\end{eqnarray}

In this way, we have eliminated the components of $\xi^{\mu}_i$ parallel to 
$p^{\mu}_s$ in a Lorentz-invariant way as in Eq.(\ref{II4}) and in the 
Hamiltonian theory associated with Eq.(\ref{II8}). The spin of each particle 
is described only by 3 Grassmann variables and the spin tensor $S_{\xi}
^{\mu\nu}$ satisfies a Weyssenhoff condition. The angular momentum tensor 
becomes

\begin{eqnarray}
&&J^{\mu \nu} = L^{\mu \nu}_s + S^{\mu \nu}_s + S^{\mu \nu}_\xi,\nonumber \\
&&S^{\mu \nu}_\xi = - i \sum_{i=1}^N\xi_i^\mu (\tau) \xi_i^\nu (\tau) ,
\nonumber \\
&&{}\nonumber \\
&&\{ J^{\mu \nu} , J^{\alpha \beta} \}^\ast_D  =
\{ J^{\mu \nu} , J^{\alpha \beta} \}^\ast =
C^{\mu \nu \alpha \beta}_{\gamma \delta} J^{\gamma \delta} , \nonumber \\
&&\{ L^{\mu \nu}_s , L^{\alpha \beta}_s \}^\ast_D  =
C^{\mu \nu \alpha \beta}_{\gamma \delta} L^{\gamma \delta}_s -
P^{\mu \nu \alpha \beta}_{\gamma \delta} S^{\gamma \delta}_\xi , \nonumber \\
&&\{ S^{\mu \nu}_\xi , S^{\alpha \beta}_\xi \}^\ast_D  =
C^{\mu \nu \alpha \beta}_{\gamma \delta} S^{\gamma \delta}_\xi -
P^{\mu \nu \alpha \beta}_{\gamma \delta} S^{\gamma \delta}_\xi ,\nonumber \\
&&\{ L^{\mu \nu}_s , S^{\alpha \beta}_\xi \}^\ast_D  =
P^{\mu \nu \alpha \beta}_{\gamma \delta} S^{\gamma \delta}_\xi ,\nonumber \\
&&\{ L^{\mu \nu}_s + S_\xi^{\mu \nu},
 L^{\alpha \beta}_s + S^{\alpha \beta}_\xi\}^\ast_D  =
C^{\mu \nu \alpha \beta}_{\gamma \delta} \Big(L_s^{\gamma \delta} +
S_\xi^{\gamma \delta} \Big) ,\nonumber \\
&&{}\nonumber \\
&&P^{\mu \nu \alpha \beta}_{\gamma \delta} \equiv
\frac{p_s^\mu p_s^\beta}{p_s^2} \eta^\nu_\gamma \eta^\alpha_\delta +
\frac{p_s^\nu p_s^\alpha}{p_s^2} \eta^\mu_\gamma \eta^\beta_\delta -
\frac{p_s^\mu p_s^\alpha}{p_s^2} \eta^\nu_\gamma \eta^\beta_\delta -
\frac{p_s^\nu p_s^\beta}{p_s^2} \eta^\mu_\gamma \eta^\alpha_\delta .
\label{IV8}
\end{eqnarray}

Since by asking the time constancy of the gauge fixings (\ref{IV1})
we get\cite{lus1} $\lambda^{\mu}(\tau ,\vec \sigma )={\tilde \lambda}
^{\mu}(\tau )+{\tilde \lambda}^{\mu}{}_{\nu}(\tau )b^{\nu}_{\check r}(\tau ) 
\sigma^{\check r}$, ${\tilde \lambda}^{\mu}(\tau )=-{\dot x}^{\mu}_s(\tau )$, 
${\tilde \lambda}^{\mu\nu}(\tau )=-{\tilde \lambda}^{\nu\mu}(\tau )={1\over 2}
\sum_{\check r}[{\dot b}^{\mu}_{\check r}b^{\nu}_{\check r}-b^{\mu}_{\check r}
{\dot b}^{\nu}_{\check r}](\tau )$, the Dirac Hamiltonian becomes

\begin{equation}
H^F_D = \tilde{\lambda}^\mu(\tau) \tilde{H}_\mu(\tau) -
\frac 12 \tilde{\lambda}^{\mu \nu}(\tau) \tilde{H}_{\mu \nu}(\tau) +
\int{d^3\sigma \Big[ - A_\tau (\tau, \vec{\sigma}) \Gamma (\tau, \vec{\sigma}) +
\mu_\tau (\tau, \vec{\sigma}) \pi^\tau (\tau, \vec{\sigma}) \Big] }
\label{IV9}
\end{equation}

\noindent and we are left with only 12 first class constraints

\begin{eqnarray}
\pi^\tau (\tau, \vec{\sigma}) &\approx & 0, \nonumber \\
\Gamma (\tau, \vec{\sigma}) &=& \partial_{\breve{r}}
 \pi^{\breve{r}} (\tau, \vec{\sigma}) -
\sum_{i=1}^N \delta^3 (\vec{\sigma} -\vec{\eta}_i(\tau)) Q_i(\tau) \approx 0,
\nonumber \\
\tilde{H}_\mu(\tau) &=&
\int{d^3\sigma {\cal H}_\mu (\tau, \vec{\sigma}) }= {p_s}_\mu -
b_{\mu \tau} \Big\{ \int{d^3\sigma \Big(\frac {\vec{\pi}^2 (\tau, \vec{\sigma})
+ \vec{B}^2(\tau, \vec{\sigma}) }{2}\Big)} + \nonumber \\
&+& \sum_{i=1}^N \eta_i \sqrt{m_i^2 +\big[\vec{k}_i(\tau) -
Q_i(\tau) \vec{A}(\tau, \vec{\eta}_i(\tau)) \big]^2} + \nonumber \\
&+& \frac 12 \sum_{i,j=1}^N
\delta^3 (\vec{\eta}_i(\tau) - \vec{\eta}_j(\tau)) \cdot \nonumber \\
&\cdot & \frac{\eta_i \eta_j Q_i(\tau) Q_j(\tau)
 \xi^\mu_i(\tau)\xi^\nu_i(\tau)}
{\sqrt{m_i^2 + \vec{k}_i^2(\tau)} \sqrt{m_j^2 + \vec{k}_j^2(\tau)}}
b_{\tau \mu} \eta_{\nu \beta} \xi^\alpha_j(\tau)\xi^\beta_j(\tau)
b_{\tau \alpha} \Big] + \nonumber \\
&+& i \sum_{i=1}^N \eta_i  \frac{ Q_i(\tau) \xi^\mu_i(\tau)\xi^\nu_i(\tau)}
{\sqrt{m_i^2 + \vec{k}_i^2(\tau)}}b_{\tau \mu} b_{\breve{s} \nu} (\tau)
\pi^{\breve{s}} (\tau, \vec{\eta}_i(\tau)) + \nonumber \\
&-& \frac i2 \sum_{i=1}^N \eta_i  \frac{ Q_i(\tau) \xi^\mu_i(\tau)
\xi^\nu_i(\tau)}
{\sqrt{m_i^2 + \vec{k}_i^2(\tau)}} b_{\breve{u} \mu} (\tau)
b_{\breve{v} \nu} (\tau) F_{\breve{u} \breve{v}} (\tau, \vec{\eta}_i(\tau))
\Big\} + \nonumber \\
&+& b_{\breve{r} \mu} (\tau) \Big\{\int {d^3\sigma \big[\vec{\pi} \wedge
\vec{B}\big]_{\breve{r}}(\tau, \vec{\sigma}) +
\sum_{i=1}^N \big[k_{i \breve{r}}(\tau) -
Q_i(\tau) A_{\breve{r}}(\tau, \vec{\eta}_i(\tau))}\big] \Big\}  \approx 0, 
\nonumber \\
\tilde{H}^{\mu \nu}(\tau) &=& b_{\breve{r}}^\mu (\tau)
\int{d^3\sigma \sigma^{\breve{r}}{\cal H}^\nu (\tau, \vec{\sigma}) } -
b_{\breve{r}}^\nu (\tau)
\int{d^3\sigma \sigma^{\breve{r}}{\cal H}^\mu (\tau, \vec{\sigma}) } = 
\nonumber \\
&=& S_s^{\mu \nu}(\tau )-\big( b_{\breve{r}}^\mu (\tau) b^\nu_\tau -
b_{\breve{r}}^\nu (\tau) b^\mu_\tau \big)
\Big\{ \int{d^3\sigma \sigma^{\breve{r}}
\frac {\vec{\pi}^2 (\tau, \vec{\sigma})
+ \vec{B}^2(\tau, \vec{\sigma}) }{2}} + \nonumber \\
&+& \frac 12 \sum_{i,j=1}^N
\delta^3 (\vec{\eta}_i(\tau) - \vec{\eta}_j(\tau)) \cdot \nonumber \\
&\cdot & \frac{\eta_i \eta_j Q_i(\tau) Q_j(\tau)
 \xi^\mu_i(\tau)\xi^\nu_i(\tau)}
{\sqrt{m_i^2 + \vec{k}_i^2(\tau)} \sqrt{m_j^2 + \vec{k}_j^2(\tau)}}
b_{\tau \mu} \eta_{\nu \beta} \xi^\alpha_j(\tau)\xi^\beta_j(\tau)
b_{\tau \alpha}  + \nonumber \\
&+& \sum_{i=1}^N \eta_i^{\breve{r}} \eta_i
\sqrt{m_i^2 + \big[\vec{k}_i(\tau) -
Q_i(\tau) \vec{A} (\tau, \vec{\eta}_i(\tau))\big]^2} + \nonumber \\
&+& i \sum_{i=1}^N \eta_i^{\breve{r}}(\tau) \eta_i
\frac{ Q_i(\tau) \xi^\mu_i(\tau)\xi^\nu_i(\tau)}
{\sqrt{m_i^2 + \vec{k}_i^2(\tau)}}b_{\tau \mu} b_{\breve{s} \nu} (\tau)
\pi^{\breve{s}} (\tau, \vec{\eta}_i(\tau)) + \nonumber \\
&-& \frac i2 \sum_{i=1}^N \eta_i^{\breve{r}}(\tau) \eta_i
\frac{ Q_i(\tau) \xi^\mu_i(\tau)\xi^\nu_i(\tau)}
{\sqrt{m_i^2 + \vec{k}_i^2(\tau)}} b_{\breve{u} \mu} (\tau)
b_{\breve{v} \nu} (\tau) F_{\breve{u} \breve{v}} (\tau, \vec{\eta}_i(\tau))
\Big\} + \nonumber \\
&+& \big( b_{\breve{r}}^\mu (\tau) b_{\breve{s}}^\nu (\tau)  -
b_{\breve{r}}^\nu (\tau) b_{\breve{s}}^\mu (\tau)  \big)
\Big\{ \int{ d^3\sigma \sigma^{\breve{r}}\big[\vec{\pi} \wedge
\vec{B}\big]_{\breve{s}}(\tau, \vec{\sigma})} + \nonumber \\
&+&\sum_{i=1}^N \eta_i^{\breve{r}}(\tau) \big[k_{i \breve{s}}(\tau) -
Q_i(\tau) A_{\breve{s}}(\tau, \vec{\eta}_i(\tau))\big]
\Big\} \approx 0.
\label{IV10}
\end{eqnarray}

The next step \cite{lus1} is to select all the configurations of the isolated 
system which are timelike, namely with $p^2_s > 0$. For them we can boost at 
rest with the standard Wigner boost $L^\mu_{.\nu}(\stackrel{o}p_s, p_s)$ for
timelike Poincar\'e orbits all the variables of the noncanonical basis
$x_s^\mu(\tau)$, $p_s^\mu$ , $b_{\breve{A}}^\mu(\tau )$ , $S_s^{\mu \nu}(\tau 
)$ , $A_{\breve{A}}(\tau ,\vec \sigma )$ , $\pi^{\breve{A}}(\tau ,\vec \sigma 
)$, $\vec{\eta}_i(\tau )$ , $\vec{k}_i(\tau )$ , $\theta_i(\tau )$ , $\theta_i
^\ast (\tau )$ , $\xi_i^\mu (\tau )$ with Lorentz indices (except $p^{\mu}_s$). 
This is a canonical transformation generated by $e^{\{ .,{\cal F}(\tau ) \} }$ 
with generating function [see Appendix A of Ref.\cite{longhi} for a similar
transformation]

\begin{eqnarray}
{\cal F}(\tau ) &=& \frac 12 \omega(p_s) I_{\mu \nu} (p_s) S_s^{\mu \nu}(\tau )
,\nonumber \\
&&{}\nonumber \\
I(p) &\equiv & \parallel I(p)^\mu_{.\nu} \parallel =
\pmatrix{0& -\frac{p_j}{|\vec{p}|}\cr \frac{p^i}{|\vec{p}|}& 0}, \nonumber \\
I_{\mu \nu}(p) &=& - I_{\nu \mu}(p),~~~~~~~~~~~~~~~~~~I^3(p) = I(p) ,
\nonumber \\
\cosh \omega(p) &=& \frac{\eta p_0}{\sqrt{p^2}},~~~~~~~~~
\sinh \omega(p) = \eta \frac{|\vec{p}|}{\sqrt{p^2}}, \nonumber \\
L^\mu_{.\nu} (p, \stackrel{o}p) &=&
\exp \big[ \omega (p) I (p) \big]^\mu_{.\nu} = \nonumber \\
&=& \big[ \cosh\big(\omega (p) I (p)\big)
+ \sinh \big(\omega (p) I (p)\big) \big]^\mu_{.\nu} = \nonumber \\
&=& \big[ \openone - I^2(p) + I^2(p)
\cosh \omega (p)
+ I(p) \sinh \omega (p) \big]^\mu_{.\nu}, \nonumber \\
L^\mu_{.\nu}(\stackrel{o}p, p) &=&
\exp \big[ - \omega (p) I (p) \big]^\mu_{.\nu}.
\label{IV11}
\end{eqnarray}

\noindent Since we have $\xi^\mu_i {p_s}_\mu = 0$, we get $I_{\mu \nu}(p_s) 
S_\xi^{\mu \nu} = 0$, so that the addition of $S^{\mu\nu}_{\xi}$ to $S^{\mu\nu}
_s$ in ${\cal F}$ is irrelevant.

The new noncanonical basis (with the same Dirac brackets) is

\begin{eqnarray}
\tilde{x}^\mu_s &=& x_s^\mu - \frac 12 \epsilon_\nu(u(p_s)) \eta_{AB}
\frac{\partial \epsilon^B_\rho(u(p_s))}{\partial {p_s}_\mu} S_s^{\nu \rho} =
\nonumber \\
&=& x^\mu_s - \frac{1}{\eta_s \sqrt{p^2_s} (p_s^0 +\eta_s \sqrt{p_s^2})}
\Big[{p_s}_\nu S_s^{\nu \mu} + \eta_s \sqrt{p^2_s} \Big( S_s^{0 \mu} -
S_s^{0 \nu} \frac{{p_s}_\nu p_s^\mu}{p_s^2} \Big) \Big] = \nonumber \\
&=& x^\mu_s - \frac{1}{\eta_s \sqrt{p^2_s}} \Big[ \eta^\mu_A
\Big(\bar{S}_s^{\bar{o} A} - \frac{\bar{S}_s^{Ar} p_s^r}
{p_s^0 +\eta_s \sqrt{p_s^2}} \Big) +
\frac{p_s^\mu + 2 \eta_s \sqrt{p_s^2} \eta^{\mu 0}}
{\eta_s \sqrt{p^2_s} (p_s^0 +\eta_s \sqrt{p_s^2})}
 \bar{S}_s^{\bar{o} r} p_s^r \Big] \nonumber \\
p_s^\mu &=& p_s^\mu,~~~\eta_i^{\breve{r}} = \eta_i^{\breve{r}}, ~~~
k_i^{\breve{r}} = k_i^{\breve{r}},~~~ A_{\breve{A}} = A_{\breve{A}} ,~~~
\pi^{\breve{A}} = \pi^{\breve{A}} \nonumber \\
\xi_i^\mu &=& \xi_i^\mu ,~~~ \theta_i^\ast = \theta_i^\ast ,~~~
\theta_i = \theta_i \nonumber \\
b^A_{\breve{B}} &=& \epsilon^A_\mu (u(p_s)) b_{\breve{B}}^\mu \nonumber \\
\tilde{S}_s^{\mu \nu} &=& S_s^{\mu \nu} + \frac 12 \epsilon^A_\rho (u(p_s))
\eta_{AB} \Big[ \frac{\partial \epsilon^B_\sigma (u(p_s))}{\partial {p_s}_\mu}
p_s^\nu - \frac{\partial \epsilon^B_\sigma (u(p_s))}{\partial {p_s}_\nu}
p_s^\mu \Big] S_s^{\rho \sigma} = \nonumber \\
&=& S_s^{\mu \nu} +
\frac{1}{\eta_s \sqrt{p^2_s} (p_s^0 +\eta_s \sqrt{p_s^2})}
\Big[{p_s}_\beta (S_s^{\beta \mu} p_s^\nu - S_s^{\beta \nu} p_s^\mu)
+ \eta_s \sqrt{p^2_s} (S_s^{0 \mu} p_s^\nu - S_s^{0 \nu} p_s^\mu)\Big] ,
\label{IV12}
\end{eqnarray}

\noindent where $u^{\mu}(p_s)=p^{\mu}_s/\eta_s\sqrt{p^2_s}=L^\mu_o
(\stackrel{o}p_s, p_s)$ [$\eta_s=\pm1$].

For later use, let us introduce the spin tensors

\begin{eqnarray}
\bar{S}_s^{AB} &=& \epsilon^A_\mu (u(p_s))\epsilon^B_\nu (u(p_s)) S_s
^{\mu \nu},\nonumber \\
\bar{S}_\xi^{AB} &=& \epsilon^A_\mu (u(p_s))
\epsilon^B_\nu (u(p_s)) S_\xi^{\mu \nu}.
\label{IV13}
\end{eqnarray}

Since $\xi_{i \mu}p^{\mu}_s\equiv 0$ implies $\xi_{i \tau}=\xi_{i \mu}u^{\mu}
(p_s)=\xi_{i \mu} L^\mu_o(\stackrel{o}p_s, p_s)\equiv 0$, we can reduce to 3 
for each particle the Grassmann variables describing spin

\begin{equation}
\xi^r_i(\tau) \equiv  \epsilon^r_\mu (u(p_s)) \xi^\mu_i (\tau)
~~~~~r=1,2,3.
\label{IV14}
\end{equation}

Then, we get

\begin{eqnarray}
&&\bar{S}^{\bar{o} B}_{\xi} = 0, \nonumber \\
&&\bar{S}^{rs}_{\xi} = -i \sum_{i=1}^N \epsilon^r_\mu (u(p_s))
\epsilon^s_\nu (u(p_s)) \xi^\mu_i \xi^\nu_i=-i \sum_{i=1}^N \xi^r_i\xi^s_i,
\nonumber \\
&&{\bar S}^r_{\xi}={1\over 2}\epsilon^{ruv} {\bar S}_{\xi}^{uv}=\sum_{i=1}^N 
{\bar S}^r_{i\xi},\nonumber \\
&&{\bar S}^r_{i\xi}=-{i\over 2} \epsilon^{ruv} \xi^u_i\xi^v_i.
\label{IV15}
\end{eqnarray}

The $\xi^r_i(\tau)$'s satisfy

\begin{eqnarray}
\{ \xi_i^r , \xi_j^s \}^\ast_D &=& - i \delta^{rs} \delta_{ij}, \nonumber \\
\{\tilde{x}_s^\mu , \xi_i^r \}^\ast_D &=&
 - \frac{\partial \epsilon^r_\nu (u(p_s))}{\partial {p_s}_\mu} \xi^\nu_i.
\label{IV16}
\end{eqnarray}

If we define

\begin{eqnarray}
&\hat{x}^\mu_s& \equiv  \tilde{x}^\mu_s - \frac 12 \epsilon^A_\nu (u(p_s))
\eta_{AB} \frac{\partial \epsilon^B_\rho (u(p_s))}{\partial {p_s}_\mu}
S_\xi^{\nu \rho} = \nonumber \\
&=& {x}^\mu_s - \frac 12 \epsilon^A_\nu (u(p_s))
\eta_{AB} \frac{\partial \epsilon^B_\rho (u(p_s))}{\partial {p_s}_\mu}
\big( S_s^{\nu \rho} +  S_\xi^{\nu \rho} \big) ,
\label{IV17}
\end{eqnarray}

\noindent we get

\begin{eqnarray}
\{ \hat{x}^\mu_s , p^\nu_s \}^\ast_D &=& - \eta^{\mu \nu}, \nonumber \\
\{ \hat{x}^\mu_s , \xi^r_i \}^\ast_D &=& 0, \nonumber \\
\{ \hat{x}^\mu_s , \hat{x}^\nu_s \}^\ast_D &=&  0.
\label{IV18}
\end{eqnarray}

Therefore, with respect to the Dirac brackets $\{ .,. \}^\ast_D$ we have
obtained a basis in which ${\hat x}^{\mu}_s(\tau )$, $p^{\mu}_s$, $A_{\check 
A}(\tau ,\vec \sigma )$, $\pi^{\check A}(\tau ,\vec \sigma )$, ${\vec \eta}
_i(\tau )$, ${\vec \kappa}_i(\tau )$, $\xi^{\check r}_i(\tau )$, $\theta
_i(\tau )$, $\theta_i^{*}(\tau )$, are canonical variables and only $b^{\mu}
_{\check A}(\tau )$, $S^{\mu\nu}_s(\tau )$, are not canonical. The new canonical
origin ${\hat x}_s^{\mu}$ of the hyperplane has the same noncovariance of the
Newton-Wigner position operator. In terms of this variable we get

\begin{eqnarray}
J^{\mu \nu} &=&  \tilde{x}^\mu_s p_s^\nu - \tilde{x}^\nu_s p_s^\mu +
\tilde{S}_s^{\mu \nu} + S_\xi^{\mu \nu}= \hat{L}_s^{\mu \nu} + \tilde{S}_s
^{\mu \nu} +\tilde{S}_\xi^{\mu \nu},\nonumber \\
&&{}\nonumber \\
\hat{L}_s^{\mu \nu} &=& \hat{x}^\mu_s p_s^\nu - \hat{x}^\nu_s p_s^\mu , 
\nonumber \\
\tilde{S}_\xi^{\mu \nu} &=& S_\xi^{\mu \nu} + \frac 12 \epsilon^A_\rho (u(p_s))
\eta_{AB} \Big[ \frac{\partial \epsilon^B_\sigma (u(p_s))}{\partial {p_s}_\mu}
p_s^\nu - \frac{\partial \epsilon^B_\sigma (u(p_s))}{\partial {p_s}_\nu}
p_s^\mu \Big] S_\xi^{\rho \sigma},\nonumber \\
&&{}\nonumber \\
\{ \hat{L}^{\mu \nu}_s , \hat{L}^{\alpha \beta}_s \}^\ast_D  &=&
C^{\mu \nu \alpha \beta}_{\gamma \delta} \hat{L}^{\gamma \delta}_s ,
\nonumber \\
\{ \tilde{S}^{\mu \nu}_\xi , \tilde{S}^{\alpha \beta}_\xi \}^\ast_D  &=&
C^{\mu \nu \alpha \beta}_{\gamma \delta} \tilde{S}^{\gamma \delta}_\xi, 
\nonumber \\
\{ \tilde{S}^{\mu \nu}_s , \tilde{S}^{\alpha \beta}_s \}^\ast_D  &=&
C^{\mu \nu \alpha \beta}_{\gamma \delta} \tilde{S}^{\gamma \delta}_s, 
\nonumber \\
\{ \tilde{S}^{\mu \nu}_s , \tilde{S}^{\alpha \beta}_\xi \}^\ast_D  &=&
\{ \tilde{S}^{\mu \nu}_s , \hat{L}^{\alpha \beta}_s \}^\ast_D =
\{ \tilde{S}^{\mu \nu}_\xi , \hat{L}^{\alpha \beta}_s \}^\ast_D = 0, 
\nonumber \\
\{ J^{\mu \nu} , J^{\alpha \beta} \}^\ast_D &=&
C^{\mu \nu \alpha \beta}_{\gamma \delta} J^{\gamma \delta}.
\label{IV19}
\end{eqnarray}

As shown in Ref.\cite{lus1}, we can restrict ourselves to the Wigner hyperplanes
$\Sigma_{\tau W}$ with $l^{\mu}=u^{\mu}(p_s)$ [i.e. orthogonal to $p^{\mu}_s$] 
with the gauge-fixings

\begin{eqnarray}
T^\mu_{\breve{A}}(\tau) = b^\mu_{\breve{A}} (\tau) -
 \epsilon^\mu_{A=\breve{A}}(u(p_s)) &\approx & 0 \nonumber \\
\Rightarrow b^A_{\breve{A}} (\tau) =
 \epsilon_\mu^A(u(p_s)) b^\mu_{\breve{A}} (\tau) &\approx & \eta^A_{\breve{A}},
\label{IV20}
\end{eqnarray}

\noindent which imply the new Dirac brackets

\begin{eqnarray}
\{A,B \}^{\ast \ast}_D &=& \{A,B \}^{\ast}_D - \frac 14
\Big[ \{A,\tilde{H}^{\gamma \delta} \}^{\ast}_D
\Big(\eta_{\gamma \sigma} \epsilon_\gamma^D(u(p_s)) -
\eta_{\delta \sigma} \epsilon_\gamma^D(u(p_s))\Big)
 \{T_D^\sigma , B \}^{\ast}_D + \nonumber \\
&+& \{ A, T_D^\sigma \}^{\ast}_D
\Big(\eta_{\sigma \nu} \epsilon_\mu^B(u(p_s)) -
\eta_{\sigma \mu} \epsilon_\nu^B(u(p_s))\Big)
 \{ \tilde{H}^{\mu \nu} ,B \}^{\ast}_D\Big] .
\label{IV21}
\end{eqnarray}

The gauge-fixings (\ref{IV20}) imply ${\tilde \lambda}^{\mu\nu}(\tau )\approx
0$ [their time constancy], $b^{\mu}_{\check A}(\tau )\equiv
L^\mu{}_A(p_s,\stackrel{o}p_s)$ and ${\tilde {\cal H}}^{\mu\nu}(\tau )\equiv 0$
[namely the determination of $S^{\mu\nu}_s$ in terms of the variables of the 
system]. The remaining variables form a canonical basis

\begin{eqnarray}
\{ \hat{x}^\mu_s(\tau), p_s^\nu \}^{\ast \ast}_D &=& - \eta^{\mu \nu}, 
\nonumber \\
\{ \eta^r_i(\tau) , k^s_j (\tau) \}^{\ast \ast}_D &=&
\delta_{ij} \delta^{rs}, \nonumber \\
\{ \xi^r_i(\tau) , \xi^s_j (\tau) \}^{\ast \ast}_D &=&
 - i \delta^{rs} \delta_{ij} ,\nonumber \\
\{ A_A(\tau ,\vec \sigma ),\pi^B(\tau ,{\vec \sigma}^{'}) \}^{\ast \ast}_D&=&
\eta^B_A \delta^3(\vec \sigma -{\vec \sigma}^{'}).
\label{IV22}
\end{eqnarray}

As shown in Ref.\cite{lus1}, the dependence of the gauge-fixing (\ref{IV20}) on
$p^{\mu}_s$ implies that the Lorentz-scalar indices $\check A$ become Wigner
indices A: i) $A_{A=\tau}(\tau ,\vec \sigma )$ is a Lorentz-scalar field; ii)
$A_{A=r}(\tau ,\vec \sigma )$, $\xi^r_i(\tau )$, $\eta^r_i(\tau )$, $\kappa
_{ir}(\tau )$, are Wigner spin 1 ${}{}$ 3-vectors which transform with Wigner 
rotations under the action of Minkowski Lorentz boosts.

On $\Sigma_{\tau W}$ the Poincar\'e generators are

\begin{eqnarray}
p_s^\mu,&&~~~~~J^{\mu \nu}_s =  \hat{x}^\mu_s p_s^\nu - \hat{x}^\nu_s p_s^\mu +
\tilde{S}^{\mu \nu},\nonumber \\
&&{}\nonumber \\
\tilde{S}^{\mu \nu} &\equiv & \tilde{S}_s^{\mu \nu}
 + \tilde{S}_\xi^{\mu \nu}, \nonumber \\
\tilde{S}^{0i} &=& -\frac{\delta^{ir} \bar{S}^{rs} p_s^s}
{p_s^0 + \eta_s \sqrt{p_s^2}},~~~~~\tilde{S}^{ij} = \delta^{ir} \delta^{js}
\bar{S}^{rs} ,
\label{IV23}
\end{eqnarray}

\noindent because one can express ${\tilde S}^{\mu\nu}$ in terms of ${\bar S}
^{AB}=\epsilon^A_{\mu}(u(p_s))\epsilon^B_{\nu}(u(p_s)) S^{\mu\nu}$.

Since ${\tilde H}^{\mu\nu}(\tau )\equiv 0$ implies

\begin{eqnarray}
S_s^{\mu \nu}  &=& \Big(\epsilon^\mu_r(u(p_s)) u^\nu(p_s) -
\epsilon^\nu_r(u(p_s)) u^\mu(p_s)\Big)
\Big\{ \int{d^3\sigma \sigma^r \frac{(\vec{\pi}^2 (\tau, \vec{\sigma}) +
\vec{B}^2 (\tau, \vec{\sigma}))}{2}} + \nonumber \\
&+& \sum_{i=1}^N \eta^r_i(\tau) \eta_i
\sqrt{m_i^2 - i Q_i(\tau) \xi^u_i(\tau) \xi^v_i(\tau) F_{uv}(\tau,
\vec{\eta}_i) + \big[\vec{k}_i(\tau) -Q_i \vec{A}(\tau, \vec{\eta}_i)\big]^2}
\Big\} + \nonumber \\
&-& \Big(\epsilon^\mu_r(u(p_s)) \epsilon^\nu_s(u(p_s)) -
\epsilon^\nu_r(u(p_s)) \epsilon^\mu_s(u(p_s)) \Big)
\Big\{ \int{d^3\sigma \sigma^r (\vec{\pi} \wedge
\vec{B})_s (\tau, \vec{\sigma})} + \nonumber \\
&+& \sum_{i=1}^N \eta_i^r(\tau) \big(k_{is}(\tau) -
Q_i(\tau) A_s(\tau, \vec{\eta}_i)\big) \Big\} ,
\label{IV24}
\end{eqnarray}

\noindent we get

\begin{eqnarray}
\bar{S}_s^{AB} &=& \big( \delta^A_r \delta^B_{\bar{o}} -
\delta^B_r \delta^A_{\bar{o}} \big)
\Big[\int{d^3\sigma \sigma^r \frac{(\vec{\pi}^2 (\tau, \vec{\sigma}) +
\vec{B}^2 (\tau, \vec{\sigma}))}{2}} + \nonumber \\
&+& \sum_{i=1}^N \eta^r_i(\tau) \eta_i
\sqrt{m_i^2 - i Q_i(\tau) \xi^u_i(\tau) \xi^v_i(\tau) F_{uv}(\tau,
\vec{\eta}_i) + \big[\vec{k}_i(\tau) -Q_i \vec{A}(\tau, \vec{\eta}_i)\big]^2}
\Big] + \nonumber \\
&-& \big( \delta^A_r \delta^B_s -
\delta^B_r \delta^A_s \big)
\Big[\int{d^3\sigma \sigma^r (\vec{\pi} \wedge
\vec{B})_s (\tau, \vec{\sigma})} + \nonumber \\
&+& \sum_{i=1}^N \eta_i^r(\tau) \big(k_{is}(\tau) -
Q_i(\tau) A_s(\tau, \vec{\eta}_i)\big) \Big]       .
\label{IV25}
\end{eqnarray}

\noindent However, ${\bar S}^{or}_s$ does not contribute to the previous
realization of the Poincar\'e generators, which defines the rest-frame
Wigner-covariant instant form of dynamics.

The original variables $z^{\mu}(\tau ,\vec \sigma )$, $\rho_{\mu}(\tau ,\vec 
\sigma )$, are reduced only to ${\hat x}^{\mu}_s$, $p^{\mu}_s$ on the
Wigner hyperplane $\Sigma_{\tau W}$. On it only 6 first class constraints 
survive

\begin{eqnarray}
&\pi^\tau  (\tau, \vec{\sigma}) &\approx  0, \nonumber \\
&\Gamma (\tau, \vec{\sigma}) &= \partial_r \pi^r (\tau, \vec{\sigma})
-\sum_{i=1}^N Q_i \delta^3 (\vec{\sigma} -\vec{\eta}_i(\tau))
\approx 0, \nonumber \\
&&{}\nonumber \\
&\tilde{H}^\mu (\tau) &= p^{\mu}_s-[u^{\mu}(p_s) H_{rel}(\tau )+\epsilon^{\mu}
_r(u(p_s)) H_{p\, r}(\tau )]=\nonumber \\
&&=u^\mu(p_s) H(\tau) + \epsilon^\mu_r(u(p_s))
H_{pr}(\tau) \approx 0, \nonumber \\
&&{}\nonumber \\
&&or\nonumber \\
&&{}\nonumber \\
&H(\tau)& = \eta_s\sqrt{p^2_s}-H_{rel}(\tau )
=\eta_s \sqrt{p_s^2} -
\Big[\int{ d^3\sigma \frac {(\vec{\pi}^2 + \vec{B}^2 )}{2}
(\tau, \vec{\sigma}) }+ \nonumber \\
&+& \sum_{i=1}^N \eta_i \cdot
\sqrt{m_i^2 - i Q_i(\tau) \xi_i^r(\tau) \xi_i^s(\tau) F_{rs}(\tau, \vec{\eta}_i)
+ \big[ \vec{k}_i(\tau) - Q_i(\tau) \vec{A}(\tau, \vec{\eta}_i) \big]^2}
\Big] \approx 0, \nonumber \\
&H_{pr}(\tau)& = \int{ d^3\sigma \big[ \vec{\pi} \wedge \vec{B} \big]_r
 (\tau, \vec{\sigma})} + \sum_{i=1}^N \big[ {k}_{ir}(\tau)
 - Q_i(\tau) {A}_r (\tau, \vec{\eta}_i) \big] \approx 0, \nonumber \\
&&{}\nonumber \\
&\{ \tilde{H}^\mu(\tau)& \!, \tilde{H}^\nu(\tau) \}^{\ast \ast}_D =
\int{ d^3\sigma \Big\{ \Big[ u^\mu (p_s) \epsilon^\nu_r(u(p_s)) -
u^\nu (p_s) \epsilon^\mu_r(u(p_s)) \Big] \pi^r (\tau, \vec{\sigma})} + 
\nonumber \\
&-& \epsilon^\mu_r(u(p_s))F_{rs}(\tau, \vec{\sigma})\epsilon^\nu_s(u(p_s))
\Big\} \Gamma (\tau, \vec{\sigma}) \approx 0.
\label{IV26}
\end{eqnarray}

Let us remark that on $\Sigma_{\tau W}$
in $H(\tau )\approx 0$ the ``spin-spin" and ``spin-electric
field" interactions have disappeared [also the quantity $\pi^r_\xi (\tau, 
\vec{\sigma})$ vanishes]. There is only the ``spin-magnetic field" interaction

\begin{equation}
- i Q_i \xi^r_i \xi^s_i F_{rs} (\tau, \vec{\eta}_i) =
- 2 Q_i {\vec{\bar S}}_{i\xi} \cdot \vec{B} (\tau, \vec{\eta}_i), \quad\quad
{\vec{\bar S}}_{i\xi} \equiv - \frac i2 \vec{\xi}_i \times \vec{\xi}_i,
\label{IV27}
\end{equation}

\noindent like in the nonrelativistic Pauli equation \cite{grandy}.

Therefore, we get a kind of ``relativistic Pauli Hamiltonian" describing the
interaction of a massive spinning particle belonging to the $({1\over 2},0)$
representation with the electromagnetic field, whose nonrelativistic limit is 
the pseudoclassical form of the ordinary Pauli Hamiltonian.

The constraints ${\vec H}_p(\tau )\approx 0$ identify the Wigner hyperplane
$\Sigma_{\tau W}$ with the intrinsic rest frame (vanishing of the total Wigner 
spin 1 ${}$ 3-momentum of the isolated system) and say that the 3-coordinate
$\vec \sigma ={\vec \eta}_{+\, system}$ of the center of mass of the isolated 
system on $\Sigma_{\tau W}$ is a gauge variable, whose natural gauge-fixing is
${\vec \eta}_{+\, system}\approx 0$ [so that it coincides with the origin of 
$\Sigma_{\tau W}$: ${}x^{\mu}_s(\tau )=z^{\mu}(\tau ,\vec \sigma =0)$].

On $\Sigma_{\tau W}$ the Dirac Hamiltonian becomes

\begin{equation}
H_D = \lambda(\tau)H(\tau ) - \vec{\lambda}(\tau) \cdot \vec{H}_p(\tau) +
\int{d^3\sigma \Big[-A_\tau (\tau, \vec{\sigma}) \Gamma (\tau, \vec{\sigma})
+ \mu_\tau(\tau, \vec{\sigma}) \pi^\tau(\tau, \vec{\sigma})\Big] } ,
\label{IV28}
\end{equation}

\noindent so that $\hat{x}^\mu_s$ has a velocity parallel to ${p_s}^\mu$ ,
namely it has no zitterbewegung as it happens to the Foldy-Wouthuysen
mean position (see Section II).

The nonrelativistic limit of $H_{rel}$ of Eqs.(76), disregarding the kinetic
term of the electromagnetic field due to the ambiguities in defining
Galilean electromagnetism (see the two independent `electric' and `magnetic'
limits of Ref.\cite{lebl}), is

\begin{eqnarray}
H_{rel}&=&\sum_{i=1}^N \eta_i c \sqrt{m^2_ic^2-2{{Q_i}\over c}{\vec {\bar S}}
_{i\xi}\cdot \vec B(\tau ,{\vec \eta}_i(\tau ))+
[{\vec \kappa}_i(\tau )-{{Q_i}\over 
c}\vec A(\tau ,{\vec \eta}_i(\tau ))]^2}+..=\nonumber \\
&{\rightarrow}_{c\, \rightarrow \infty}& \sum_{i=1}^N\eta_im_ic^2+
\sum_{i=1}^N \{ {1\over {2m_i}} [{\vec \kappa}_i(\tau )-{{Q_i}\over c}
\vec A(\tau ,{\vec \eta}_i(\tau ))]^2-{{Q_i}\over {m_ic}} {\vec {\bar S}}
_{i\xi}\cdot \vec B(\tau ,{\vec \eta}_i(\tau )) \} +....,\nonumber \\
&&{}\nonumber \\
\sum_{i=1}^N {\vec \kappa}_i(\tau ) &\approx& 0.
\label{IV29}
\end{eqnarray}

Therefore we recover the classical basis of the nonrelativistic Pauli theory
 in the center of mass frame, with the same properties under parity
(see for instance Ref.\cite{grandy}, p.97).
See Ref.\cite{gomis} for an approach to this theory with Grassmann
variables (the fibered spin structure) and first class constraints, generating 
by quantization the Pauli theory in the form of the Levy-Leblond
nonrelativistic spin equation\cite{levy}.

\vfill\eject

\section{Dirac's observables and equations of motion.}

As shown in Ref.\cite{lusa}, the Dirac observables of the electromagnetic field 
are the transverse quantities ${\vec A}_{\perp r}(\tau, \vec{\sigma})$,
${\vec \pi}^r_{\perp}(\tau, \vec{\sigma})$, defined by the decomposition

\begin{eqnarray}
A_r (\tau, \vec{\sigma}) &=& \partial_r \eta (\tau, \vec{\sigma}) +
A_{\perp r} (\tau, \vec{\sigma}), \nonumber \\
\pi^r (\tau, \vec{\sigma}) &=& \pi^r_{\perp} (\tau, \vec{\sigma}) +
\frac{\partial^r}{\triangle_\sigma}
\Big[\Gamma (\tau, \vec{\sigma}) - \sum_{i=1}^N Q_i(\tau)
\delta^3(\vec{\sigma} - \vec{\eta}_i(\tau)) \Big] , \nonumber \\
\eta (\tau, \vec{\sigma}) &=& - \frac{\vec{\partial}}{\triangle_\sigma}
\cdot \vec{A} (\tau, \vec{\sigma}),
\label{V1}
\end{eqnarray}

\noindent while the gauge variables are $A_\tau (\tau, \vec{\sigma})$ and
$\eta (\tau, \vec{\sigma})$, being conjugated to the first class constraints
$\pi^{\tau}(\tau, \vec{\sigma})\approx 0$, $\Gamma (\tau, \vec{\sigma})\approx
0$.

Concerning the particle variables, we have that $k^r_i(\tau)$, $\theta_i(\tau)$,
$\theta_i^\ast(\tau)$, are not gauge invariant because

\begin{eqnarray}
\{ k^r_i(\tau) , \Gamma (\tau, \vec{\sigma}) \}^{\ast \ast}_D &=&
Q_i \frac{\partial}{\partial \eta_i^r}
\delta^3(\vec{\sigma} - \vec{\eta}_i(\tau)), \nonumber \\
\{ \theta_i(\tau) , \Gamma (\tau, \vec{\sigma}) \}^{\ast \ast}_D &=&
i e_i \theta_i(\tau)
\delta^3(\vec{\sigma} - \vec{\eta}_i(\tau)), \nonumber \\
\{ \theta_i^\ast(\tau) , \Gamma (\tau, \vec{\sigma}) \}^{\ast \ast}_D &=&
- i e_i \theta_i^\ast(\tau)
\delta^3(\vec{\sigma} - \vec{\eta}_i(\tau)) .
\label{V2}
\end{eqnarray}

Instead, the position variables $\eta^r_i(\tau )$ and the spin variables
$\xi^r_i(\tau )$ are gauge invariant. The Dirac observables for the particles
are obtained through a dressing with a Coulomb cloud

\begin{eqnarray}
\check{\theta}_i(\tau) &=&
 e^{i e_i \eta(\tau, \vec{\eta}_i)} \theta_i(\tau), \nonumber \\
\check{\theta}_i^\ast(\tau) &=&
 e^{- i e_i \eta(\tau, \vec{\eta}_i)} \theta_i^\ast(\tau), \nonumber \\
 \check{\vec{k}_i}(\tau) &=& \vec{k}_i(\tau) - Q_i(\tau)
\vec{\partial}\eta (\tau, \vec{\sigma}) \Rightarrow
\check{\vec{k}_i}(\tau) - Q_i \vec{A}_\perp (\tau, \vec{\eta}_i) =
\vec{k}_i (\tau) - Q_i \vec{A} (\tau, \vec{\eta}_i) .
\label{V3}
\end{eqnarray}

The electric charges are gauge invariant

\begin{equation}
{\check Q}_i=e_i \check{\theta}_i^\ast(\tau) \check{\theta}_i(\tau)=
Q_i = e_i \theta_i^\ast(\tau) \theta_i(\tau)  ~~~~~
\big[ \dot{Q}_i(\tau) = 0 \Rightarrow Q_i(\tau) \equiv Q_i \big] .
\label{V4}
\end{equation}

When the Gauss law is satisfied, $\Gamma (\tau ,\vec \sigma )=0$, Eq.(\ref{V1})
implies

\begin{equation}
\int{d^3\sigma \vec{\pi}^2 (\tau, \vec{\sigma})  } =
\int{d^3\sigma \vec{\pi}^2_\perp (\tau, \vec{\sigma})  }
+ \sum_{i \neq j}^{1...N} \frac{Q_i Q_j}{4\pi \mid \vec{\eta}_i(\tau)
-\vec{\eta}_j(\tau) \mid} ,
\label{V5}
\end{equation}

\noindent so that we get

\begin{eqnarray}
H(\tau) &=& \eta_s \sqrt{p_s^2} -
\Big\{ \int{d^3\sigma  \frac{(\vec{\pi}_{\perp}^2 (\tau, \vec{\sigma}) +
\vec{B}^2 (\tau, \vec{\sigma}))}{2}}
+ \sum_{i \neq j}^{1...N} \frac{Q_i Q_j}{4\pi \mid \vec{\eta}_i(\tau)
-\vec{\eta}_j(\tau) \mid} + \nonumber \\
&+& \sum_{i =1}^{N} \eta_i
\sqrt{m_i^2 - i Q_i \xi^r_i(\tau ) \xi^s_i(\tau ) F_{rs}(\tau,
\vec{\eta}_i) + \big[\check{\vec{k}}_i(\tau) -
Q_i \vec{A}_\perp (\tau, \vec{\eta}_i)\big]^2}
\Big\} \approx 0   .
\label{V6}
\end{eqnarray}

We see \cite{lus1} the emergence of the Coulomb potential from field theory 
and the regularization of the Coulomb self-energy (the $\sum_{i\not= j}$ rule)
due to the Grassmann character of the electric charges, $Q^2_i=0$. In this way
all the effects of order $Q^2_i$ are eliminated, but not those of order 
$Q_iQ_j$, $i\not= j$ \cite{lus2}. This means the elimination of all the effects
connected with pair production, consistently with the description of only one
branch of the mass spectrum of the N-body system (all particles have only one 
sign of the energy).

There is no odd first class constraint, because massive 2-spinors do not satisfy
any spinor equation. The quantization of this Hamiltonian in the free case
gives a nonlocal Schroedinger equation with the kinetic square root operator
\cite{lamm} for a 2-spinor, which corresponds to the upper (or lower) part of
positive (or negative) energy Dirac spinors boosted at rest [so that they also
coincide with the corresponding parts of the positive (or negative) energy
Foldy-Wouthuysen spinors boosted at rest; see the Appendix]. These 2-spinors 
are parity eigenstates in the rest frame. All these facts shows the similarities
and the differences of this rest-frame representation from the Chakrabarti one
discussed in Section II.

The three constraints defining the rest frame become

\begin{equation}
\vec{H}_p(\tau) = \sum_{i=1}^N \check{\vec{k}_i}(\tau) +
\int{d^3\sigma \vec{\pi}_\perp (\tau, \vec{\sigma}) \wedge
\vec{B} (\tau, \vec{\sigma}) }\approx 0 ,
\label{V7}
\end{equation}

\noindent and are independent from the interactions as expected in an instant 
form of dynamics, like the expression for the spin\cite{lus2} implied by
Eq.(\ref{IV25})

\begin{eqnarray}
{\bar S}^{rs}_s&=&\sum_{i=1}^N[\eta_i^r(\tau ){\check \kappa}^s_i(\tau )-
\eta^s_i(\tau ){\check \kappa}^r_i(\tau )]+\int d^3\sigma [\sigma^r({\vec \pi}
_{\perp}\times \vec B)^s-\sigma^s({\vec \pi}_{\perp}\times \vec B)^r](\tau 
,\vec \sigma ),\nonumber \\
{\bar S}^{rs}&=&{\bar S}^{rs}_s+{\bar S}^{rs}_{\xi}={\bar S}^{rs}_s-i
\sum_{i=1}^N \xi^r_i\xi^s_i.
\label{V7a}
\end{eqnarray}

The Pauli-Lubanski 4-vector and the spin Poincar\'e Casimir are [$\epsilon_s=
\eta_s\sqrt{p^2_s}$]

\begin{eqnarray}
W^{\mu}_s&=&{1\over 2} \epsilon^{\mu\nu\rho\sigma}p_{s \nu} J_{s \rho\sigma}=
{1\over 2} \epsilon^{\mu\nu\rho\sigma}p_{s \nu} {\tilde S}_{\rho\sigma}=
\nonumber \\
&=& ({\vec p}_s\cdot {\vec {\bar S}} ; \epsilon_s {\vec {\bar S}}+{{ {\vec p}_s
\cdot {\vec {\bar S}} }\over {p^o_s+\epsilon_s}} {\vec p}_s )=\nonumber \\
&=&{1\over 2} \epsilon^{\mu\nu\rho\sigma}p_{s \nu} ({\tilde S}_{s \rho\sigma}+
{\tilde S}_{\xi \rho\sigma})\, {\buildrel {def} \over =}\, W^{(L) \mu}_s +
\Sigma^{\mu}_s,\nonumber \\
W^2_s&=&-{1\over 2} p^2_s {\tilde S}_{\mu\nu}{\tilde S}^{\mu\nu}=-p^2_s
{\vec {\bar S}}^2.
\label{V7b}
\end{eqnarray}

\noindent This shows that ${\vec {\bar S}}={\vec {\bar S}}_s+{\vec {\bar S}}
_{\xi}$ is the rest-frame Thomas spin: $W_{s \mu}(p_s)=W_{s \nu} 
L^{\nu}{}_{\mu}(p_s;{\buildrel \circ \over p}_s)= (0; \epsilon_s {\vec 
{\bar S}})$.

As shown in Ref.\cite{lus1}, it is possible to separate the relative variables 
from the center-of-mass ones and from the invariant mass $\pm \sqrt{p^2_s}$
with the following canonical transformation [$T_s$ is the Lorentz-scalar time 
of the rest frame]

\begin{eqnarray}
T_s &=& \frac{p_s^\mu x_{\mu s}}{\eta_s\sqrt{p_s^2}},~~~~~~~~
\epsilon_s = \eta_s\sqrt{p_s^2}, \nonumber \\
\vec{z}_s &=& \eta_s\sqrt{p_s^2} \Big( \vec{\hat{x}_s} -
\frac{\vec{p_s}}{p_s^0} \hat{x}_s^0 \Big) ,~~~~~
\vec{k}_s = \frac{\vec{p}_s}{\eta_s\sqrt{p_s^2}}, \nonumber \\
\vec{\eta}_+ &=& \frac{1}{N} \sum_{i=1}^N \vec{\eta}_i, ~~~~~~~~
{\check {\vec \kappa}}_{+} = \sum_{i=1}^N {\check {\vec \kappa}}_i, \nonumber \\
\vec{\rho}_a &=& \sqrt{N} \sum_{i=1}^N \hat{\gamma}_{ai}\vec{\eta}_i, ~~~~~
{\check {\vec \pi}}_a = \frac{1}{\sqrt{N}} \sum_{i=1}^N \hat{\gamma}_{ai}
{\check {\vec \kappa}}_i,\nonumber \\
&&a=1,...,N-1,\nonumber \\
&&{}\nonumber \\
\sum_{i=1}^N \hat{\gamma}_{ai} &=& 0, ~~~~~
\sum_{i=1}^N \hat{\gamma}_{ai} \hat{\gamma}_{bi} = \delta_{ab},~~~~~
\sum_{a=1}^{N-1} \hat{\gamma}_{ai} \hat{\gamma}_{aj} =
\delta_{ij} -\frac 1N .
\label{V8}
\end{eqnarray}

The inverse canonical transformation is

\begin{eqnarray}
\hat{x}^0_s &=& \sqrt{1-\vec{k}_s}
\Big( T_s + \frac{\vec{k}_s \cdot \vec{z}_s}{\epsilon_s} \Big) ,\nonumber \\
\vec{\hat{x}_s} &=& \frac{\vec{z}_s}{\epsilon_s} +
\Big( T_s + \frac{\vec{k}_s \cdot \vec{z}_s}{\epsilon_s} \Big) \vec{k}_s, 
\nonumber \\
p_s^0 &=& \epsilon_s \sqrt{1+\vec{k}_s^2} ,\nonumber \\
\vec{p}_s &=& \epsilon_s \vec{k}_s, \nonumber \\
\vec{\eta}_i &=& \vec{\eta}_+  + \frac{1}{\sqrt{N}}
\sum_a \hat{\gamma}_{ai} \vec{\rho}_a, \nonumber \\
{\check {\vec \kappa}}_i &=& \frac 1N {\check {\vec \kappa}}_+ +
\sqrt{N} \sum_a \hat{\gamma}_{ai} {\check {\vec \pi}}_a  .
\label{V9}
\end{eqnarray}

The constraints (\ref{V7}) take the form

\begin{equation}
\vec{H}(\tau) = \check{\vec{k}}_+(\tau) +
\int{d^3\sigma \vec{\pi}_\perp (\tau, \vec{\sigma}) \wedge
\vec{B}_\perp (\tau, \vec{\sigma}) }\approx 0 .
\label{V10}
\end{equation}

\noindent where the second term is the total electromagnetic 3-momentum, while
the spin tensor may be written as ${\bar S}^{rs}_s \approx \sum_{a=1}^{N-1}
[\rho^r_a(\tau ){\check \pi}^s_a(\tau )-\rho^s_a(\tau ){\check \pi}^r_a(\tau )]
+\int d^3\sigma [(\sigma^r-\eta^r_{+})({\vec \pi}_{\perp}\times \vec B)^s-
(\sigma^s-\eta^s_{+})({\vec \pi}_{\perp}\times \vec B)^r](\tau ,\vec \sigma )$.

In
absence of the electromagnetic field the natural gauge fixing to ${\vec \kappa}
_{+}\approx 0$ is ${\vec \eta}_{+}\approx 0$. As said in Ref.\cite{lus1}, this 
construction suggests the existence of a decomposition in center-of-mass and 
relative variables also of the electromagnetic field [see Ref.\cite{lm} for a
solution of this problem for Klein-Gordon fields]. When it will be available, 
one will be able to find the natural gauge fixing to be associated with the 
constraints (\ref{V10}) and to express $H(\tau )\approx 0$ only in terms of 
relative variables. For the time being, this constraint can be put in the 
form

\begin{eqnarray}
H(\tau) &=&\epsilon_s-H_{rel}=\nonumber \\
&=& \epsilon_s - \sum_{i=1}^N \eta_i
\Big[m_i^2 +2Q_i{\vec {\bar S}}_{i\xi}(\tau )\cdot \vec B
(\tau, \vec{\eta}_+ (\tau) + \frac{1}{\sqrt{N}}
\sum_{a=1}^{N-1} \hat{\gamma}_{ai} \vec{\rho}_a(\tau) ) + \nonumber \\
&+& \Big(\frac 1N \check{\vec{k}_+}(\tau) +
\sqrt{N} \sum_{a=1}^{N-1} \hat{\gamma}_{ai} \vec{\pi}_a -
 Q_i \vec{A}_\perp (\tau, \vec{\eta}_+ (\tau) + \frac{1}{\sqrt{N}}
\sum_{a=1}^{N-1} \hat{\gamma}_{ai} \vec{\rho}_a(\tau) ) \Big) \Big]^{1/2} + 
\nonumber \\
&+&\frac 12 \sum_{i \neq j}^{1...N} {{Q_i Q_j}\over {4\pi}} 
{|  \frac{1}{\sqrt{N}}
\sum_{a=1}^{N-1} (\hat{\gamma}_{ai} - \hat{\gamma}_{aj})
 \vec{\rho}_a(\tau) ) |}^{-1} - \frac 12 \int{d^3\sigma (\vec{\pi}^2_\perp
+ \vec{B}^2 ) (\tau, \vec{\sigma}) }\approx 0 
\label{V11}
\end{eqnarray}

We see that by putting ${\vec A}_{\perp}(\tau ,\vec \sigma )={\vec \pi}_{\perp}
(\tau ,\vec \sigma )=0$ by hand as second class constraints, one gets a subspace
of the reduced phase space describing N spinning particles with a mutual 
Coulomb interaction.

By adding the gauge-fixing $T_s - \tau \approx 0$ [which identifies the 
rest-frame time $T_s$ with the parameter $\tau$ of the foliation of Minkowski 
spacetime with the Wigner hyperplanes associated with the isolated system], 
whose time constancy implies $\lambda(\tau) = -1$, we get the Dirac Hamiltonian

\begin{equation}
\hat{H}_D = H_{rel}(\tau) - \vec{\lambda}(\tau) \cdot \vec{H}_p(\tau),
\label{V12}
\end{equation}

\noindent where [the nonrelativistic limit would give Pauli theory in the
center of mass frame and in the Coulomb gauge for the electromagnetic field]

\begin{eqnarray}
H_{rel}(\tau) &=& \sum_{i=1}^N \eta_i
\sqrt{m_i^2 -i Q_i \xi_i^r(\tau)  \xi_i^s(\tau) F_{rs}(\tau, \vec{\eta}_i
(\tau)) + \big[ \check{\vec{k}}_i(\tau) -
 Q_i \vec{A}_\perp  (\tau, \vec{\eta}_i(\tau)) \big]^2} + \nonumber \\
&+& \sum_{i \neq j}^{1...N} Q_i Q_j \frac{1}{4\pi \mid \vec{\eta}_i(\tau)
-\vec{\eta}_j(\tau) \mid} +
\int{d^3\sigma \Big(\frac{\vec{\pi}^2_\perp (\tau, \vec{\sigma}) +
\vec{B}^2 (\tau, \vec{\sigma})}{2}\Big) }  = \nonumber \\
&=& \sum_{i=1}^N \eta_i \sqrt{m_i^2 - 2 Q_i {\vec {\bar S}}_{i\xi} \cdot \vec{B}
(\tau, \vec{\eta}_i) + \big[ \check{\vec{k}_i} (\tau) - Q_i
\vec{A}_\perp (\tau, \vec{\eta}_i) \big]^2}+ \nonumber \\
&+& \sum_{i \neq j}^{1...N} Q_i Q_j \frac{1}{4\pi \mid \vec{\eta}_i(\tau)
-\vec{\eta}_j(\tau) \mid} +
\int{d^3\sigma \frac{\big(\vec{\pi}^2_\perp +
\vec{B}^2\big)}{2} (\tau, \vec{\sigma}) }
\label{V13}
\end{eqnarray}

Therefore, we have a free point ${\vec z}_s$ , ${\vec k}_s$, decoupled from 
the system describing the canonical noncovariant origin of the Wigner 
hyperplane plus a description of the isolated system in terms of relative 
variables [but it is more convenient to work with the particle positions
${\vec \eta}_i(\tau )$].

We get the following Hamilton equations [as in Ref.\cite{lus2} it is convenient
to write $\vec \lambda (\tau )={\dot {\vec g}}(\tau )$; ${\partial}_{\eta_i}^r 
\equiv \frac{\partial}{\partial \eta_{ir}}$ ; $P^{rs}_{\perp}(\vec \sigma )=
\delta^{rs}+\partial^r\partial^s/\triangle$, $\triangle =-{\vec \partial}^2$]

\begin{eqnarray}
\dot{\vec{\eta}_i} (\tau)+{\dot {\vec g}}(\tau ) 
\, &{\buildrel \circ \over =}\,&
\eta_i \frac {\big[ \check{\vec{k}}_i(\tau) -
 Q_i \vec{A}_\perp  (\tau, \vec{\eta}_i(\tau)) \big]}
{\sqrt{m_i^2 - 2 Q_i {\vec {\bar S}}_{i\xi} (\tau) \cdot
\vec{B}  (\tau, \vec{\eta}_i(\tau))+ \big[ \check{\vec{k}}_i(\tau) -
 Q_i \vec{A}_\perp  (\tau, \vec{\eta}_i(\tau)) \big]^2}} = \nonumber \\
&=& \eta_i \Big[ \frac{ \check{\vec{k}}_i(\tau) -
 Q_i \vec{A}_\perp  (\tau, \vec{\eta}_i(\tau)) }
{\sqrt{m_i^2 +\big[\check{\vec{k}}_i(\tau) -
 Q_i \vec{A}_\perp  (\tau, \vec{\eta}_i(\tau)) \big]^2}} +\nonumber \\
&+&\frac{ Q_i \check{\vec{k}}_i(\tau)\big( {\vec {\bar S}}_{i\xi} (\tau) \cdot
\vec{B} (\tau, \vec{\eta}_i(\tau))\big)}
{\big[m_i^2 + \check{\vec{k}^2}_i(\tau)\big]^{3/2}} \Big] ,\nonumber \\
\dot{\check{\vec{k}}}_i(\tau) \, &{\buildrel \circ \over =}\,&
-\sum_{j=1}^N  \frac {Q_i Q_j}{4\pi \mid \vec{\eta}_i(\tau)
-\vec{\eta}_j(\tau) \mid^2} {{\vec{\eta}_i(\tau)
-\vec{\eta}_j(\tau)}\over {|\vec{\eta}_i(\tau)
-\vec{\eta}_j(\tau)|}} + \nonumber \\
&+& [\dot{\eta}^r_i(\tau)+{\dot g}^r(\tau )] Q_i \vec{\partial}_{\eta_i}
A^r_\perp  (\tau, \vec{\eta}_i(\tau)) + \eta_i
\frac{Q_i \vec{\partial}_{\eta_i} \big[ {\vec {\bar S}}_{i\xi} (\tau) \cdot
\vec{B}  (\tau, \vec{\eta}_i(\tau))\big]}
{\sqrt{m_i^2 + \check{\vec{k}^2}_i(\tau)}},\nonumber \\
&&{}\nonumber \\
{\dot \xi}^r_i(\tau )\, &{\buildrel \circ \over =}\,&
\eta_i {{Q_i \xi^s_i(\tau )F_{sr}(\tau ,{\vec \eta}
_i(\tau ))}\over {\sqrt{m^2_i-2Q_i{\vec {\bar S}}_{i\xi}(\tau )\cdot\vec 
B(\tau ,{\vec \eta}_i(\tau ))+[{\check {\vec \kappa}}_i(\tau )-Q_i{\vec A}
_{\perp}(\tau ,{\vec \eta}_i(\tau ))]^2}}}=\nonumber \\
&=&\eta_i {{Q_i \xi^s_i(\tau )F_{sr}(\tau ,{\vec \eta}
_i(\tau ))}\over {\sqrt{m^2_i+{\check {\vec \kappa}}^2_i(\tau )}}},\nonumber \\
{\dot {\bar S}}^r_{i\xi}(\tau )\, &{\buildrel \circ \over =}\,&
\eta_i {{Q_i \epsilon^{rst}F^{ut}(\tau 
,{\vec \eta}_i(\tau )) i\xi_i^u(\tau )\xi^s_i(\tau )}\over {\sqrt{m_i^2+{\check
{\vec \kappa}}^2_i(\tau )}}}=\nonumber \\
&=&\eta_i {{Q_i [\vec B(\tau ,{\vec \eta}_i(\tau ))\times {\vec {\bar S}}_{i\xi}
(\tau )]^r}\over {\sqrt{m^2_i+{\check {\vec \kappa}}^2_i(\tau )}}},\nonumber \\
{\dot {\check \theta}}_i(\tau )\, &{\buildrel \circ \over =}\,&
i\eta_ie_i{\check \theta}_i(\tau ){{{\vec {\bar S}}
_{i\xi}(\tau )\cdot \vec B(\tau ,{\vec \eta}_i(\tau ))+{\check {\vec \kappa}}
_i(\tau )\cdot {\vec A}_{\perp}(\tau ,{\vec \eta}_i(\tau ))}\over {\sqrt{m^2_i+
{\check {\vec \kappa}}^2_i(\tau )}}}-\nonumber \\
&-&2ie_i {\check \theta}_i(\tau ) \sum_{j\not= i} {{Q_j}\over {|{\vec \eta}
_i(\tau )-{\vec \eta}_j(\tau )|}},\nonumber \\
{\dot {\check Q}}_i\, &{\buildrel \circ \over =}\,& 0,\nonumber \\
&&{}\nonumber \\
\dot{A}_{\perp r} (\tau, \vec{\sigma}) \, &{\buildrel \circ \over =}\,&
- \pi_{\perp r} (\tau, \vec{\sigma})-[{\dot {\vec g}}(\tau )\cdot \vec
\partial ] A_{\perp r}(\tau ,\vec \sigma ), \nonumber \\
\dot{\pi}_{\perp}^r (\tau, \vec{\sigma}) \, &{\buildrel \circ \over =}\,&
\triangle_\sigma A_{\perp}^r (\tau, \vec{\eta}_i(\tau)) -[{\dot {\vec g}}
(\tau )\cdot \vec \partial ] \pi^r_{\perp}(\tau ,\vec \sigma )-\nonumber \\
&-&\sum_{i=1}^N Q_i [\dot{\eta}^s_i (\tau) +{\dot g}^s(\tau )]
P^{rs}_{\perp}(\vec{\sigma})
\delta^3(\vec{\sigma} - \vec{\eta}_i(\tau)) + \nonumber \\
&+& \sum_{i=1}^N \eta_i \frac{i Q_i \xi_i^u(\tau)
\xi_i^s(\tau) P^{rs}_{\perp}(\vec{\sigma})}
{\sqrt{m_i^2 + \check{\vec{k}^2}_i(\tau)}} \partial_u
\delta^3(\vec{\sigma} - \vec{\eta}_i(\tau)),\nonumber \\
&&{}\nonumber \\
&&\Downarrow \nonumber \\
&&{}\nonumber \\
(\, ( {{\partial}\over {\partial \tau}}+{{d\vec g(\tau )}\over {d\tau}}\cdot
{{\partial}\over {\partial \vec \sigma}})^2 &-&
({{\partial}\over  {\partial \vec \sigma}})^2 \, )
A_{\perp}^r (\tau, \vec \sigma) \, {\buildrel \circ \over =}\nonumber \\
&{\buildrel \circ \over =}\,&
P_{\perp}^{rs}(\vec{\sigma}-\vec g(\tau )) \sum_{i=1}^N Q_i
\Big[ [\dot{\eta}^s_i (\tau)+{\dot g}^s(\tau )]
\delta^3(\vec{\sigma} - \vec{\eta}_i(\tau)) -\nonumber \\
&-& i \eta_i \frac{ \xi_i^u(\tau)\xi_i^s(\tau) \partial_u}
{\sqrt{m_i^2 + \check{\vec{k}^2}_i(\tau)}}
\delta^3(\vec{\sigma} - \vec{\eta}_i(\tau)) \Big] .
\label{V14}
\end{eqnarray}

The last equation shows that, besides the standard term for scalar particles 
\cite{lus2}, the nonlocal (due to the projector) particle current contains also
a dipole term $P_{\perp}^{rs}(\vec \sigma ) \sum_{i=1}^N \eta_i {{({\vec {\bar 
S}}_{i\xi} \times \vec \partial )^s}\over {\sqrt{m_i^2+{\check {\vec \kappa}}
^2_i} }} \delta^3(\vec \sigma -{\vec \eta}_i(\tau ))$ in accord with the fact 
that the spinning particle has a pole-dipole structure\cite{g1} according to
Papapetrou's classification\cite{g2} [see Refs.\cite{g3} for other pole-dipole
models and Ref.\cite{g4} for their influence on the energy momentum tensor of
action-at-a-distance models].

For N=1 [${\vec \eta}_i \mapsto \vec \eta$, $m_i \mapsto m$,...] we have: i)
${\bar S}_{\xi}^{rs}=\epsilon^{rst}{\bar S}_{\xi}^t=-i \xi^r\xi^s$,
${\bar S}^{rs}_s\approx \int d^3\sigma [(\sigma^r-\eta^r)({\vec \pi}_{\perp}
\times \vec B)^s-(\sigma^s-\eta^s)({\vec \pi}_{\perp}\times \vec B)^r](\tau
,\vec \sigma )$; ii) $\Sigma_{s \mu}={1\over 2}\epsilon_{\mu\nu\rho\sigma}p_s
^{\nu}{\tilde S}_{\xi}^{\rho\sigma}=({1\over 2}\epsilon_{\mu\nu ij}\delta^{ir}
\delta^{js}-\epsilon_{\mu\nu oi}{{\delta^{ir}p^s_s}\over {p^o_s+\epsilon_s}})
p^{\nu}_s {\bar S}^{rs}_{\xi}$. The Bargmann-Michel-Telegdi equation\cite{bmt}
${\dot \Sigma}_{\mu}\, {\buildrel \circ \over =}\, {e\over m} F_{\mu\nu}
\Sigma^{\nu}$ for a spinning particle in an external electromagnetic field
[$\Sigma_{\mu}={1\over 2}\epsilon_{\mu\nu\rho\sigma}P^{\nu}S^{\rho\sigma}$
with ${\dot P}^{\mu}\not= 0$] is replaced by the following equation for the
spin part of the Pauli-Lubanski 4-vector of Eq.(88) in the canonical
realization (73) of the Poincar\'e group in the rest-frame instant form
for the isolated system of a positive energy spinning particle plus the
electromagnetic field [${\dot p}_s^{\mu}=0$]

\begin{eqnarray}
{\dot \Sigma}_{s \mu}&=&({1\over 2}\epsilon_{\mu\nu ij}\delta^{ir}\delta^{js}-
\epsilon_{\mu \nu oi}{{\delta^{ir}p^s_s}\over {p^o_s+\epsilon_s}}) p^{\nu}_s 
\epsilon^{rst} {\dot {\bar S}}^t_{\xi}\, {\buildrel \circ \over =}\nonumber \\
&{\buildrel \circ \over =}\,& {Q\over {\eta \sqrt{m^2+{\check {\vec \kappa}}
^2(\tau )}}}({1\over 2}\epsilon_{\mu\nu ij}\delta^{ir}\delta^{js}-
\epsilon_{\mu \nu oi}{{\delta^{ir}p^s_s}\over {p^o_s+\epsilon_s}}) p^{\nu}_s 
{1\over 2} \epsilon^{rst} [\vec B(\tau ,{\vec \eta}(\tau ))\times 
{\vec {\bar S}}_{\xi}]^t.
\label{V15}
\end{eqnarray}

Let us remark that the distribution function on the Grassmann variables
introduced in Ref.\cite{bm} to recover classical results as an expectation
value of the pseudoclassical ones , allows to arrive to a classical theory
with a classical electric charge but without spin.

Finally, the Lagrangian corresponding to the Dirac Hamiltonian of 
Eq.(\ref{V13}) is

\begin{eqnarray}
L_R(\tau )&=&\sum_{i=1}^N \Big( {i\over 2}[{\check \theta}^{*}_i(\tau )
{\dot {\check \theta}}_i(\tau )-{\dot {\check \theta}}^{*}_i(\tau ){\check
\theta}_i(\tau )]-{i\over 2}\xi_{ir}(\tau ){\dot \xi}^r_i(\tau )-\nonumber \\
&-&\eta_i \sqrt{m^2_i-2Q_i(\tau ) {\vec {\bar S}}_{i\xi}(\tau )\cdot 
\vec B(\tau ,{\vec \eta}_i(\tau ))} \sqrt{1-({\dot {\vec \eta}}_i(\tau )+
{\vec \lambda}(\tau )\, )^2}+\nonumber \\
&+&Q_i [{\dot {\vec \eta}}_i(\tau )+{\vec \lambda}(\tau )]\cdot {\vec A}
_{\perp}(\tau ,{\vec \eta}_i(\tau ))+{1\over 2}\sum_{j\not= i}{{Q_iQ_j}\over 
{4\pi |{\vec \eta}_i(\tau )-{\vec \eta}_j(\tau )|}} \Big) +\nonumber \\
&+& {1\over 2} \int d^3\sigma [({\dot {\vec A}}_{\perp}+[\vec \lambda (\tau )
\cdot \vec \partial ]{\vec A}_{\perp})^2 -{\vec B}^2 ](\tau ,\vec \sigma ).
\label{V16}
\end{eqnarray}

\vfill\eject

\section{The Lienard-Wiechert potentials.}

The last of Eqs.(95) [with the gauge condition $\vec \lambda (\tau )={\dot 
{\vec g}}(\tau )=0$] can be solved by using the retarded Green function
\cite{lus2}

\begin{eqnarray}
G_{ret}(\tau;\vec{\sigma})&=&(1/2\pi)\theta(\tau)\delta[\tau^{2}-
\vec{\sigma}^{2}],\nonumber \\
\Rightarrow && \Box G_{ret}(\tau,\vec{\sigma})=\delta(\tau)\delta^3
(\vec{\sigma}),
\label{VI1}
\end{eqnarray}

and one obtains

\begin{eqnarray}
{\check A}^{r}_{\perp RET}(\tau,\vec{\sigma})\, &{\buildrel \circ \over =}&\,
{\check A}^r_{\perp IN}(\tau,\vec{\sigma})+\nonumber \\
&+&\sum_{i=1}^N\frac{Q_i}{2\pi}P_{\perp}^{rs}(\vec{\sigma}) \{
\int d\tau'\theta(\tau-\tau')\delta[(\tau-\tau')^{2}-(\vec{\sigma}-\vec{\eta}
_{i}(\tau^{'}))^{2}]\dot{\eta}^{s}_{i}(\tau')+\nonumber \\
&+&\int d\tau^{'} d^3\sigma^{'} \theta (\tau -\tau^{'}) \delta [(\tau -\tau
^{'})^2-(\vec \sigma -{\vec \sigma}^{'})^2] \nonumber \\
&&{{\epsilon^{suv}{\bar S}^u_{i\xi}
(\tau^{'})}\over {\eta_i \sqrt{m^2_i+{\check {\vec \kappa}}^2_i(\tau^{'})}}}
{{\partial}\over {\partial \sigma^{{'}\, v}}} \delta^3({\vec \sigma}^{'}-
{\vec \eta}_i(\tau^{'})) \}=\nonumber \\
&=&{\check A}^{r}_{\perp IN}(\tau,
\vec{\sigma})+\sum_{i=1}^N\frac{Q_{i}}{4\pi}P^{rs}_{\perp}(\vec{\sigma}) \{
\frac{{\dot \eta}^{s}_{i}(\tau_{i+}(\tau ,\vec \sigma ))}{\varrho_{i+}
(\tau_{i+}(\tau ,\vec \sigma ),\vec \sigma )}+\nonumber \\
&+&2\int d\tau^{'} d^3\sigma^{'} \theta (\tau -\tau^{'}) \delta [(\tau -\tau
^{'})^2-(\vec \sigma -{\vec \sigma}^{'})^2] \nonumber \\
&&{{\epsilon^{suv}{\bar S}^u_{i\xi}
(\tau^{'})}\over {\eta_i \sqrt{m^2_i+{\check {\vec \kappa}}^2_i(\tau^{'})}}}
{{\partial}\over {\partial \sigma^{{'}\, v}}} \delta^3({\vec \sigma}^{'}-
{\vec \eta}_i(\tau^{'})) \} .
\label{VI2}
\end{eqnarray}

\noindent where $A_{\perp IN}(\tau ,\vec \sigma )\,\,$ [$\Box A_{\perp IN}(\tau,
\vec{\sigma})=0$] is a homogeneous solution describing arbitrary incoming 
radiation.

Here we introduced the following notations.
Let $(\tau ,\vec \sigma )$ be the coordinates of a point $z^{\mu}(\tau ,\vec 
\sigma )$ of Minkowski spacetime lying on the Wigner hyperplane $\Sigma_W(\tau 
)$, on which the locations of the particles are $(\tau ,{\vec \eta}_i(\tau ))$
[i.e. $x^{\mu}_i(\tau )=z^{\mu}(\tau ,{\vec \eta}_i(\tau ))$]. The rest-frame
distance between $z^{\mu}(\tau ,\vec \sigma )$ and $x^{\mu}_i(\tau )$ is
${\vec r}_i(\tau ,\vec \sigma )=\vec \sigma -{\vec \eta}_i(\tau )$; let ${\hat
{\vec r}}_i(\tau ,\vec \sigma )=(\vec \sigma -{\vec \eta}_i(\tau ))/|\vec \sigma
-{\vec \eta}_i(\tau )|$ be the associated unit vector, ${\hat {\vec r}}_i^2
(\tau ,\vec \sigma )=1$.

Let $\tau_{i+}(\tau ,\vec \sigma )$ [the retarded times] denote the retarded 
solutions of the equations

\begin{equation} 
(\tau-\tau_{i+})^{2}=(\vec{\sigma}-\vec{\eta}_{i}(\tau_{i+}))^{2},\quad\quad
i=1,..,N.
\label{VI6}
\end{equation}

\noindent The point $z^{\mu}(\tau ,\vec \sigma )$ lies on the lightcones 
emanating from the particle worldlines at their points $x^{\mu}_i(\tau_{i+}
(\tau ,\vec \sigma ))=z^{\mu}(\tau_{i+}(\tau ,\vec \sigma ),{\vec \eta}_i
(\tau_{i+}(\tau ,\vec \sigma )))$, lying on the Wigner hyperplanes $\Sigma_W
(\tau_{i+}(\tau ,\vec \sigma ))$ respectively. The point $z^{\mu}(\tau ,\vec 
\sigma )$ on $\Sigma_W(\tau )$ will define points $z^{\mu}(\tau_{i+}(\tau ,
\vec \sigma ),\vec \sigma )$ on the Wigner hyperplanes $\Sigma_W(\tau_{i+}
(\tau ,\vec \sigma ))$ by orthogonal projection [since $(z^{\mu}(\tau ,\vec 
\sigma )-x^{\mu}_i(\tau_{i+}(\tau ,\vec \sigma ))^2=0$, we have $R_{i+}
(\tau ,\vec \sigma )=\sqrt{(z(\tau ,\vec \sigma )-z(\tau_{i+}(\tau ,\vec 
\sigma ),\vec \sigma ))^2}=\sqrt{-(z(\tau_{i+}(\tau ,\vec \sigma ),\vec \sigma )
-x_i(\tau_{i+}(\tau ,\vec \sigma ))^2}$ and $z^{\mu}(\tau ,\vec \sigma )-x^{\mu}
_i(\tau_{i+}(\tau ,\vec \sigma ))=R_{i+}(\tau ,\vec \sigma )(t^{\mu}_{i+}
(\tau ,\vec \sigma )+s^{\mu}_{i+}(\tau ,\vec \sigma ))$ with $t^{\mu}_{i+}
(\tau ,\vec \sigma )$ and $s^{\mu}_{i+}(\tau ,\vec \sigma )$ being the timelike
and spacelike unit vectors associated with $z^{\mu}(\tau ,\vec \sigma )-
z^{\mu}(\tau_{i+}(\tau ,\vec \sigma ),\vec \sigma )$ and $z^{\mu}(\tau_{i+}
(\tau ,\vec \sigma ),\vec \sigma )-x^{\mu}_i(\tau_{i+}(\tau ,\vec \sigma ))$
respectively; $R_{i+}(\tau ,\vec \sigma )$ is the Minkowski retarded distance
between $z^{\mu}(\tau ,\vec \sigma )$ and $x^{\mu}_i(\tau_{i+}(\tau ,\vec 
\sigma ))$].

Let ${\vec r}_{i+}(\tau_{i+}(\tau ,\vec \sigma ),\vec \sigma )=\vec \sigma -
{\vec \eta}_i(\tau_{i+}
(\tau ,\vec \sigma ))$ denote the rest-frame retarded distance between the 
points $z^{\mu}(\tau_{i+}(\tau ,\vec \sigma ),\vec \sigma )$ and the points
$x^{\mu}_i(\tau_{i+}(\tau ,\vec \sigma ))$ of the worldlines belonging to 
$\Sigma_W(\tau_{i+}(\tau ,\vec \sigma ))$ [with ${\hat {\vec r}}_{i+}
(\tau_{i+}(\tau ,\vec \sigma ),\vec \sigma )$ being the unit vector, ${\hat 
{\vec r}}^2_{i+}=1$]. Let us denote the length of the vectors ${\vec r}_{i+}
(\tau_{i+}(\tau ,\vec \sigma ),\vec \sigma )$ with

\begin{eqnarray}
r_{i+}(\tau_{i+}(\tau ,\vec \sigma ),\vec \sigma )&=&|{\vec r}_{i+}(\tau
_{i+}(\tau ,\vec \sigma ),\vec \sigma )|=|\vec \sigma
-{\vec \eta}_i(\tau_{i+}(\tau ,\vec \sigma ))|=\nonumber \\
&=&\tau -\tau_{i+}(\tau ,\vec \sigma ) > 0.
\label{VI7}
\end{eqnarray}

Then, we have

\begin{eqnarray}
\theta (\tau -\tau^{'})&& \delta [(\tau -\tau^{'})^2-(\vec \sigma -{\vec \eta}_i
(\tau^{'}))^2]={{\delta (\tau^{'}-\tau_{i+}(\tau ,\vec \sigma ))}\over
{2 \rho_{i+}(\tau_{i+}(\tau ,\vec \sigma ),\vec \sigma )}},\nonumber \\
&&{}\nonumber \\
\rho_{i+}(\tau_{i+}(\tau ,\vec \sigma ),\vec \sigma )&=&\tau -\tau_{i+}(\tau ,
\vec \sigma )-{\dot {\vec \eta}}_i(\tau_{i+}(\tau ,\vec \sigma ))\cdot [\vec 
\sigma -{\vec \eta}_i(\tau_{i+}(\tau ,\vec \sigma ))]=\nonumber \\
&=&r_{i+}(\tau_{i+}(\tau ,\vec \sigma ),\vec \sigma ) [1-{\dot {\vec \eta}}_i
(\tau_{i+}(\tau ,\vec \sigma ))\cdot {\hat {\vec r}}_{i+}(\tau_{i+}(\tau ,\vec 
\sigma ),\vec \sigma )].
\label{VI8}
\end{eqnarray}

Since the quantities $Q_i\epsilon^{suv}{\bar S}^u_{i \xi}/\eta_i\sqrt{m^2_i
+{\check {\vec \kappa}}^2_i}$ are constants of the motion as implied by 
Eqs.(95) [since ${\dot {\vec {\bar S}}}_{i\xi}$ and ${\dot {\vec \kappa}}_i$
are proportional to $Q_i$ and $Q^2_i=0$, ${\dot Q}_i\, {\buildrel \circ \over 
=}\, 0$],  Eq.(\ref{VI2}) may be rewritten as

\begin{eqnarray}
{\check A}^{r}_{\perp RET}(\tau,\vec{\sigma})\, &{\buildrel \circ \over =}&\,
{\check A}^{r}_{\perp IN}(\tau,
\vec{\sigma})+\sum_{i=1}^N\frac{Q_{i}}{4\pi}P^{rs}_{\perp}(\vec{\sigma}) \{
\frac{{\dot \eta}^{s}_{i}(\tau_{i+}(\tau ,\vec \sigma ))}{\varrho_{i+}
(\tau_{i+}(\tau ,\vec \sigma ),\vec \sigma )}+\nonumber \\
&+&2 {{\epsilon^{suv}{\bar S}^u_{i\xi}
(\tau )}\over {\eta_i \sqrt{m^2_i+{\check {\vec \kappa}}^2_i(\tau )}}} \cdot
\nonumber \\
&&\int d\tau^{'} d^3\sigma^{'} \theta (\tau -\tau^{'}) \delta [(\tau -\tau
^{'})^2-(\vec \sigma -{\vec \sigma}^{'})^2]
{{\partial}\over {\partial \sigma^{{'}\, v}}} \delta^3({\vec \sigma}^{'}-
{\vec \eta}_i(\tau^{'})) \} .
\label{VI3}
\end{eqnarray}

After an integration by parts, one has

\begin{eqnarray}
\int d\tau^{'} d^3\sigma^{'}&& \theta (\tau -\tau^{'}) \delta [(\tau -\tau
^{'})^2-(\vec \sigma -{\vec \sigma}^{'})^2]
{{\partial}\over {\partial \sigma^{{'}\, v}}} \delta^3({\vec \sigma}^{'}-
{\vec \eta}_i(\tau^{'}))  =\nonumber \\
&=&2\pi \int d\tau^{'} d^3\sigma^{'} G_{RET}[\tau -\tau^{'}, |\vec \sigma 
-{\vec \sigma}^{'}|] 
{{\partial}\over {\partial \sigma^{{'}\, v}}} \delta^3({\vec \sigma}^{'}-
{\vec \eta}_i(\tau^{'}))  =\nonumber \\
&=&-2\pi \int d\tau^{'} {{\partial}\over {\partial \eta^v_i(\tau^{'})}}
G_{RET}[\tau -\tau^{'}, |\vec \sigma -{\vec \eta}_i(\tau^{'})|] =\nonumber \\
&=&{{\partial}\over {\partial \sigma^v}} \int d\tau^{'} \theta (\tau -\tau^{'})
\delta [(\tau -\tau^{'})^2-|\vec \sigma -{\vec \eta}_i(\tau^{'})|^2]=
\nonumber \\
&=&{{\partial}\over {\partial \sigma^v}} {1\over {2 \rho_{i+}(\tau_{i+}(\tau
,\vec \sigma),\vec \sigma )}}. 
\label{VI4}
\end{eqnarray}

Therefore the retarded Lienard-Wiechert potential of the spinning particle is

\begin{eqnarray}
{\check A}^{r}_{\perp RET}(\tau,\vec{\sigma})\, &{\buildrel \circ \over =}&\,
{\check A}^r_{\perp IN}(\tau,\vec{\sigma})+
\sum_{i=1}^N\frac{Q_{i}}{4\pi}P^{rs}_{\perp}(\vec{\sigma}) \{
\frac{{\dot \eta}^{s}_{i}(\tau_{i+}(\tau ,\vec \sigma ))}{\varrho_{i+}
(\tau_{i+}(\tau ,\vec \sigma ),\vec \sigma )}-\nonumber \\
&-&{{\epsilon^{suv}{\bar S}^u_{i \xi}(\tau )}\over {\eta_i\sqrt{m^2_i+
{\check {\vec \kappa}}^2_i(\tau )}}} {{\partial_v \rho_{i+}(\tau_{i+}(\tau
,\vec \sigma ),\vec \sigma )}\over {[\rho_{i+}(\tau_{i+}(\tau ,\vec \sigma ),
\vec \sigma ) ]^2}} \} .
\label{VI5}
\end{eqnarray}

Eq.(\ref{VI5}) can be rewritten as

\begin{eqnarray}
{\check A}^{r}_{\perp RET}(\tau,\vec{\sigma})&=&{\check A}^{r}_{\perp IN}(\tau,
\vec{\sigma})+\sum_{i=1}^N\frac{Q_{i}}{4\pi}P^{rs}_{\perp}(\vec{\sigma})
\Big( \frac{{\dot \eta}^{s}_{i}(\tau_{i+}(\tau ,\vec \sigma ))}{\varrho_{i+}
(\tau_{i+}(\tau ,\vec \sigma ),\vec \sigma )}+\nonumber \\
&+&{{[{\vec {\bar S}}_{i\xi} \times \vec \partial \varrho_{i+}(\tau_{i+}(\tau 
,\vec \sigma ),\vec \sigma )]^s}\over {\varrho^2_{i+}(\tau_{i+}(\tau 
,\vec \sigma ),\vec \sigma )\, \eta_i\sqrt{m^2_i+{\vec \kappa}^2_i(\tau )}}}
\Big) =\nonumber\\
&=&{\check A}^{r}_{\perp IN}(\tau,\vec{\sigma})+\sum_{i=1}^N
\frac{Q_{i}}{4\pi}[\, \Big(
\frac{{\dot \eta}^r_{i}(\tau_{i+}(\tau ,\vec \sigma ))}
{\varrho_{i+}(\tau_{i+}(\tau ,\vec \sigma ),\vec \sigma )}+\nonumber\\
&+&{{[{\vec {\bar S}}_{i\xi} \times \vec \partial \varrho_{i+}(\tau_{i+}(\tau 
,\vec \sigma ),\vec \sigma )]^r}\over {\varrho^2_{i+}(\tau_{i+}(\tau 
,\vec \sigma ),\vec \sigma )\, \eta_i\sqrt{m^2_i+{\vec \kappa}^2_i(\tau )}}}
\Big) -\nonumber \\
&-&{1\over {4\pi}}
\int d^{3}\sigma'\frac{\pi^{rs}(\vec{\sigma}-\vec{\sigma}')}
{\mid\vec{\sigma}-\vec{\sigma}'\mid^{3}} \Big( \frac
{{\dot \eta}^{s}_{i}(\tau_{i+}(\tau,\vec{\sigma'}))}{\varrho_{i+}(\tau_{i+}
(\tau ,\vec \sigma^{'}),\vec{\sigma}')}\, +\nonumber \\
&+&{{[{\vec {\bar S}}_{i\xi} \times \vec \partial \varrho_{i+}(\tau_{i+}(\tau 
,{\vec \sigma}^{'}),{\vec \sigma}^{'})]^s}\over {\varrho^2_{i+}(\tau_{i+}(\tau 
,{\vec \sigma}^{'}),{\vec \sigma}^{'})\, \eta_i\sqrt{m^2_i+{\vec \kappa}^2
_i(\tau )}}} \Big) \, ] =\nonumber\\
&=&{\check A}^{r}_{\perp IN}(\tau,\vec{\sigma})+\sum_{i=1}^N\check{A}^{r}
_{\perp (i+)}(\tau_{i+}(\tau ,\vec \sigma ),\vec{\sigma})=\nonumber \\
&=&{\check A}^{r}_{\perp IN}(\tau,\vec{\sigma})+\sum_{i=1}^NQ_i\tilde{A}^{r}
_{\perp (i+)}(\tau_{i+}(\tau ,\vec \sigma ),\vec{\sigma}),\nonumber \\
&&{}\nonumber \\
\pi^{rs}(\vec \sigma )&=&\delta^{rs}-{{3\sigma^r\sigma^s}\over {{\vec \sigma}
^2}},
\label{VI9}
\end{eqnarray}

\noindent where ${\check {\vec A}}_{\perp (i+)}(\tau_{i+}(\tau ,\vec \sigma )
,\vec \sigma )$ is the 
rest-frame form of the Lienard-Wiechert retarded potential produced by particle 
i [its Minkowski analogue, i.e. the relativistic generalization of the Coulomb
potential, is $A^{\mu}_{(i+)}(z)={{Q_i}\over {4\pi}} {{{\dot x}^{\mu}_i(\tau
_{i+})}\over {{\dot x}_i(\tau_{i+})\cdot [z-x_i(\tau_{i+})]}}={{Q_i}\over 
{4\pi}} {{{\dot x}^{\mu}_i(\tau_{i+})}\over {R_{i+}[1-{\hat {\vec R}}_{i+}\cdot
{\dot {\vec x}}_i(\tau_{i+})]}}$ in the case of scalar particles]. 
Since we are in the rest-frame Coulomb gauge
with only transverse Wigner-covariant vector potentials, ${\check {\vec A}}
_{\perp (i+)}(\tau_{i+}(\tau ,\vec \sigma ),\vec \sigma )$ has a first standard 
term generated at the retarded time $\tau_{i+}(\tau ,\vec \sigma )$ at $x^{\mu}
_i(\tau_{i+}(\tau ,\vec \sigma ))$, which is, however, accompanied by a 
nonlocal term receiving contributions from all the retarded times $-\infty < 
\tau_{i+}(\tau ,{\vec \sigma}^{'}) \leq \tau$., which is due to the elimination
of the electromagnetic gauge degrees of freedom [this is the origin of the
transverse projector]. If we put the ${\check {\vec A}}_{\perp (i+)}(\tau_{i+}
(\tau ,\vec \sigma ),\vec \sigma )$'s in the particle equations (95), 
with ${\check {\vec A}}_{\perp IN}(\tau ,\vec \sigma )=0$, then the equations 
of motion become integro-differential equations like the ones generated by 
a Fokker action.

To evaluate the electric [${\check {\vec E}}_{\perp}=-{\dot {\check {\vec A}}}
_{\perp}$] and magnetic [${\check {\vec B}}=-\vec \partial \times
{\check {\vec A}}_{\perp}$] fields produced by ${\check {\vec A}}_{\perp (i+)}
(\tau ,\vec \sigma )$, we need the rule of derivation of `retarded' functions
$g(\tau ,\vec \sigma; \tau_{i+}(\tau ,\vec \sigma ))$. From Eq.(\ref{VI6})
we get $(\tau -\tau_{i+})(d\tau -d\tau_{i+})=
r_{i+}(d\tau -d\tau_{i+})=[\vec \sigma -{\vec \eta}_i(\tau_{i+})]\cdot [d\vec 
\sigma -{\dot {\vec \eta}}_i(\tau_{i+})d\tau_{i+}]={\vec r}_{i+}\cdot
[d\vec \sigma -{\dot {\vec \eta}}_i(\tau_{i+})d\tau_{i+}]$. Therefore, by
introducing the notation

\begin{eqnarray}
\vec{v}_{i+}(\tau_{i+}(\tau ,\vec \sigma ),\vec \sigma )&=&\frac{\vec{r}_{i+}
(\tau_{i+}(\tau ,\vec \sigma ),\vec \sigma )}{\varrho_{i+}(\tau_{i+}(\tau ,\vec 
\sigma ),\vec \sigma )}=\frac{\hat{\vec{r}}_{i+}(\tau_{i+}(\tau ,\vec \sigma ),
\vec \sigma )}{1-\dot{\vec{\eta}}_{i}(\tau_{i+}(\tau ,\vec \sigma ))\cdot
\hat{\vec{r}}_{i+}(\tau_{i+}(\tau ,\vec \sigma ),\vec \sigma )},
\quad {\hat {\vec r}}_{i+}={{{\vec v}_{i+}}\over {|{\vec v}_{i+}|}},\nonumber \\
\:\mid\vec{v}_{i+}(\tau_{i+}(\tau ,\vec \sigma ),\vec \sigma )\mid &=& \frac{1}
{1-\dot{\vec{\eta}}{i}(\tau_{i+}(\tau ,\vec \sigma ))\cdot\hat{\vec{v}}_{i+}
(\tau_{i+}(\tau ,\vec \sigma ),\vec \sigma )}=\frac{\tau-\tau_{i+}(\tau ,\vec 
\sigma )}{\varrho_{i+}(\tau_{i+}(\tau ,\vec \sigma ),\vec \sigma )},
\label{VI10}
\end{eqnarray}

\noindent we get

\begin{eqnarray}
{{\partial \tau_{i+}(\tau ,\vec \sigma )}\over {\partial \tau}}&=&|{\vec v}
_{i+}(\tau_{i+}(\tau ,\vec \sigma ),\vec \sigma )|,\nonumber \\
{{\partial \tau_{i+}(\tau ,\vec \sigma )}\over {\partial \sigma^s}}&=&{\hat r}
_{i+\, s}(\tau_{i+}(\tau ,\vec \sigma ),\vec \sigma )\, |{\vec v}_{i+}(\tau
_{i+}(\tau ,\vec \sigma ),\vec \sigma )|=v_{i+\, s}(\tau_{i+}(\tau ,\vec 
\sigma ),\vec \sigma ),\nonumber \\
{{\partial g(\tau ,\vec \sigma ;\tau_{i+}(\tau ,\vec \sigma )))}\over {\partial
\tau}}&=&[({{\partial}\over {\partial \tau}}{|}_{\tau^{'}}+|{\vec v}_{i+}
(\tau_{i+}(\tau ,\vec \sigma ),\vec \sigma )|{{\partial}\over {\partial \tau
^{'}}})g(\tau ,\vec \sigma ;\tau^{'})]{|}_{\tau^{'}=\tau_{i+}(\tau ,\vec 
\sigma )},\nonumber \\
{{\partial g(\tau ,\vec \sigma ;\tau_{i+}(\tau ,\vec \sigma )))}\over {\partial
\sigma^s}}&=&[({{\partial}\over {\partial \sigma^s}}{|}_{\tau^{'}}+v_{i+\, s}
(\tau_{i+}(\tau ,\vec \sigma ),\vec \sigma ){{\partial}
\over {\partial \tau^{'}}})g(\tau ,\vec \sigma ;\tau^{'})]{|}_{\tau^{'}=
\tau_{i+}(\tau ,\vec \sigma )}.
\label{VI11}
\end{eqnarray}

\noindent so that, using the derived equations

\begin{eqnarray}
&&{{\partial {\vec r}_{i+}(\tau_{i+},\vec \sigma )}\over
{\partial \tau_{i+}}}=-{\dot {\vec \eta}}_i(\tau_{i+}),\nonumber \\ 
&&{{\partial r_{i+}(\tau_{i+} ,\vec \sigma )}\over {\partial \tau_{i+}}}
=-{\dot {\vec \eta}}_i(\tau_{i+})\cdot {\hat 
{\vec r}}_{i+}(\tau_{i+},\vec \sigma ), \nonumber \\
&&{{\partial \rho_{i+}(\tau_{i+},\vec \sigma )}\over {\partial \tau_{i+}}}
={\dot {\vec \eta}}_i^2(\tau_{i+})-(\, {\dot {\vec 
\eta}}_i(\tau_{i+})+r_{i+}(\tau_{i+},\vec \sigma ){\ddot {\vec \eta}}_i(\tau
_{i+})\, )\cdot{\hat {\vec r}}_{i+}(\tau_{i+},\vec \sigma ), \nonumber \\
&&{{\partial}\over 
{\partial \sigma^s}}{|}_{\tau_{i+}}\, r^r_{i+}(\tau_{i+},\vec \sigma )=\delta
^r_s, \nonumber \\
&&{{\partial}\over {\partial \sigma^s}}{|}_{\tau_{i+}}\, r_{i+}(\tau
_{i+},\vec \sigma )={\hat r}^s_{i+}(\tau_{i+},\vec \sigma ), \nonumber \\
&&{{\partial}\over
{\partial \sigma^s}}{|}_{\tau_{i+}}\, \rho_{i+}(\tau_{i+},\vec \sigma )=-({\dot
\eta}_i^s(\tau_{i+})+{\hat r}^s_{i+}(\tau_{i+},\vec \sigma )\, )
\label{VI12}
\end{eqnarray}

\noindent we get

\begin{eqnarray}
{\check A}^{r}_{\perp RET}(\tau,\vec{\sigma})&=&{\check A}^{r}_{\perp IN}(\tau,
\vec{\sigma})+\sum_{i=1}^N\frac{Q_{i}}{4\pi}P^{rs}_{\perp}(\vec{\sigma})
\Big( \frac{{\dot \eta}^{s}_{i}(\tau_{i+}(\tau ,\vec \sigma ))}{\varrho_{i+}
(\tau_{i+}(\tau ,\vec \sigma ),\vec \sigma )}+\nonumber \\
&+&{{[{\vec {\bar S}}_{i\xi} \times ({\dot {\vec \eta}}_i(\tau_{i+}(\tau 
,\vec \sigma ))+{\hat {\vec r}}_{i+}(\tau_{i+}(\tau 
,\vec \sigma ),\vec \sigma ) )]^s}\over {\varrho^2_{i+}(\tau_{i+}(\tau 
,\vec \sigma ),\vec \sigma )\, \eta_i\sqrt{m^2_i+{\check {\vec \kappa}}
^2_i(\tau )}}} \Big) =\nonumber\\
&=&{\check A}^{r}_{\perp IN}(\tau,\vec{\sigma})+\sum_{i=1}^N\check{A}^{r}
_{\perp (i+)}(\tau_{i+}(\tau ,\vec \sigma ),\vec{\sigma})=\nonumber \\
&=&{\check A}^{r}_{\perp IN}(\tau,\vec{\sigma})+\sum_{i=1}^NQ_i\tilde{A}^{r}
_{\perp (i+)}(\tau_{i+}(\tau ,\vec \sigma ),\vec{\sigma}),\nonumber \\
&&{}\nonumber \\
{\check E}^{r}_{\perp RET}(\tau ,\vec \sigma )&=&-{{\partial}\over {\partial 
\tau}}{\check A}^r_{\perp RET}(\tau_{i+}(\tau ,\vec \sigma ),\vec \sigma )
\,{\buildrel \circ \over =}\,
{\check E}^{r}_{\perp IN}(\tau,\vec{\sigma})-\nonumber\\
&-&P^{rs}_{\perp}(\vec{\sigma})\sum_{i=1}^N\frac{Q_{i}}{4\pi}
|{\vec v}_{i+}(\tau_{i+}(\tau ,\vec \sigma ),\vec \sigma )|\, \Big(
{{{\ddot \eta}^s_i(\tau_{i+}(\tau ,\vec \sigma ))}\over {\rho_{i+}(\tau
_{i+}(\tau ,\vec \sigma ),\vec \sigma )}}-\nonumber \\
&-&{{{\dot \eta}^s_i(\tau_{i+}(\tau ,\vec \sigma ))}\over {\rho^2_{i+}
(\tau_{i+}(\tau ,\vec \sigma ),\vec \sigma )}} (\, {\dot {\vec \eta}}^2
_i(\tau_{i+}(\tau ,\vec  \sigma ))-\nonumber \\
&-&({\dot {\vec \eta}}_i(\tau_{i+}(\tau ,\vec \sigma ))+r_{i+}(\tau
_{i+}(\tau ,\vec \sigma ),\vec \sigma ) {\ddot {\vec \eta}}_i(\tau_{i+}(\tau ,
\vec \sigma ))\, )\cdot {\hat {\vec r}}_{i+}(\tau_{i+}(\tau ,\vec \sigma ),
\vec \sigma ))\, +\nonumber \\
&+&{{({\vec {\bar S}}_{i\xi} \times {\vec C}_i(\tau_{i+}(\tau ,\vec \sigma ),
\vec \sigma )\, )^s}\over {\eta_i\sqrt{m^2_i+{\check {\vec \kappa}}
_i^2(\tau )}}}\Big) =\nonumber \\
&=&{\check E}^r_{\perp IN}(\tau ,\vec \sigma )+\sum_{i=1}^N{\check E}^r_{\perp 
(i+)}(\tau_{i+}(\tau ,\vec \sigma ),\vec \sigma )=\nonumber \\
&=&{\check E}^{r}_{\perp IN}(\tau,\vec{\sigma})+\sum_{i=1}^NQ_{i}\tilde{E}
^{r}_{\perp (i+)}(\tau_{i+}(\tau ,\vec \sigma ),\vec{\sigma}),\nonumber \\
&&{}\nonumber \\
with&& {\vec C}_i(\tau_{i+}(\tau ,\vec \sigma ),\vec \sigma )=
{1\over {\rho^2_{i+}(\tau_{i+}(\tau ,\vec \sigma ),\vec \sigma )}} \Big(
{\ddot {\vec \eta}}_i(\tau_{i+}(\tau ,\vec \sigma ))+\nonumber \\
&+&{{({\dot {\vec \eta}}_i(
\tau_{i+}(\tau ,\vec \sigma )) \cdot {\hat {\vec r}}_{i+}(\tau_{i+}(\tau ,\vec 
\sigma ),\vec \sigma )\, ) {\hat {\vec r}}_{i+}(\tau_{i+}(\tau ,\vec \sigma ),
\vec \sigma )-{\dot {\vec \eta}}_i(\tau_{i+}(\tau ,\vec \sigma ))}\over
{r_{i+}(\tau_{i+}(\tau ,\vec \sigma ),\vec \sigma )}}-\nonumber \\
&-&2{{({\dot {\vec \eta}}_i(\tau_{i+}(\tau ,\vec \sigma ))+{\hat {\vec r}}
_{i+}(\tau_{i+}(\tau ,\vec \sigma ),\vec \sigma ) )}\over {\rho_{i+}
(\tau_{i+}(\tau ,\vec \sigma ),\vec \sigma )}} [{\dot {\vec \eta}}^2_i(\tau
_{i+}(\tau ,\vec \sigma ))-\nonumber \\
&-&({\dot {\vec \eta}}_i(\tau_{i+}(\tau ,\vec \sigma ))+
r_{i+}(\tau_{i+}(\tau ,\vec \sigma ),\vec \sigma ) {\ddot {\vec \eta}}
_i(\tau_{i+}(\tau ,\vec \sigma ))\, )\cdot {\hat {\vec r}}_{i+}(\tau_{i+}(\tau 
,\vec \sigma ),\vec \sigma ) ] \Big),\nonumber\\
&&{}\nonumber \\
{\check B}^{r}_{RET}(\tau,\vec{\sigma})&=&-\epsilon^{rsu}(\partial^{s}
{\check A}^{u}_{\perp RET}(\tau,\vec{\sigma}))\, {\buildrel \circ \over =}\,
={\check B}^{r}_{IN}(\tau,\vec{\sigma})+\nonumber \\
&+&\sum_{i=1}^N{{Q_i}\over {4\pi}} \epsilon^{rsu}P^{uv}_{\perp}(\vec \sigma )
({{\partial}\over {\partial \sigma^s}}{|}_{\tau_{i+}}+v_{i+,\, s}(\tau_{i+}
(\tau ,\vec \sigma ),\vec \sigma){{\partial}\over {\partial \tau_{i+}}})\Big(
{{{\dot \eta}_i^v(\tau_{i+}(\tau ,\vec \sigma ))}\over {\rho_{i+}(\tau_{i+}
(\tau ,\vec \sigma ),\vec \sigma )}}+\nonumber \\
&+&{{[{\vec {\bar S}}_{i\xi} \times ({\dot {\vec \eta}}_i(\tau_{i+}(\tau 
,\vec \sigma ))+{\hat {\vec r}}_{i+}(\tau_{i+}(\tau 
,\vec \sigma ),\vec \sigma ) )]^v}\over {\varrho^2_{i+}(\tau_{i+}(\tau 
,\vec \sigma ),\vec \sigma )\, \eta_i\sqrt{m^2_i+{\check {\vec \kappa}}
^2_i(\tau )}}} \Big) = \nonumber \\
&=&{\check B}^r_{IN}(\tau ,\vec \sigma )+\sum_{i=1}^N [{\hat {\vec r}}_{i+}
(\tau_{i+}(\tau ,\vec \sigma ),\vec \sigma ) \times {\check {\vec E}}_{\perp 
(i+)}(\tau_{i+}(\tau ,\vec \sigma ),\vec \sigma )\, ]^r+\nonumber \\
&+&\sum_{i=1}^N{{Q_i}\over {4\pi}} \epsilon^{rsu}P^{uv}_{\perp}(\vec \sigma )
\Big( {{[{\dot \eta}^s_{i+}(\tau_{i+}(\tau \vec \sigma ))+
{\hat r}^s_{i+}(\tau_{i+}(\tau ,\vec \sigma ),\vec \sigma )] {\dot \eta}_i^v
(\tau_{i+}(\tau ,\vec \sigma ))}\over {\rho^2_{i+}(\tau_{i+}(\tau ,\vec \sigma )
,\vec \sigma )}}+\nonumber \\
&+&{{\epsilon^{vnm} {\bar S}^n_{i\xi}(\tau )}\over {\eta_i\sqrt{m^2_i+
{\check {\vec \kappa}}^2_i(\tau )}}} C^{ms}_i(\tau_{i+}(\tau ,\vec \sigma ),
\vec \sigma ) \Big) =\nonumber \\
&=&{\check B}^r_{IN}(\tau ,\vec \sigma )+\sum_{i=1}^N{\check B}^r_{(i+)}(\tau
_{i+}(\tau ,\vec \sigma ),\vec \sigma )=\nonumber \\
&=&{\check B}^r_{IN}(\tau ,\vec \sigma )+\sum_{i=1}^NQ_i{\tilde B}^r_{(i+)}
(\tau_{i+}(\tau ,\vec \sigma ),\vec \sigma ),\nonumber \\
&&{}\nonumber \\
with&& C^{ms}_i(\tau_{i+}(\tau ,\vec \sigma ),\vec \sigma )={1\over
{\rho^2_{i+}(\tau_{i+}(\tau ,\vec \sigma ),\vec \sigma )}} \Big(
{{\delta^{ms}-
({\hat r}^m_{i+}{\hat r}^s_{i+})(\tau_{i+}(\tau ,\vec \sigma ),\vec \sigma )}
\over {r_{i+}(\tau_{i+}(\tau ,\vec \sigma ),\vec \sigma )}}+\nonumber \\
&+&2 {{({\dot \eta}^m_i(\tau_{i+}(\tau ,\vec \sigma ))+{\hat r}^m_{i+}(\tau
_{i+}(\tau ,\vec \sigma ),\vec \sigma ) )({\dot \eta}^s_i(\tau_{i+}(\tau 
,\vec \sigma ))+{\hat r}^s_{i+}(\tau_{i+}(\tau ,\vec \sigma ),\vec \sigma ) )}
\over {\rho_{i+}(\tau_{i+}(\tau ,\vec \sigma ),\vec \sigma )}} \Big) .
\label{VI13}
\end{eqnarray}

\noindent The particle equations of motion contained in Eqs.(95), the 
definition of the rest frame, Eq,(91), and the conserved relative energy 
$H_{rel}$ of Eq.(94) have now
the following form 

\begin{eqnarray}
\frac{d}{d\tau}(\eta_{i}&&\sqrt{m^2_i-2Q_i{\vec {\bar S}}_{i\xi}(\tau )\cdot 
[{\check {\vec B}}_{IN}(\tau ,{\vec \eta}_i(\tau ))+\sum_{j\not= i}Q_j{\vec
{\tilde B}}_{(j+)}(\tau_{j+}(\tau ,{\vec \eta}_i(\tau )),{\vec \eta}_i(\tau ))}
\frac{\dot{\vec{\eta}}_{i}(\tau)}{\sqrt{
1-\dot{\vec{\eta}}_{i}^{2}(\tau)}})\, {\buildrel \circ \over =}\, \nonumber \\
&{\buildrel \circ \over =}\,& -\sum
_{k\neq i}\frac{Q_{i}Q_{k}[{\vec \eta}_i(\tau )-{\vec \eta}_k(\tau )]}
{4\pi\mid\vec{\eta}_{i}(\tau )-\vec{\eta}_{k}(\tau )\mid^{3}}
+{{Q_i}\over {m_i}} \eta_i\sqrt{1-{\dot {\vec \eta}}^2_i(\tau )} {\bar S}
_{i\xi}^u(\tau ) \cdot \nonumber \\
&&[{{\partial {\check B}_{IN}^u(\tau ,{\vec \eta}_i(\tau ))}\over {\partial
{\vec \eta}_i}}+\sum_{j\not= i} Q_j {{\partial {\tilde B}_{(j+)}^u(\tau_{j+}
(\tau ,{\vec \eta}_i(\tau )),{\vec \eta}_i(\tau )) }\over {\partial
{\vec \eta}_i}} ]+\nonumber \\
&+&Q_{i}[\check{\vec{E}}_{\perp IN}(\tau,\vec{\eta}_{i}(\tau ))+
\dot{\vec{\eta}}_{i}(\tau)\times\check{\vec{B}}_{IN}(\tau,\vec{\eta}_{i}(\tau )
)]+\nonumber\\
&+&\sum_{k\neq i}Q_{i}Q_{k}[\vec{\tilde{E}}_{\perp (k+)}(\tau_{i+}(\tau ,
{\vec \eta}_i(\tau )),\vec{\eta}_{i}(\tau ))+\nonumber \\
&+&\dot{\vec{\eta}}_{i}(\tau)\times
\vec{\tilde{B}}_{(k+)}(\tau_{i+}(\tau ,{\vec \eta}_i(\tau )),\vec{\eta}_{i}
(\tau ))],\nonumber \\
&&{}\nonumber \\
\sum_{i=1}^N[\eta_{i}&&\sqrt{m^2_i-2Q_i{\vec {\bar S}}_{i\xi}(\tau )\cdot 
[{\check {\vec B}}(\tau ,{\vec \eta}_i(\tau )) +\sum_{j\not= i}Q_j{\vec
{\tilde B}}_{(j+)}(\tau_{j+}(\tau ,{\vec \eta}_i(\tau )),{\vec \eta}_i(\tau ))}
\frac{{\dot {\vec \eta}}_i(\tau )}{\sqrt{
1-\dot{\vec{\eta}}^{2}_{i}(\tau)}}+\nonumber \\
&+&Q_{i}{\check {\vec A}}_{\perp IN}(\tau,
\vec{\eta}_{i}(\tau ))]+
\sum_{i\neq j}Q_{i}Q_{j}\vec{\tilde{A}}_{\perp (j+)}(\tau_{j+}(\tau ,{\vec 
\eta}_i(\tau ),\vec{\eta}_{i}(\tau ))+\nonumber\\
&+&\int d^{3}\sigma\:
[\check{\vec{E}}_{\perp IN}\times \check{\vec{B}}_{IN}
+\sum_{i=1}^NQ_{i}(\check{\vec{E}}_{\perp IN}\times\vec{\tilde{B}}_{\perp (i+)}
(\tau_{i+},\vec \sigma )+\nonumber \\
&+&\vec{\tilde{E}}_{\perp (i+)}(\tau_{i+},\vec \sigma )
\times\check{\vec{B}}_{IN})+\nonumber\\
&+&\sum_{i\neq j}Q_{i}Q_{j}\vec{\tilde{E}}_{\perp (i+)}(\tau_{i+},\vec \sigma )
\times\vec{\tilde{B}}_{(j+)}(\tau_{j+},\vec \sigma )](\tau,\vec{\sigma})\,
{\buildrel \circ \over =}\, 0,\nonumber \\
&&{}\nonumber \\    
E_{rel}&=&\sum_{i}\frac{\eta_{i}
\sqrt{m^2_i-2Q_i{\vec {\bar S}}_{i\xi}(\tau )\cdot 
[{\check {\vec B}}(\tau ,{\vec \eta}_i(\tau )) +\sum_{j\not= i}Q_j{\vec
{\tilde B}}_{(j+)}(\tau_{j+}(\tau ,{\vec \eta}_i(\tau )),{\vec \eta}_i(\tau ))}
}{\sqrt{1-\dot{\vec{\eta}}_{i}^{2}(\tau)}}+\nonumber \\
&+&\sum_{i\neq j}\frac{Q_{i}Q_{j}}
{4\pi\mid\vec{\eta}_{i}(\tau )-\vec{\eta}_{j}(\tau )\mid}+
\int d^{3}\sigma\:[
\frac{\check{\vec{E}}^{2}_{\perp IN}+\check{\vec{B}}^{2}_{IN}}{2}
+\nonumber\\
&+&\sum_{i=1}^NQ_{i}(\check{\vec{E}}_{\perp IN}\cdot\vec{\tilde{E}}_{\perp
(i+)}(\tau_{i+},\vec \sigma )+\check{\vec{B}}_{IN}\cdot\vec{\tilde{B}}_{(i+)}
(\tau_{i+},\vec \sigma ))+\nonumber\\
&+&\sum_{i>j}Q_{i}Q_{j}(\vec{\tilde{E}}_{\perp (i+)}(\tau_{i+},\vec \sigma )
\cdot\vec{\tilde{E}}_{\perp (j+)}(\tau_{j+},\vec \sigma )+\nonumber \\
&+&\vec{\tilde{B}}_{(i+)}(\tau_{i+},\vec \sigma )\cdot\vec{\tilde{E}}_{\perp 
(j+)}(\tau_{j+},\vec \sigma ))\, ](\tau,\vec{\sigma})=const.
\label{VI14}
\end{eqnarray}
  
The property $Q^2_i=0$ has been used in these equations and it will be used also
in what follows. Besides the divergent Coulomb self-interaction it eliminates
other divergent terms.

In the nonrelativistic limit $|{\dot {\vec \eta}}_i(\tau )| < < 1$ and in wave 
zone [$\tau_{i+}(\tau ,\vec \sigma )\rightarrow \tau$, $\rho_{i+}(\tau_{i+}
(\tau ,\vec \sigma ),\vec \sigma )\rightarrow r(\tau ,\sigma )\approx |\vec
\sigma |\rightarrow \infty$] with ${\check {\vec A}}_{\perp IN}(\tau ,\vec
\sigma )=0$, since the `spin' contribution is of  order $|\vec \sigma|^{-2}$,
the asymptotic limit of the retarded fields is like for scalar particles

\begin{eqnarray}
{\check E}^{r}_{\perp RET,AS}(\tau ,\vec \sigma )&\approx&-P^{rs}_{\perp}(\vec
\sigma )\sum_{i=1}^N{{Q_i}\over {4\pi}} {{{\ddot \eta}^s_i(\tau )}\over
{|\vec \sigma |}},\nonumber \\ 
{\check B}^{r}_{RET,AS}(\tau ,\vec \sigma )&\approx&-P^{rs}_{\perp}(\vec \sigma
)\sum_{i=1}^N {{Q_i}\over {4\pi}} {{[\vec \sigma \times {\ddot {\vec \eta}}
_i(\tau )]^s}\over {|\vec \sigma |}},
\label{VI15}
\end{eqnarray}

\noindent so that the ``Larmor formula" for the radiated energy become
[$\vec n={\hat {\vec r}}=\vec \sigma /|\vec \sigma |$]

\begin{eqnarray}
\frac{dE}{d\tau}&\approx&
\int_{S}d\Sigma\:\vec{n}\cdot(\check{\vec{E}}_{\perp RET,AS}\times 
\check{\vec{B}}_{RET,AS})(\tau ,\vec \sigma )=\nonumber\\
&=&\sum_{i\not= j}{{Q_iQ_j}\over {(4\pi )^2}}\int d\Omega
\:\vec{n}\cdot ({\ddot {\vec \eta}}_i(\tau )\times[\vec{n}\times
\ddot{\vec{\eta}}_j(\tau )])=\nonumber \\
&=&\sum_{i\not= j} {{Q_iQ_j}\over {(4\pi )^2}} \int d\Omega (\vec n\times 
{\ddot {\vec \eta}}_i(\tau ))\cdot (\vec n\times {\ddot {\vec \eta}}_j(\tau ))=
\nonumber \\
&=&\frac{2}{3}\sum_{i\not= j} {{Q_iQ_j}\over {(4\pi )^2}} {\ddot {\vec \eta}}_i
(\tau )\cdot {\ddot {\vec \eta}}_j(\tau ).
\label{VI16}
\end{eqnarray}

The usual terms ${{Q^2_i}\over {(4\pi )^2}} {2\over 3} {\ddot {\vec \eta}}_i^2
(\tau )$ are absent due to the pseudoclassical conditions $Q^2_i=0$. Therefore,
at the pseudoclassical level, there is no radiation coming from single charges,
but only  interference radiation due to terms $Q_iQ_j$ with $i\not= j$. Since 
it is not possible to control whether the source is a single elementary charged
particle (only macroscopic sources are testable), this result is in accord
with macroscopic experimental facts.

For a single particle, N=1, the pseudoclassical equations (\ref{VI14}) 
become

\begin{eqnarray}
\frac{d}{d\tau}&&(\eta \sqrt{m^2-2Q {\vec {\bar S}}_{\xi}\cdot {\check 
{\vec B}}_{IN}(\tau ,\vec \eta (\tau ))}\frac{\dot{\vec{\eta}}(\tau)}{\sqrt
{1-\dot{\vec{\eta}}^2(\tau) } })\, {\buildrel \circ \over =}\, \nonumber \\
&{\buildrel \circ \over =}\,&Q[
\check{\vec{E}}_{\perp IN}(\tau,\vec{\eta}(\tau ))+\dot{\vec{\eta}}(\tau)\times
\check{\vec{B}}_{IN}(\tau,\vec{\eta})]+\eta {{\sqrt{1-{\dot {\vec \eta}}
^2(\tau )}}\over m} Q {\bar S}_{\xi}^u{{\partial {\check B}^u_{IN}(\tau ,\vec 
\eta (\tau ))}\over {\partial \vec \eta}},\nonumber \\
&&{}\nonumber \\
\eta && \sqrt{m^2-2Q {\vec {\bar S}}_{\xi}\cdot {\check 
{\vec B}}_{IN}(\tau ,\vec \eta (\tau ))}
\frac{\dot{\vec \eta}(\tau)}{\sqrt{1-\dot{\vec{\eta}}(\tau)}}
+Q\check{\vec{A}}_{\perp IN}(\tau,\vec{\eta}(\tau ))+\nonumber\\
&+&\int d^{3}\sigma
[(\check{\vec{E}}_{\perp IN}\times\check{\vec{B}}_{IN})(\tau,\vec{\sigma})+
Q(\check{\vec{E}}_{\perp IN}(\tau,\vec{\sigma})
\times\vec{\tilde{B}}_{(+)}(\tau_{+}(\tau,\vec{\sigma}),\vec \sigma )+
\nonumber \\
&+&\vec{\tilde{E}}_{\perp (+)}(\tau_{+}(\tau,\vec{\sigma}),\vec \sigma )\times
\check{\vec{B}}_{IN}(\tau,\vec{\sigma}))]
\, {\buildrel \circ \over =}\, 0,\nonumber \\
&&{}\nonumber \\
E_{rel}&=&\frac{\eta \sqrt{m^2-2Q {\vec {\bar S}}_{\xi}\cdot {\check 
{\vec B}}_{IN}(\tau ,\vec \eta (\tau ))}}
{\sqrt{1-\dot{\vec{\eta}}(\tau)^{2}}}+
\int d^{3}\sigma\:[\frac{\check{\vec{E}}^{2}_{\perp IN}+\check{\vec{B}}^{2}
_{IN}}{2}+\nonumber\\
&+&Q(\check{\vec{E}}_{\perp IN}\cdot\vec{\tilde{E}}_{\perp (+)}(\tau_{+},\vec
\sigma )+\check{\vec{B}}_{IN}\cdot\vec{\tilde{B}}_{(+)}(\tau_{+},\vec \sigma ))
](\tau,\vec{\sigma})=const. ,\nonumber \\
\frac{d}{d\tau}&&\frac{\eta \sqrt{m^2-2Q {\vec {\bar S}}_{\xi}\cdot {\check 
{\vec B}}_{IN}(\tau ,\vec \eta (\tau ))}}
{\sqrt{1-\dot{\vec{\eta}}^{2}(\tau)}}\, {\buildrel \circ \over =}\nonumber \\ 
&{\buildrel
\circ \over =}&\, Q\dot{\vec{\eta}}(\tau)\cdot\check{\vec{E}}_{\perp IN}(\tau,
\vec{\eta}(\tau ))+\int_{S_{as}}d\Sigma\:\vec{n}\cdot[
\check{\vec{E}}_{\perp IN}\times\check{\vec{B}}_{IN}+\nonumber\\
&+&Q(\check{\vec{E}}_{\perp IN}\times\vec{\tilde{B}}_{(+)}(\tau_{+},\vec 
\sigma )+\vec{\tilde{E}}_{\perp (+)}(\tau_{+},\vec \sigma )\times\check{\vec{B}
}_{IN})](\tau,\vec{\sigma}).
\label{VI17}
\end{eqnarray}

The first of Eqs.(\ref{VI17}) replaces the Abraham-Lorentz-Dirac equation 
[see for instance Ref.\cite{itz}] for
an electron in an external electromagnetic field [$Q=e\theta^{*}\theta$].

\vfill\eject

\section{Conclusions}

We have obtained the pseudoclassical description of the positive (or negative)
energy solutions of the Dirac equation on spacelike hypersurfaces and then
in the rest-frame instant form on Wigner hyperplanes. The coupling to the
electromagnetic field of these spinning particles on spacelike hypersurfaces is
such to be consistent with only a ``spin-magnetic field" interaction in the
intrinsic rest-frame on Wigner hyperplanes like in the nonrelativistic Pauli
theory, which is recovered in the limit $c\, \rightarrow \infty$. These results
are consistent with the elimination of the classical effects at the basis of
pair production and of all the effects of the same order in the electric
charge, which are obstructions to the diagonalization of the Dirac Hamiltonian
with the Foldy-Wouthuysen transformation.

The reduction to the rest-frame Wigner-covariant Coulomb gauge is done for the
system on N charged spinning particles plus the electromagnetic field. 
Also the Lienard-Wiechert potential of spinning particles was studied by
following Ref.\cite{lus2}. 

Then one has to find a 
connection with the existing literature on two-body equations for relativistic 
bound states starting from two coupled spinning particles with
various kinds of potentials \cite{crater,saz}. Also the coupling of colored
spinning particles to the SU(3) Yang-Mills field along the lines of
Ref.\cite{lus3} has to be done.

The quantization of this spinning particle (which will be studied elsewhere)
should be done following the scheme of Ref.\cite{lamm}, giving rise to a 
nonlocal  Schroedinger equation with the kinetic square root operator for a 
2-spinor, which corresponds to the SU(2) spinor in the positive (or negative)
energy solutions of the Dirac equation after a boost to the rest frame.

However, as pointed out in Section II, it is not clear how to recover the 
classical fibration describing the spin structure. In the pseudoclassical
theory of Eq.(1) one has the constraints $\chi =p^2-m^2\approx 0$,
describing a scalar particle, and $\chi_D=p_{\mu}\xi^{\mu}-m\xi_5\approx 0$,
restricting the $G_5$ spin fiber over $x^{\mu}(\tau )$ to a Grassmann algebra
$G_4$, with the consistency relation $\{ \chi_D,\chi_D \} =i\chi$. For a
scalar particle with only $\chi =p^2-m^2\approx 0$, quantization produces the
Klein-Gordon equation $(\Box +m^2) \phi (x)=0$. Therefore, for the spinning
particle one would expect a double role of the constraint $\chi \approx 0$:
i) $\chi =p^2-m^2\approx 0$ going to $(\Box +m^2) \phi (x)=0$; ii) $\chi_D=
p_{\mu}\xi^{\mu}-m\xi_5\approx 0$ going to $\gamma_5(i\partial_{\mu}\gamma
^{\mu}-m) \psi (x)=0$ with $\{ \chi_D,\chi_D \}=i\chi$ replaced by $(\Box +m^2)
\psi (x)=0$. This would replace the superfield $X^{\mu}=x^{\mu}+\theta \xi
^{\mu}$ of Eq.(6) not with a quantum superfield $\phi (x) +\theta \psi (x)$
(the supersymmetric scalar multiplet with a Grassmann valued Dirac field)
but with a first quantization fibration [$\psi (x)\,\, over\,\, \phi (x)$]
(with $\psi (x)$ a Dirac wave function) describing the spin structure. Here,
in [$\psi (x)\,\, over\,\, \phi (x)$] the Klein-Gordon field $\phi (x)$ should
be restricted to a special class of configurations peaked on a particle
worldline, the trace in Minkowski spacetime of the average position of the
electric current produced by the lepton. In the free case this worldline
should be a straightline, projection to Minkowski spacetime of the 
Foldy-Wouthuysen mean position. Therefore, the allowed configurations
of $\phi (x)$ should depend on only 8 degrees of freedom like for a scalar
particle. This can be obtained by using the new canonical decomposition in
center-of-mass and relative variables of a Klein-Gordon field \cite{lm,mater}
and by selecting its monopole configurations
[this decomposition is obtained starting from canonical action-angle variables, 
which do not exist for a Dirac wave function: this again points at the lack of
completeness in our description of fermions]. This restriction on the $\phi
(x)$ configurations is also implied by the fact that also bound states of
fermions should have the same fibered structure over restricted bound states
of scalar particles, otherwise one should have contradiction with the
experimental fact that the fermion bound state in the fiber is sufficient to
explain the spectra of bound states.

Let us remark that the same pattern should exist in the nonrelativistic Pauli
theory with [$\psi (x)\,\, over\,\, \phi (x)$] replaced by [$\psi_P(x)\,\, 
over\,\, \tilde \psi (x)$], where $\tilde \psi (x)$ is a Schroedinger wave
function and $\psi_P(x)$ a Pauli spinor. Bound states of these objects are
needed in the explaination of superconductivity by means of Cooper pairs.

As we shall see in Ref.\cite{bilu}, devoted to the pseudoclassical basis of QED
on spacelike hypersurfaces, the use of Grassmann valued Dirac fields will 
create a geometrical problem, which again points towards the necessity of a
 fibration to describe the spin structure both at the classical and at the
second quantized (functional Schroedinger equations \cite{hat}) level.

Instead bosonic fields like the electromagnetic one already contain the 
fibration: in the limit of geometrical optics one has a null ray of light with
the spin structure (the Stokes parameters) over it (see the pseudoclassical
photon \cite{photon}).

Finally, massless spinning particles and fermion fields on spacelike
hypersurfaces will be treated in a future paper, because they require the
reformulation of the front form of dynamics in the instant form.

\vfill\eject

\appendix

\section{Foldy-Wouthuysen, Cini-Touschek and chiral representations.}

In the Foldy-Wouthuysen representation
\cite{fw}, the momentum space Dirac equation in the noncovariant
form $(i\partial^o -H) \psi (p)=0$ with 
$\hat H=\beta m+\vec \alpha \cdot \vec p$ is
transformed to $(i\partial^o-\gamma^o \sqrt{{\vec p}^2+m^2}) \psi_{(FW)}(p)=0$
with $\psi_{(FW)}(p)=\left( \begin{array}{c} \chi_{(FW)} \\ \eta_{(FW)}
\end{array} \right) (p)={\hat U}_{(FW)} \psi (p)={\hat U}_{(FW)} \left(
\begin{array}{c} \chi \\ \eta \end{array} \right) (p)$ [$\chi_{(FW)}={{
(\sqrt{{\vec p}^2+m^2}+m)\chi +\vec p\cdot \vec \sigma \eta}\over 
{\sqrt{2 \sqrt{{\vec p}^2+m^2}(\sqrt{{\vec p}^2+m^2}+m)} }}$
and $\eta_{(FW)}={{(\sqrt{{\vec p}^2+m^2}+m)\eta -\vec p\cdot \vec \sigma \chi}
\over {\sqrt{2 \sqrt{{\vec p}^2+m^2}(\sqrt{{\vec p}^2+m^2}+m)} }}$]
and ${\hat U}_{(FW)}={1\over 
{\sqrt{ 2 \sqrt{{\vec p}^2+m^2}(\sqrt{{\vec p}^2+m^2}+m)} }}
(\beta \hat H+\sqrt{{\vec p}^2+m^2})=\sqrt{
{ {\sqrt{{\vec p}^2+m^2}+m}  \over{2\sqrt{{\vec p}^2+m^2}}}  } \left(
\begin{array}{cc} \openone & {{\vec p\cdot \vec \sigma}\over {\sqrt{{\vec p}^2
+m^2}+m}}\\ {{-\vec p\cdot \vec \sigma}\over {\sqrt{{\vec p}^2+m^2}+m}}&
\openone \end{array} \right)$. In this representation the Poincar\'e 
generators are ${\hat p}^{\mu}_{(FW)}=(\beta \sqrt{{\hat {\vec p}}^2+m^2};
{\hat {\vec p}})$, ${\hat {\vec J}}_{(FW)}={\hat {\vec x}}\times {\hat {\vec 
p}}+{1\over 2}\vec \sigma$, ${\hat J}^{oi}_{(FW)}={\hat x}^o\, {\hat p}^i
-{1\over 2}\beta ({\hat x}^i\sqrt{{\hat {\vec p}}^2+m^2}+\sqrt{{\hat {\vec p}}
^2+m^2}{\hat x}^i)+\beta {{{1\over 2}\vec \sigma \times {\hat {\vec p}}}\over 
{\sqrt{{\hat {\vec p}}^2+m^2}+m}}$;
see table I of Ref.\cite{fw} for the form of the position operator
(suffering zitterbewegung) in the new representation and for the definition of
the mean position (with free motion without zitterbewegung).

Following the notation of Ref.\cite{gros}, the positive energy solutions of the
Dirac equations are connected with the spinors $u(\vec p,s)=\sqrt{p^o+m}
\left( \begin{array}{c} \chi^{(s)}(p) \\ {{\vec p\cdot \vec \sigma}\over 
{p^o+m}}\chi^{(s)}(p) \end{array} \right)$, which under the free 
Foldy-Wouthuysen transformation are sent in the spinors $u_{(FW)}(\vec p,s)= 
\sqrt{2p^o}\left( \begin{array}{c} \chi^{(s)}(p) \\ 0 \end{array} \right)$. The 
2-spinors $\chi^{(s)}(p)$ describe the two levels of spin 1/2. The
negative energy spinors are $v(\vec p,s)=\sqrt{p^o+m} \left( \begin{array}{c}
{{\vec p\cdot \vec \sigma}\over {p^o+m}}(-i\sigma^2 \chi^{(s)})(p) \\
(-i\sigma^2 \chi^{(s)}(p)) \end{array} \right)$ and the antiparticles of 
positive mass, same spin and opposite electric charge are described by the 
charge conjugate spinor $C\gamma^o v^{*}(-\vec p,-s)$; $v(-\vec p,-s)$ is sent
by the free Foldy-Wouthuysen transformation into $\sqrt{2p^o} \left(
\begin{array}{c} 0\\ (-i\sigma^2 \chi^{(-s)}(p)) \end{array} \right)$.
The spin projector in the spacelike direction $s^{\mu}$, $s^2=-1$, is
$P(s)={1\over 2}(1+\gamma_5s_{\mu}\gamma^{\mu})={1\over 2}[1=s_o\gamma_5\gamma
^o+\gamma^o\vec s\cdot \vec \Sigma ]$ with $\vec \sigma =\gamma_5\vec \alpha =
\left( \begin{array}{cc} \vec \sigma &0\\ 0&\vec \sigma \end{array} \right)$
[the rest-frame Pauli-Lubanski four-vector is ${\buildrel \circ \over W}^{\mu}=
L^{\mu}{}_{\nu}({\buildrel \circ \over p},p) W^{\nu}=-{1\over 4}
L^{\mu}{}_{\nu}({\buildrel \circ \over p},p) \epsilon^{\nu \alpha\beta\gamma}
p_{\alpha}\sigma_{\beta\gamma}=-{1\over 2}L^{\mu}{}_{\nu}({\buildrel \circ 
\over p},p) \gamma_5\gamma^{\nu}p_{\alpha}\gamma^{\alpha}= (0; {m\over 2}
\vec \Sigma )$]. Instead the (constant of the motion) helicity operator $h(\vec
p)=\vec p\cdot \vec \Sigma /|\vec p|$ has different eigenstates; if $\Lambda
_{\pm}(p)={{\pm p_{\mu}\gamma^{\mu}+m}\over {2m}}$ is the energy projection
operator and if one chooses $s^{\mu}_p=({{|\vec p|}\over m}; {{p^o}\over m}
{{\vec p}\over {|\vec p|}})$ (helicity basis) one gets the following
relation between spin and helicity states: $P(s_p) \Lambda_{\pm}(p)=[1\pm
h(\vec p)] \Lambda_{\pm}(p)$.

In Ref.\cite{sp3} there is a study of the pseudoclassical Foldy-Wouthuysen
transformation: the generator ${\hat S}_{(FW)}=-i\beta \vec \alpha \cdot \vec p
\theta (\vec p)=-i\vec p\cdot \vec \gamma \theta (\vec p)$, $tg\, 2|\vec p| 
\theta (\vec p)=|\vec p|/m$, ${\hat U}_{(FW)}=e^{{\hat S}_{(FW)}}$ 
of the quantum unitary Foldy-Wouthuysen transformation in the free case 
[realizing the diagonalization of the
Hamiltonian: $\hat H=\vec \alpha \cdot \vec p +\beta m\, \mapsto {\hat H}
_{(FW)}=p_o-\sqrt{{\vec p}^2+m^2}\gamma_o$] corresponds to the
generator $S_{(FW)}=2i \vec p\cdot \vec \xi \xi_5 \theta (\vec p)$ of a
pseudoclassical canonical transformation $e^{{\tilde S}_{(FW)}}f=f+\{ f,S_{(FW)}
\} {}^{*}+{1\over {2!}} \{ \, \{ f,S_{(FW)} \} {}^{*}\, \} {}^{*}+...$.
Therefore, one finds the new canonical basis

\begin{eqnarray}
&&x^o_{(FW)}=e^{{\tilde S}_{(FW)}} x^o=x^o,\nonumber \\
&&x^i_{(FW)}=e^{{\tilde S}_{(FW)}} x^i=x^i-{{ [p^ip_j+\delta^i_j
\sqrt{{\vec p}^2+m^2}(\sqrt{{\vec p}
^2+m^2}+m)](-i\xi^i\xi_5)}\over {({\vec p}^2+m^2)(\sqrt{{\vec p}^2+m^2}+m)}}-
\nonumber \\
&-&{{(-i\xi^i\xi^j)p_j}\over {\sqrt{{\vec p}^2+m^2}(\sqrt{{\vec p}^2+m^2}+m)}},
\nonumber \\
&&p^{\mu}_{(FW)}=e^{{\tilde S}_{(FW)}} p^{\mu}=p^{\mu},\nonumber \\
&&\xi^o_{(FW)}=e^{{\tilde S}_{(FW)}} \xi^o=\xi^o,\nonumber \\
&&\xi^i_{(FW)}=e^{{\tilde S}_{(FW)}} \xi^i=\xi^i+ {{p^i}\over {|\vec p|}}\xi_5
sin\, 2|\vec p|\theta (\vec p)+ {{p^i \vec p\cdot \vec \xi}\over {{\vec p}^2}}
(cos\, 2|\vec p| \theta (p)-1)=\nonumber \\
&&=\xi^i+{{p^i}\over {\sqrt{{\vec p}^2+m^2}}} \xi_5-(1-{m\over {\sqrt{{\vec p}^2
+m^2}}}) {{p^i\vec p\cdot \vec \xi}\over {{\vec p}^2}},\nonumber \\
&&\xi_{5\, (F)W}=e^{{\tilde S}_{(FW)}} \xi_5=cos\, 2|\vec p| \theta (\vec p) 
\xi_5 -sin\, 2|\vec p| \theta (\vec p) {{\vec p\cdot \vec \xi}\over {|\vec p|}}
={{\vec p\cdot \vec \xi +m\xi_5}\over {\sqrt{{\vec p}^2+m^2}}},
\nonumber \\
&&\{ x^{\mu}_{(FW)},p^{\nu} \} {}^{*}=-\eta^{\mu\nu},\quad \{ \xi^{\mu}_{(FW)},
\xi^{\nu}_{(FW)} \} {}^{*}=i\eta^{\mu\nu},\quad \{ \xi_{5\, FW},\xi_{5\, FW} \}
{}^{*}=-i,\nonumber \\
&&{}\nonumber \\
&&e^{{\tilde S}_{(FW)}}(p_{\mu}\xi^{\mu}-m\xi_5)=p_{\mu}\xi^{\mu}_{(FW)}-m\xi
_{5\, (FW)}=p_o\xi^o -\sqrt{{\vec p}^2+m^2} \xi_5 \nonumber \\
&&\mapsto \gamma_5\gamma_o(p_o-\sqrt{{\vec p}^2+m^2}\gamma_o),\nonumber \\
&&x^{\mu}_{M(FW)}=e^{{\tilde S}_{(FW)}}x^{\mu}_M=x^{\mu},\quad\quad
S^{ij}_{M(FW)}=e^{{\tilde S}_{(FW)}}S^{ij}_M=S^{ij}.
\label{II5}
\end{eqnarray}

In quantization the Grassmann variables $\xi^{\mu}_{(FW)}$, $\xi_{5 (FW)}$,
should be replaced by the standard expression with the Dirac matrices in
the Foldy-Wouthuysen representation $\gamma_{(FW)}={\hat U}_{(FW)} \gamma 
{\hat U}_{(FW)}^{-1}$ [$\gamma^o_{(FW)}=\gamma^o {{m-\vec p\cdot \vec \gamma}
\over {p^o}}$, ${\vec \alpha}_{(FW)}=\gamma^o_{(FW)}{\vec \gamma}_{(FW)}=\gamma
^o(\vec \gamma -{{\vec p \vec p\cdot \vec \gamma}\over {p^o(p^o+m)}}+{{\vec p}
\over {p^o}})$, $\gamma_{5 (FW)}=\gamma_5 {{m-\vec p\cdot \vec \gamma}\over 
{p^o}}$, with $p^o=\sqrt{{\vec p}^2+m^2}$].

Instead the Cini-Touschek transformation \cite{cini} antidiagonalizes $\hat H$,
sending it into ${\hat H}_{(CT)}=p^o {{\vec p\cdot \vec \alpha}\over 
{|\vec p|}}=p^o \left( \begin{array}{cc} 0& {{\vec p\cdot \vec \sigma}\over 
{|\vec p|}}\\  {{\vec p\cdot \vec \sigma}\over {|\vec p|}}& 0 \end{array}
\right)$.  It is generated by ${\hat S}_{(CT)}=-
{i\over {2m}}\beta \vec \alpha \cdot \vec p w(\vec p)$, $cotg\, {{|\vec p|}
\over m} w(\vec p)=-{{|\vec p|}\over m}$, ${\hat U}_{(CT)}=e^{{\hat S}_{(CT)}}=
{1\over {\sqrt{2p^o}}}(\sqrt{p^o+|\vec p|}+\sqrt{p^o-|\vec p|}{{\vec p\cdot 
\vec \gamma}\over {|\vec p|}}=\sqrt{ {{p^o+|\vec p|}\over {2p^o}} } \left(
\begin{array}{cc} 1& {{p^o-|\vec p|}\over m}{{\vec p\cdot \vec \sigma}\over
{|\vec p|}}\\ -{{p^o-|\vec p|}\over m}{{\vec p\cdot \vec \sigma}\over
{|\vec p|}} &1 \end{array} \right)$, $\psi_{(CT)}=\left( \begin{array}{c}
\chi_{(CT)}\\ \eta_{(CT)} \end{array} \right)={\hat U}_{(CT)}\psi =
\sqrt{ {{p^o+|\vec p|}\over {2p^o}} } \left( \begin{array}{c} \chi +{{p^o-|\vec
p|}\over m}{{\vec p\cdot \vec \sigma}\over {|\vec p|}}\eta \\ \eta -{{p^o-
|\vec p|}\over m}{{\vec p\cdot \vec \sigma}\over {|\vec p|}} \chi \end{array}
\right)$,
and its pseudoclassical generator is $S_{(CT)}={i\over m}\vec p\cdot 
\vec \xi \xi_5 w(\vec p)$. It could be used to find a new pseudoclassical
basis $x^{\mu}_{(CT)}$, $p^{\mu}$, $\xi^{\mu}_{(CT)}$, $\xi_{5\, (CT)}$.

The chiral-Weyl representation of $\gamma$ matrices (which is systematically 
used in Ref.\cite{scheck,peskin} instead of the Dirac representation even in 
the massive case) is done with the unitary transformation ${\hat U}_{(chi)}=
e^{-{{\pi}\over 4}\gamma_5\gamma_o}={1\over {\sqrt{2}}}
(1-\gamma_5\gamma_o)={1\over {\sqrt{2}}} \left( \begin{array}{cc} 1&-1\\ 1&1
\end{array} \right)$, so that $\psi_{(chi)}=\left( \begin{array}{c} \chi_L \\
\eta_R \end{array} \right) ={\hat U}_{(chi)} \psi ={1\over {\sqrt{2}}}
\left( \begin{array}{c} \chi -\eta \\ \chi+\eta \end{array} \right)$, where
$\chi_L= (\phi_a) \in ({1\over 2},0)$, $\eta_R=(\phi^{\dot a})\in (0,{1\over 
2})$ are the left and right components belonging to the corresponding 
representations of SL(2,C) \cite{ramond,scheck} . The components of $\psi$,
the $({1\over 2},{1\over 2})$ representation, are $\chi ={1\over {\sqrt{2}}}
(\eta_R+\chi_L)$, $\eta ={1\over {\sqrt{2}}}(\eta_R-\chi_L)$. The Hamiltonian
becomes ${\hat H}_{(chi)}={\hat U}_{(chi)}\hat H{\hat U}^{\dagger}_{(chi)}=
\left( \begin{array}{cc} -\vec p\cdot \vec \sigma &m\\ m& \vec p\cdot \vec
\sigma \end{array} \right)=\gamma_5(\vec p\cdot \vec \gamma +m)$.
${\hat H}_{(chi)}$ may be antidiagonalized to 
$-{m\over {|\vec p|}} \sqrt{{\vec p}^2+m^2} \left( \begin{array}{cc}
0& 1\\ 1&0 \end{array} \right)= {m\over {|\vec p|}} \sqrt{{\vec p}^2+m^2}
\gamma^o_{(chi)}$ with $\hat U=\sqrt{ {{p^o+m}\over 
{2p^o}}} \left( \begin{array}{cc} 1& \sqrt{ {{p^o-m}\over {p^o+m}} }
{{\vec p\cdot
\vec \sigma}\over {|\vec p|}}\\ -\sqrt{ {{p^o-m}\over {p^o+m}} }{{\vec p\cdot
\vec \sigma}\over {|\vec p|}} &1 \end{array} \right)$ and diagonalized to
$\sqrt{{\vec p}^2+m^2} \left( \begin{array}{cc}
-{{\vec p\cdot \vec \sigma}\over {|\vec p|}}& 0\\ 0& {{\vec p\cdot \vec \sigma}
\over {|\vec p|}} \end{array} \right)$ with 
$\hat U=\sqrt{ {{|\vec p|+\sqrt{{\vec p}^2+m^2}}\over {2\sqrt{{\vec p}^2+m^2}}
} } \left( \begin{array}{cc} 1& -{{\vec p\cdot \vec \sigma}\over {|\vec p|}}
\sqrt{ {{\sqrt{{\vec p}^2+m^2}-|\vec p|}\over {\sqrt{{\vec p}^2+m^2}+|\vec p|}
}} \\ {{\vec p\cdot \vec \sigma}\over {|\vec p|}}
\sqrt{ {{\sqrt{{\vec p}^2+m^2}-|\vec p|}\over {\sqrt{{\vec p}^2+m^2}+|\vec p|}
}} & 1 \end{array} \right)$.
Since $\gamma_5\gamma_o=-i\gamma_1\gamma_2\gamma_3=
i\gamma_5(\gamma_5\gamma_1)(\gamma_5\gamma_2)(\gamma_5\gamma_3)$, the
pseudoclassical generator is $S_{(chi)}=\pi \xi_5\xi_1\xi_2\xi_3$ and the
pseudoclassical chiral basis is $x^{\mu}_{(chi)}=x^{\mu}$, $p^{\mu}$,
$\xi^{o}_{(chi)}=\xi^o$, $\xi^i_{(chi)}=\xi^i+i{{\pi}\over 2} \epsilon^{ijk}
\xi_5\xi^i\xi^j$, $\xi_{5\, (chi)}=\xi_5-i\pi \xi_1\xi_2\xi_3$.

\vfill\eject

\end{document}